\PassOptionsToPackage{unicode}{hyperref}
\PassOptionsToPackage{hyphens}{url}
\PassOptionsToPackage{dvipsnames,svgnames,x11names}{xcolor}
\documentclass[
  12pt]{article}

\usepackage{amsmath,amssymb}
\usepackage{amsthm}
\usepackage{iftex}
\usepackage{upgreek}
\ifPDFTeX
  \usepackage[T1]{fontenc}
  \usepackage[utf8]{inputenc}
  \usepackage{textcomp} 
\else 
  \usepackage{unicode-math}
  \defaultfontfeatures{Scale=MatchLowercase}
  \defaultfontfeatures[\rmfamily]{Ligatures=TeX,Scale=1}
\fi
\usepackage{lmodern}
\usepackage{verbatim}
\usepackage{booktabs}
\usepackage{multirow}
\usepackage{makecell}
\usepackage{threeparttable}
\usepackage{diagbox}
\usepackage{array}
\usepackage{siunitx}
\usepackage{rotating}
\usepackage{tikz}
\usepackage{arydshln}
\ifPDFTeX\else  
\fi
\IfFileExists{upquote.sty}{\usepackage{upquote}}{}
\IfFileExists{microtype.sty}{
  \usepackage[]{microtype}
  \UseMicrotypeSet[protrusion]{basicmath} 
}{}
\makeatletter
\@ifundefined{KOMAClassName}{
  \IfFileExists{parskip.sty}{%
    \usepackage{parskip}
  }{
    \setlength{\parindent}{0pt}
    \setlength{\parskip}{6pt plus 2pt minus 1pt}}
}{
  \KOMAoptions{parskip=half}}
\makeatother
\usepackage{xcolor}
\makeatletter
\ifx\paragraph\undefined\else
  \let\oldparagraph\paragraph
  \renewcommand{\paragraph}{
    \@ifstar
      \xxxParagraphStar
      \xxxParagraphNoStar
  }
  \newcommand{\xxxParagraphStar}[1]{\oldparagraph*{#1}\mbox{}}
  \newcommand{\xxxParagraphNoStar}[1]{\oldparagraph{#1}\mbox{}}
\fi
\ifx\subparagraph\undefined\else
  \let\oldsubparagraph\subparagraph
  \renewcommand{\subparagraph}{
    \@ifstar
      \xxxSubParagraphStar
      \xxxSubParagraphNoStar
  }
  \newcommand{\xxxSubParagraphStar}[1]{\oldsubparagraph*{#1}\mbox{}}
  \newcommand{\xxxSubParagraphNoStar}[1]{\oldsubparagraph{#1}\mbox{}}
\fi

\makeatother

\usepackage{longtable,booktabs,array}
\usepackage{calc} 
\usepackage{etoolbox}
\makeatletter
\patchcmd\longtable{\par}{\if@noskipsec\mbox{}\fi\par}{}{}
\makeatother
\IfFileExists{footnotehyper.sty}{\usepackage{footnotehyper}}{\usepackage{footnote}}
\makesavenoteenv{longtable}
\usepackage{graphicx}
\makeatletter
\def\maxwidth{\ifdim\Gin@nat@width>\linewidth\linewidth\else\Gin@nat@width\fi}
\def\maxheight{\ifdim\Gin@nat@height>\textheight\textheight\else\Gin@nat@height\fi}
\makeatother
\setkeys{Gin}{width=\maxwidth,height=\maxheight,keepaspectratio}
\makeatletter
\def\fps@figure{htbp}
\makeatother

\makeatletter
\@ifpackageloaded{caption}{}{\usepackage{caption}}
\AtBeginDocument{%
\ifdefined\contentsname
  \renewcommand*\contentsname{Table of contents}
\else
  \newcommand\contentsname{Table of contents}
\fi
\ifdefined\listfigurename
  \renewcommand*\listfigurename{List of Figures}
\else
  \newcommand\listfigurename{List of Figures}
\fi
\ifdefined\listtablename
  \renewcommand*\listtablename{List of Tables}
\else
  \newcommand\listtablename{List of Tables}
\fi
\ifdefined\figurename
  \renewcommand*\figurename{Figure}
\else
  \newcommand\figurename{Figure}
\fi
\ifdefined\tablename
  \renewcommand*\tablename{Table}
\else
  \newcommand\tablename{Table}
\fi
}
\@ifpackageloaded{float}{}{\usepackage{float}}
\floatstyle{ruled}
\@ifundefined{c@chapter}{\newfloat{codelisting}{h}{lop}}{\newfloat{codelisting}{h}{lop}[chapter]}
\floatname{codelisting}{Listing}

\makeatother
\makeatletter
\@ifpackageloaded{caption}{}{\usepackage{caption}}
\@ifpackageloaded{subcaption}{}{\usepackage{subcaption}}
\makeatother

\ifLuaTeX
  \usepackage{selnolig}  
\fi
\usepackage[]{natbib}
\usepackage{bookmark}

\IfFileExists{xurl.sty}{\usepackage{xurl}}{} 
\hypersetup{
  pdftitle={Title},
  pdfauthor={Author 1; Author 2},
  pdfkeywords={3 to 6 keywords, that do not appear in the title},
  colorlinks=true,
  linkcolor={blue},
  filecolor={Maroon},
  citecolor={Blue},
  urlcolor={Blue},
  pdfcreator={LaTeX via pandoc}}

\def\P{{ \mathrm{\scriptscriptstyle P} }}
\def\S{{ \mathrm{\scriptscriptstyle S} }}
\def\v{{\varepsilon}}

\DeclareMathOperator{\var}{Var} 
\DeclareMathOperator*{\E}{\mathbb{E}} 
\DeclareMathOperator{\cov}{Cov} 
\DeclareMathOperator*{\R}{\mathbb{R}}

\newtheorem{proposition}{Proposition}

\newtheorem{corollary}{Corollary}
\newtheorem{property}{Property}

\theoremstyle{definition}
\newtheorem{definition}{Definition}

\theoremstyle{remark}
\newtheorem{remark}{Remark}

\newcommand{\anon}{1}

\begin{document}

\def\spacingset#1{\renewcommand{\baselinestretch}%
{#1}\small\normalsize} \spacingset{1}


\if1\anon
{
  \title{\bf Dynamic cross-scale wavelet coherence}
  \author{Haibo Wu \thanks{Correspondence: haibo.wu@kaust.edu.sa}\hspace{.2cm}\\
    Statistics Program, King Abdullah University of Science and Technology\\
    and \\
    Marina I. Knight \hspace{.2cm}\\
    Department of Mathematics, University of York \\and \\ Hernando Ombao \hspace{.2cm}\\
    Statistics Program, King Abdullah University of Science and Technology}
  \maketitle
} \fi

\if0\anon
{
  \bigskip
  \bigskip
  \bigskip
  \begin{center}
    {\LARGE\bf Title}
\end{center}
  \medskip
} \fi

\bigskip
\begin{abstract}
This paper develops a novel statistical approach that allows for the {\em first time} the {\em cross}-oscillatory characterisation of temporally localised interactions between channels in a functional brain network. Brain signals are often nonstationary and the proposed framework uses wavelets as an effective tool for capturing (i) single-scale channel transient features, due to their adaptiveness to the dynamic signal properties, and (ii) cross-scale channel interactions, due to their multiscale nature. Our approach introduces scale-specific {\em subprocesses} and {\em cross-scale (CS) dependencies} for a new class of multivariate locally stationary (MvLSW) wavelet processes that we refer to as CS-MvLSW. Under this new model, we develop two consistent estimation procedures for the {\em localised} single- and cross-scale channel dependence. 
Extensive simulation studies demonstrate that the theoretically established properties hold in practice. The proposed CS-MvLSW framework remains accurate under pronounced cross-scale dependence, whereas existing MvLSW coherence estimates dramatically deteriorate even for single-scales when such complex structure is present. The proposed approach was used for electroencephalogram (EEG) data to study alterations in the functional connectivity structure in children diagnosed with attention deficit hyperactivity disorder (ADHD), and identified novel clinically pertinent cross-scale interactions in the functional brain network across the left and right hemispheres, differentiating brain connectivity between control and ADHD groups.
\end{abstract}

\noindent%
\vfill

\newpage
\spacingset{1.8} 

\section{Introduction}
\label{s:intro}

This paper is motivated by the prevailing interest in the neuroscience community to identify alterations in brain functional networks due to attention deficit hyperactivity disorder (ADHD). We focus on functional connectivity derived from electroencephalograms (EEGs), which are recordings from the scalp that capture cortical brain electrical activity. Functional networks derived from multichannel EEGs are often characterized by dependence measures including (partial) cross-correlation, cross-coherence, spectral dependence \citep{OmbaoPinto2024}, mutual information and transfer entropy \citep{RedondoHuserOmbao2025}. Such measures are of substantial interest because functional brain connectivity has been studied as a potential biomarker for neurological diseases, such as epilepsy and Alzheimer disease, and mental disorders, such as depression, obsessive-compulsive disorder and ADHD \citep{cribben2016estimating}. However, analyzing EEGs is challenging because of their inherent nonstationarity \citep{knight24:jcgs}. In particular, their statistical properties, such as spectrum, covariance, coherence and correlation, may evolve over time \citep{ombao2005}. This motivates an approach that can naturally adapt to nonstationarity and extract complex dependence structures that are currently not accessible with existing approaches.

Inspired by \cite{nason2000wavelet} and \cite{ombao2014}, we develop a stochastic representation of multichannel EEGs, denoted by $\{\mathbf{X}_{t;T}\}$, that uses wavelets as building blocks. Let $\{\psi_{j,k}(t)\}$ be a collection of wavelet functions defined over multiple scales $j$ and time shifts $k$; see Sections~\ref{wavelets} and~\ref{s:brief_LSW} for their formal introduction. The representation of the multichannel EEGs in the current framework is
\begin{eqnarray}\label{Eq:MvLSWModel}
\mathbf{X}_{t;T}
=
\sum_{j=1}^{\infty}
\sum_{k \in \mathbb{Z}}
\mathbf{V}_{j}(k/T)\psi_{j,k}(t)\mathbf{z}_{j,k},
\end{eqnarray}
where $\mathbf{V}_{j}(k/T)\mathbf{z}_{j,k}$ is a random coefficient matrix corresponding to the specific wavelet $\psi_{j,k}(t)$. From this representation, \cite{ombao2014} developed scale-specific time-varying coherence and partial coherence. Intuitively, coherence between two channels $X_t^{(p)}$ and $X_t^{(q)}$ at scale $j$ measures the local dependence between their scale-$j$ components, denoted by $X_{j,t}^{(p)}$ and $X_{j,t}^{(q)}$. Thus, the existing multivariate locally stationary wavelet (MvLSW) framework provides a natural way to quantify time-varying \underline{same-scale} dependence between multichannel nonstationary signals.

The key limitation of the model in equation~\eqref{Eq:MvLSWModel} is that it can capture only same-scale dependence between component waveforms. For example, it can quantify coherence, or synchronicity, between fine-resolution millisecond activity in a pair of channels, such as the left prefrontal and right occipital channels, and it can also quantify coherence between their coarse-resolution long-term activity. However, it \underline{cannot} provide coherence between millisecond-scale activity in the left prefrontal channel and long-term-scale activity in the right occipital channel. This is a serious limitation because cross-frequency coupling (CFC) is known to link dispersed cortical regions during the execution of cognitive tasks including working memory, attention and sensory processing \citep{Canolty2010,Siebenhuhner2016}. CFC is also known as a potential biomarker for pathophysiology, as abnormal coupling has been linked to Parkinson's disease, where increased beta-gamma phase-amplitude coupling has been observed in motor regions \citep{deHemptinne2013,Tanaka2022}.

The inability of equation~\eqref{Eq:MvLSWModel} to capture such cross-scale interactions is a direct consequence of the standard orthogonality assumption in the MvLSW representation. Specifically, the random components $\{\mathbf{z}_{j,k}\}$ are assumed to be uncorrelated across scales $j$ and temporal shifts $k$. Hence, under equation~\eqref{Eq:MvLSWModel},
$\cov(X_{j,t}^{(p)},X_{j',t'}^{(q)})=0$
for different scales $j\ne j'$ and time points $t, \, t'$. The goal of this paper is to remove this stringent orthogonality condition and propose a new model that for the {\em first time allows for cross-scale dependence} between the components of multichannel nonstationary time series.

There has been a long history of statistical models for nonstationary time series. To generalize the Cram\`er representation, \cite{Priestley} and \cite{Dahlaus} proposed a linear mixture of Fourier waveforms with random amplitudes that vary across time. In \cite{ombao2005}, a model based on the library of smooth localised complex exponentials (SLEX) was proposed. As an alternative to the Fourier representation of signals, \cite{nason2000wavelet} introduced a stochastic process that uses discrete wavelets as its building blocks. Under the locally stationary wavelet model, the univariate evolutionary wavelet spectrum was introduced, most recently for continuous-time processes \citep{palasciano2025continuous}. A special framework in \cite{sanderson} was developed to estimate wavelet coherence for bivariate nonstationary time series. In \cite{ombao2014}, the multivariate locally stationary wavelet process was developed and new cross-channel dependence measures, including coherence and partial coherence, were introduced. The framework in \cite{fiecas2016modeling} extends the Dahlhaus model to multiple trials, while assuming that the signals are uncorrelated across trials. The multiple-trials locally stationary wavelet alternative was developed in \cite{embleton2022multiscale} and additionally allowed for dependence across the timeline of ordered replicates. These Fourier and wavelet stochastic representations share a common feature: the random coefficients are uncorrelated across frequencies, in the Fourier case, or across scales and time shifts, in the wavelet case. This poses a serious limitation when the objective is to study dependence between different scales, namely short-term and long-term process dynamics.

\begin{figure}[htbp]
    \centering
    \includegraphics[width=\textwidth]{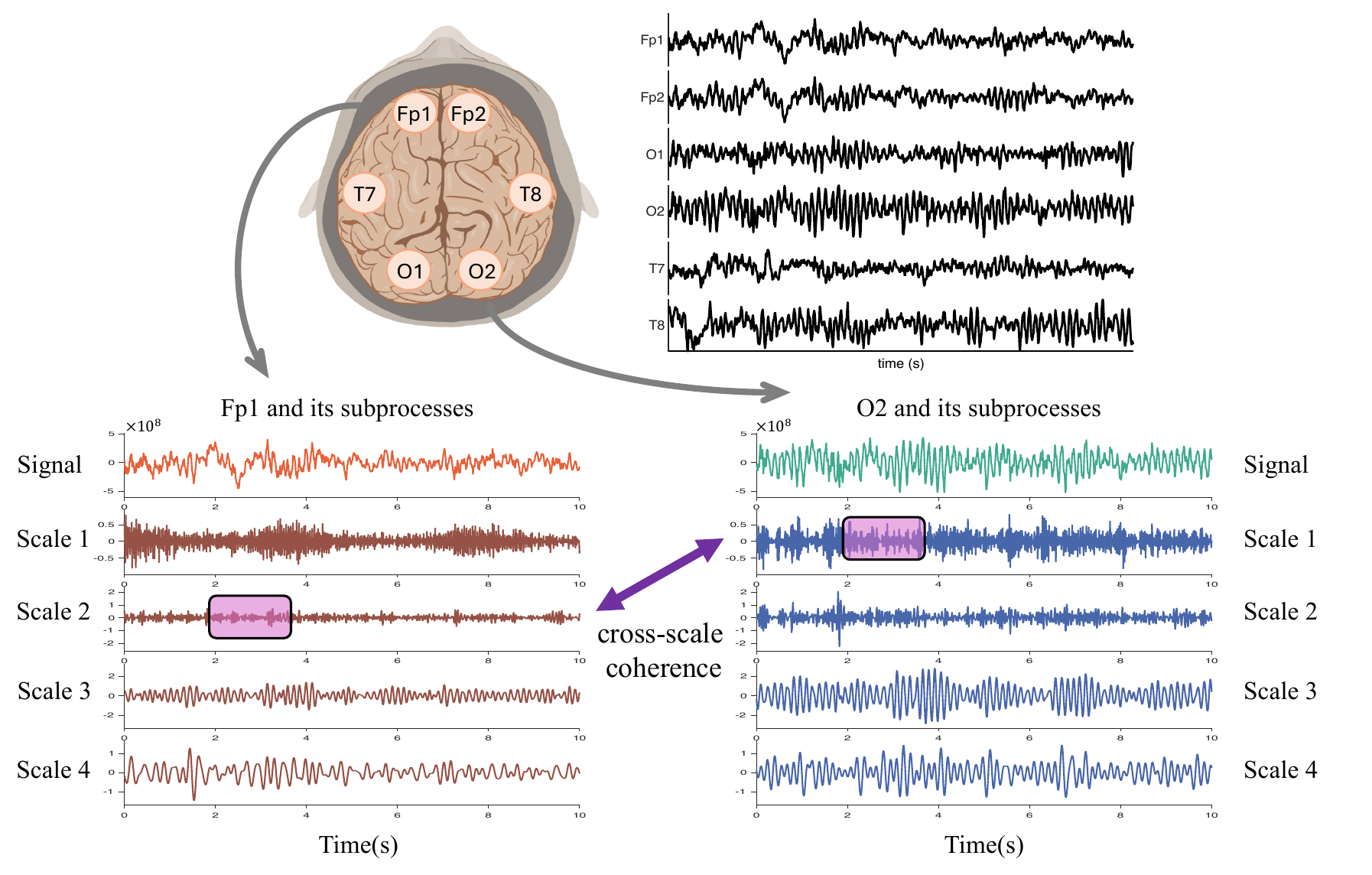}
\caption{
Top: EEG scalp topography and traces at six channels (left and right):  prefrontal (Fp1 and Fp2), temporal (T7 and T8), and occipital (O1 and O2). Bottom: wavelet subprocess decomposition of the EEG signals at the Fp1 and O2 channels. (Section~\ref{EEG analysis} identifies time-varying cross-scale coherence between scale-2 Fp1 and scale-1 O2 signals.)}
    \label{fig:eegplot}
\end{figure}

One contribution of this paper is the decomposition of a multichannel nonstationary wavelet process into scale-specific stochastic processes, which we refer to as \underline{subprocesses}. Under our proposed framework, the time series $\{X_t^{(p)}\}_t$ is decomposed into subprocesses $\{X_{j,t}^{(p)}\}_{j,t}$, so that
$X_t^{(p)}=\sum_{j=1}^{\infty} X_{j,t}^{(p)}$.
Here, the scale-1 subprocess $X_{1,t}^{(p)}$ captures high-frequency, finer features of the time series, while the larger scale-$j$ subprocess captures low-frequency, coarser features. This subprocess representation is central to our construction as it allows us to examine multiscale dependence across different components of multichannel EEG signals and to identify how fluctuations in longer-term dynamics in one channel may impact the amplitude of shorter-term dynamics, or vice versa, in another channel. Thus, the proposed framework establishes the concept of \underline{evolving cross-scale dependence} in multivariate time series (see Figure \ref{fig:eegplot}).

Relaxing the orthogonality constraint to allow dependence between $X_{j,t}^{(p)}$ and $X_{j',t}^{(q)}$ at scales $j\ne j'$ may appear to be a simple extension, but it presents several theoretical challenges. Once cross-scale orthogonality is removed, the usual scale-specific spectra and covariances are no longer sufficient to characterize the dependence structure. It is therefore necessary to provide rigorous definitions of the dual-scale local wavelet spectrum, dual-scale local cross-covariance and coherency. Moreover, because the target dependence structure is enlarged, one has to overcome the challenge of building mean-square consistent estimators capable to extract fundamentally more information from the same amount of data than currently existing same-scale estimators. 

The proposed framework therefore provides both a new model and new estimation procedures for cross-scale dependence, which have not been realized by previous approaches. By measuring cross-scale dependence structures between different brain channels, the method provides new inference about localised interactions among brain regions. 
This is useful for studying associations between brain connectivity and cognitive function, including the role of altered brain connectivity in the development and progression of neurological and mental disorders. Although the main motivation is EEG-based brain connectivity, our proposed model can be used for time series data from other fields. For example, systemic risk in financial systems has received substantial attention \citep{stock,cen2025inference}, and the proposed framework provides a new approach to identifying inter-connectedness  among stocks and the time scales that contain crucial information. 

The format of the paper is as follows. Sections~\ref{wavelets} and~\ref{s:brief_LSW} give a brief introduction to wavelets and  multivariate locally stationary wavelet processes. Sections~\ref{scale-specific} and~\ref{coherence} introduce the new scale-specific multivariate locally stationary wavelet process with cross-scale dependence structure and define the associated coherence measures. Estimation theory is addressed in Section~\ref{sec: estimation}, with results validated through the simulation studies in Section~\ref{simulation study} and illustrating the practical advantages of the proposed modeling framework. Section~\ref{EEG analysis} investigates the EEG-based findings of the proposed methodology for differentiating brain connectivity in children with ADHD from that in a healthy control group. Section~\ref{s:discuss} concludes the paper.

\section{Novel scale-specific nonstationary subprocesses}\label{s:model}

In this section, we propose a framework for measuring cross-dependence among channels of multivariate nonstationary time series at different resolution scales, and for capturing the frequency information of the subprocesses at the scales responsible for high dependence. Multiresolution analysis (see e.g., \cite{daubechies1992ten}) provides the theoretical foundation for deriving the subprocesses at every resolution scale. These ideas suggest that a time series $X(t)$ can be represented as a sum of `smooth' and `detail' random process components. This construction will be examined in detail in the following subsections.

\subsection{Brief introduction of wavelets}
\label{wavelets}
Fourier analysis is a fundamental tool in studying stationary time series because it identifies the frequencies of random oscillations that dominate the signal. 
However, due to the global support of Fourier waveforms, it cannot be directly used to study signals whose oscillatory content changes with time. For nonstationary signals, the alternative is to consider wavelets, which are known to have good time-localisation properties. Wavelets are generated from some special functions, one typically referred to as the father wavelet (denoted $\phi$) which integrates to one and which is used to capture the smooth, low frequency nature of the time series; the other is the mother wavelet (denoted $\psi$) which integrates to zero and is used to capture the detailed, high frequency information of the data. Intuitively, the mother wavelets are compressed or dilated, and then shifted to produce `children' wavelets. Formally, the wavelet functions at scale $a\in \R^\star$ and shift $b\in \R$, denoted by $\psi_{a,b}$, are generated from the mother wavelet, $\psi$, and are defined as,
\begin{align*}
    \psi_{a,b}(t) = |a|^{-1/2} \psi\left(\frac{t-b}{a}\right). 
\end{align*}
Typical choices generating wavelet functions $\psi_{j,k}$ may be $a=2^{-j}, \, b=2^{-j}k$ for decimated wavelets, and $a=2^{-j}, \, b=k$ for non-decimated wavelets at signal resolution $j$. Similar constructions for the father wavelet, denoted by $\phi_{j,k}$, are typically known as scaling functions. Further properties, such as orthonormality are discussed in  \cite{daubechies1992ten}.

\subsection{Summary of locally stationary wavelet (LSW) processes}
\label{s:brief_LSW}
First, a brief description of key features of the \textit{locally stationary wavelet} (LSW) framework is presented. Suppose that $X(t)$ is a univariate nonstationary time series of length $T=2^J$. A basic formulation for its orthonormal wavelet decomposition in the notation above is 
\begin{align*}
   X(t) = \sum\limits_{k \in \mathbb{Z}} s_{J,k}\phi_{J,k}(t)+ \sum\limits_{k \in \mathbb{Z}} d_{J,k} \psi_{J,k}(t)
    + \sum\limits_{k \in \mathbb{Z}} d_{J-1,k} \psi_{J-1,k}(t) + \ldots + \sum\limits_{k \in \mathbb{Z}} d_{1,k} \psi_{1,k}(t)
\end{align*}
where $s_{J,k}$ and $d_{j,k}$ are the coefficients of father wavelet and mother wavelet respectively at the corresponding scales and shifts. The LSW model proposed by \cite{nason2000wavelet} gives a discrete non-decimated wavelets representation of a discretely sampled nonstationary time series with time-varying second order structures, where the non-decimated wavelet is shift-invariant since it eliminates the downsampling and consequently, is more appropriate for identifying both stationary and nonstationary behaviors in signals \citep{TCAM1070}. This framework provides a time- and scale-localised wavelet spectrum which is estimated using a wavelet periodogram.
Specifically, a sequence of (doubly-indexed) stochastic processes, $\{X_{t;T}\}_{t=1,\ldots,T}$, is defined to be an LSW process if it has the following representation
\begin{align*}
X_{t ; T}=\sum_{j=1}^{\infty} \sum_{k \in \mathbb{Z}} W_j(k/T) \psi_{j, k}(t) \xi_{j, k}
\end{align*}
where for scale $j$ and shift $k$, $W_j(k/T)$ is a smoothly varying transfer function corresponding to the discrete non-decimated wavelet $\psi_{j,k}(t)$  with $\psi_{j,k}(t)=\psi_{j,k-t}$, and $\{\xi_{j,k}\}$ are a collection of zero-mean, unit-variance uncorrelated random variables. Here, $W_j(\cdot)$ are assumed to be  smoothly varying continuous Lipschitz functions ensuring that transfer functions can be estimated locally. The transfer function $W_j(\cdotp)$ measures the time-varying contribution to the variance of the overall process. Consequently, the evolutionary wavelet spectrum is defined as $S_j(u) = \left| W_j(u) \right|^2$, to describe the power at given scale $j$ and rescaled time points $u$, where $u=k/T \in (0,1)$. The multivariate LSW (MvLSW) process of \cite{ombao2014} is based on the above original single-replicate formulation and allows us to estimate the dependence -- {\em at the same scale} -- between time series data recorded from different channels. The $P$-dimensional MvLSW process $\{\mathbf{X}_{t;T}\}_{t=1,\ldots,T}$ with $\mathbf{X}_{t;T}=[X_{t;T}^{(1)},X_{t;T}^{(2)},\ldots,X_{t;T}^{(P)}]^{\top}$ is
\begin{align*}
                \mathbf{X}_{t;T}=\sum_{j=1}^{\infty} \sum_{k \in \mathbb{Z}} \mathbf{V}_{j}(k / T) \psi_{j, k}(t) \mathbf{z}_{j, k} 
\end{align*}
where $^\top$ denotes matrix transposition; $\mathbf{V}_j(k/T)$ is a $P \times P$ transfer function matrix assumed to have a lower-triangular form; $\{\psi_{j,k}\}_{j,k}$ is a set of discrete non-decimated wavelets of \cite{nason2000wavelet}; $\{\mathbf{z}_{j,k}\}_{j,k}$ is a set of uncorrelated random vectors with (column) mean vector $\boldsymbol{0}$ and $P \times P$ identity covariance matrix. The covariance of these random innovations in the MvLSW model is defined as $\cov(z_{j,k}^{(i)},z_{j',k'}^{(i')})= \delta_{i,i'}\delta_{j,j'} \delta_{k,k'}$, which means that the uncorrelation holds not only across different times $k \neq k'$, but also at different scales $j \neq j'$ and channels $i \neq i'$. In particular, if $j=j'$, $i=i'$ and $k=k'$, then the covariance of the random innovations is the identity matrix. In this work, we abandon these strong assumptions and introduce a dependence function between the innovations that will enable us to capture the dependence between channels at \underline{different} scales, as next discussed.

\subsection{Novel cross-scale multivariate LSW process and subprocesses}\label{scale-specific}

A significant limitation of existing models for multivariate nonstationary time series is the assumption that the random innovations at different scales are uncorrelated. The major \underline{novelty} of our framework is to provide a measure for potential cross-scale dependence structure between  channels, although the cross-scale dependence could be naturally weak in many cases. Interchangeably, we refer to cross-scale activity as dual-scale.

\noindent 
\begin{definition}\label{def1}
A $P$-variate stochastic time series, $\{\mathbf{X}_{t;T}\}_{t=1,\ldots,T}$, where $T=2^{J}$ is defined to be a {\em multivariate locally stationary wavelet process with cross-scale dependence (CS-MvLSW process)} if it has the representation,
\begin{align} \label{(3)}
    \mathbf{X}_{t;T}=\sum\limits_{j=1}^{\infty}\sum\limits_{k \in \mathbb{Z}} \mathbf{V}_{j}(k/T)\psi_{j,k}(t) \mathbf{z}_{j,k}
\end{align}
where $\{\psi_{j,k}\}_{j,k}$ is a set of discrete non-decimated wavelets; $\mathbf{V}_j(k/T)$ is the scale-specific and time-varying $P \times P$ transfer function matrix, with lower triangular structure and such that each of its elements is a Lipschitz continuous function; $\{\mathbf{z}_{j,k}\}=\{[z_{j,k}^{(1)},\ldots,z_{j,k}^{(P)}]^{\top}\}$ is a collection of $P \times 1$ random innovation vectors that satisfy (a)
 for all $j$, $k$, $\E[\mathbf{z}_{j,k}]=\boldsymbol{0}$ and $\var(\mathbf{z}_{j,k})=\boldsymbol{I}$, the $P \times P$ identity matrix; (b) the cross-scale covariance $ \cov(\mathbf{z}_{j,k},\mathbf{z}_{j',k'})= \delta_{k,k'} \boldsymbol{Q}_{j,j'}(k/T)$ is determined by the $P \times P$ matrices $\{\boldsymbol{Q}_{j,j'}(k/T)\}_k$ at scales $j$ and $j'$, across times $k$. The cross-scale dependence function satisfies  $Q^{(p,q)}_{j,j'}(k/T) = Q^{(q,p)}_{j',j}(k/T)$ and $\left| Q_{j,j'}^{(p,q)}(k/T) \right| \leq 1$ for all $j$, $j'$, $k$, $(p,q)$.
 The special case of uncorrelated scales is given by $\boldsymbol{Q}_{j,j'}(k/T)=\delta_{j,j'}\boldsymbol{I}$, and non-identity cross-scale structure is assumed to exist at scales satisfying $h=|j-j'|<\infty$.
\end{definition}

\noindent  
\begin{remark}
Although Definition~\ref{def1} is valid for any discrete non-decimated wavelets, note that we employ the Haar family across our entire development below. 
\end{remark}

\noindent 
\begin{definition} \label{def2}
For the CS-MvLSW process in \eqref{(3)}, the \underline{scale-$j$ subprocess} with cross-scale dependence is defined to be the $P$-variate locally stationary wavelet process, $\mathbf{X}_{j,t}=[X_{j,t}^{(1)},\ldots,X_{j,t}^{(P)}]^\top$, with the following representation,
\begin{align*} 
    \mathbf{X}_{j,t}= \sum_{k \in \mathbb{Z}} \mathbf{V}_j(k/T) \psi_{j,k}(t) \mathbf{z}_{j,k}
\end{align*} 
where $\mathbf{V}_j(k/T), \{\psi_{j,k}\}$ and $\{\mathbf{z}_{j,k}\}$ follow Definition~\ref{def1}. Here, $\{X_{j,t}^{(p)}\}$ gives the scale-$j$ contribution to the original channel $p$ process, $\{X_{t;T}^{(p)}\}$, and from equation~\eqref{(3)} we have $X_{t;T}^{(p)} = \sum\limits_{j=1}^{\infty} X_{j,t}^{(p)}$. 
\end{definition}

The subprocesses correspond to single- and cross-scale spectral structures introduced next.

\begin{definition} \label{def3}
   Let $\{\mathbf{X}_{t;T}\}$ be a CS-MvLSW process with cross-scale dependence structure as in Definition 1. Suppose that $\mathbf{V}_j(u)$ and $\mathbf{V}_{j'}(u)$ are scale-specific time-dependent ($u=k/T$) transfer function matrices corresponding to scale-$j$ and scale-$j'$ components of $\mathbf{X}_{t;T}$ respectively. Let $\mathbf{S}_{jj'}(u)$ denote the $P \times P$ \textit{cross-scale local wavelet spectral matrix} (cross-scale LWS) for dual scales $(j,j')$ and rescaled time $u \in (0,1)$, defined as,
\begin{align*} 
    \mathbf{S}_{jj'}(u)&=\mathbf{V}_j(u) \boldsymbol{Q}_{jj'}(u)\mathbf{V}^{\top}_{j'}(u). 
\end{align*}
If $j=j'$, then $\mathbf{S}_{jj'}(u)=\mathbf{S}_j(u)$ and, for the special case of uncorrelated scales, it coincides with the definition of \cite{ombao2014}. 
\end{definition}

\noindent 
\begin{remark}
We shall next see that the cross-scale LWS matrix provides a measure of local contribution to cross-scale covariance between channels at a given rescaled time $u$ and pair of scales $(j,j')$. Compared with the single scale LWS matrix, $\mathbf{S}_j(u)$, the cross-scale LWS matrix $\mathbf{S}_{jj'}(u)$ does not have a symmetric structure if
$j \neq j'$. The diagonal elements of the cross-scale LWS matrix are the individual channel spectra and are denoted ${S}_{jj'}^{(p,p)}(u)$. The off-diagonal terms, $S_{jj'}^{(p,q)}(u)$, describe the cross-spectrum between channels $p$ and $q$ at a dual scales $(j, j')$. When $j \neq j'$, $\mathbf{S}_{jj'}^{\top}(u) =\mathbf{S}_{j'j}(u)$ as $\boldsymbol{Q}_{jj'}^{\top}(u)=\boldsymbol{Q}_{j'j}(u)$ from property (b) in Definition 1. If  $\boldsymbol{Q}_{jj'}(u)$ is invertible, then the invertibility of the cross-spectrum  follows.
\end{remark} 

\subsection{Cross-scale local covariance and coherence} \label{coherence}

In this section we develop novel quantities that measure dependence between CS-MvLSW subprocesses at different scales. We will introduce and discuss the connection between the cross-scale LWS matrix and the local auto- and cross-scale covariance and coherence.

For any pair of scales $j,j' \in \mathbb{N}$, time $k$ and lag $\tau \in \mathbb{Z}$, the cross-scale autocorrelation wavelets (see e.g., \cite{killick2020}) are defined as,
\begin{align*}
    \boldsymbol{\Psi}_{jj'}(\tau)= \sum\limits_{k \in \mathbb{Z}} \psi_{j,k}(0)\psi_{j',k}(\tau),
\end{align*}
and form the building blocks for the following operators.

\noindent 
\begin{definition} \label{def6}
   Define the operator $A_{jj';ll'}^{(\delta)}$ with $\delta \in \mathbb{Z}$ by
\begin{align*}
    A_{jj';ll'}^{(\delta)}:= \sum\limits_{\tau \in \mathbb{Z}} \boldsymbol{\Psi}_{jj'}(\tau
    ) \boldsymbol{\Psi}_{ll'}(\tau+\delta).
\end{align*} 
\end{definition}

\noindent
\begin{remark}
    If $j=j'$ (single scale), then $\boldsymbol{\Psi}_{jj}(\tau)=\boldsymbol{\Psi}_j(\tau)$, where $\boldsymbol{\Psi}_j(\tau)$ is the autocorrelation wavelet defined by \cite{nason2000wavelet}. When $\delta=0$, denote $A_{jj';ll'}:=A_{jj';ll'}^{(0)}$, and we have $A_{ll';jj'}=A_{jj';ll'}=A_{jl;j'l'}$.  Moreover, since $\Psi_{jl}(\tau)=\Psi_{lj}(-\tau)$ for any lag $\tau$ \citep{killick2020}, we directly obtain $A_{jj';ll'}=A_{j'j;l'l}$. For any $\delta$,  $A_{jj';ll'}^{(\delta)}=A_{jl;j'l'}^{(\delta)}$ and $A_{jj';ll'}^{(-\delta)}=A_{j'j;l'l}^{(\delta)}$ (the proof follows in the vein of Lemma 3 in \cite{embleton2022multiscale}, thus omitted here). 
\end{remark}

\noindent 
\begin{definition} \label{def7}
   For a given scale $j$ and rescaled time $u$, denote $c_{j}^{(p,p)}(u,\tau)$ to be the \underline{scale-specific local autocovariance} of channel $p$ at lag $\tau$, and $c_{j}^{(p,q)}(u,\tau)$ be the scale-specific local cross-covariance between channels $p$ and $q$. We define these in terms of the elements of the spectral matrix and of the autocorrelation wavelets, as follows,
\begin{equation*} \label{eqn7}
 \begin{split}
     c_{j}^{(p,p)}(u,\tau) &=S_{j}^{(p,p)}(u)\boldsymbol{\Psi}_{j}(\tau), \\
    c_{j}^{(p,q)}(u,\tau) &=S_{j}^{(p,q)}(u)\boldsymbol{\Psi}_{j}(\tau) .
 \end{split}
\end{equation*} 
\end{definition}

\noindent 
\begin{definition} \label{def8}
For a given pair of scales $(j,j')$ and rescaled time $u$, let $c_{jj'}^{(p,p)}(u,\tau)$ denote the \underline{dual-scale local autocovariance} of channel $p$ at lag $\tau$ and $c_{jj'}^{(p,q)}(u,\tau)$ be the dual-scale local  cross-covariance between channels $p$ and $q$. We define these functions in terms of the elements of cross-scale LWS matrix and the cross-scale autocorrelation wavelets, as 
\begin{equation*} \label{eqn8}
 \begin{split}
     c_{jj'}^{(p,p)}(u,\tau) &=S_{jj'}^{(p,p)}(u)\boldsymbol{\Psi}_{jj'}(\tau), \\
    c_{jj'}^{(p,q)}(u,\tau) &=S_{jj'}^{(p,q)}(u)\boldsymbol{\Psi}_{jj'}(\tau).
 \end{split}
\end{equation*}
\end{definition}

\noindent 
\begin{remark}\label{rem:scalecov}   For identical scales $j':=j$, we have $c_{jj'}^{(p,p)}(u,\tau)=c_{j}^{(p,p)}(u,\tau)$ and $c_{jj'}^{(p,q)}(u,\tau)=c_{j}^{(p,q)}(u,\tau)$. For a cross-scale pair $(j,j')$, using the cross-scale autocorrelation wavelet property $\boldsymbol{\Psi}_{jj'}(\tau)=\boldsymbol{\Psi}_{j'j}(-\tau)$, we can easily show that $c_{jj'}^{(p,q)}(u,\tau)=c_{j'j}^{(q,p)}(u,-\tau)$. 
\end{remark}

For reasons that will become obvious, we now  introduce the following notation for any channel pair $(p,q)$ at a given scale $j$, rescaled time $u$ and lag $\tau$,
\begin{align*}
    \tilde{c}^{(p,q)}(u,\tau)&=\sum_{j=1}^{\infty}\sum_{j'=1}^{\infty} c_{jj'}^{(p,q)}(u,\tau),\\
    \tilde{c}_{j}^{(p,q)}(u,\tau)&=\sum_{j'=1}^{\infty} c_{jj'}^{(p,q)}(u,\tau).\notag
\end{align*}
Note that $\tilde{c}^{(p,q)}(u,\tau)=\tilde{c}^{(q,p)}(u,-\tau)$ and additionally, the properties in Remark~\ref{rem:scalecov} imply that $\sum_j{c}_{jj'}^{(p,q)}(u,\tau)=\tilde{c}_{j'}^{(q,p)}(u,-\tau)$ for any $u\in (0,1)$ and lag $\tau\in \mathbb{Z}$.

We next establish the correspondence, given the definitions above, between the dual-scale local auto/cross-covariance functions for the CS-MvLSW process and its subprocesses, and the scale-specific subprocesses covariances.

\begin{proposition} \label{proposition1}
   Let $\{\mathbf{X}_{t;T}\}$ be a CS-MvLSW process with cross-scale dependence structure as in Definition 1 and suppose the elements of its dual-scale local wavelet spectral matrices, $\{S_{jj'}^{(p,q)}(\cdotp)\}_{j,j'}$, are Lipschitz continuous functions of rescaled time $u$ whose corresponding Lipschitz constants, $L_{jj'}^{(p,q)}$ for any channel pairs $(p,q)$, collectively admit $\sum\limits_{j,j'=1}^{\infty}2^{j+j'}L_{jj'}^{(p,q)}< \infty$. Let $c_j^{(p,q)}(u,\tau)$ denote the scale-$j$ specific local cross-covariance from Definition~\ref{def7}, and $c_{jj'}^{(p,q)}(u,\tau)$ denote the dual-scale $(j,j')$ local cross-covariance from Definition~\ref{def8}.  Then, for any rescaled time $u \in (0,1)$ and lag $\tau \in \mathbb{Z}$, these functions can be asymptotically represented in terms of the covariance between the scale-specific subprocesses, namely,
\begin{align}
    \left|\cov(X_{j,[uT]}^{(p)}, X_{j,[uT]+\tau}^{(q)})-c_j^{(p,q)}(u,\tau) \right| &= \mathcal{O}(2^{-j}T^{-1}), \notag \\
    \left|\cov(X_{j,[uT]}^{(p)}, X_{j',[uT]+\tau}^{(q)})-c_{jj'}^{(p,q)}(u,\tau)\right| &=\mathcal{O}(2^{-(j+j')/2}T^{-1}), \mbox{  for any channels }(p,q). \notag
\end{align}
\end{proposition}

\noindent \textsc{Proof:} See Appendix \ref{app:cov}.

\noindent

\begin{corollary} \label{corllary1}
For a CS-MvLSW process $\{\mathbf{X}_{t;T}\}$ as in Proposition~\ref{proposition1}, its (sub)process covariance structure at rescaled time $u \in (0,1)$ and lag $\tau \in \mathbb{Z}$ can be approximated by
\begin{align}
    \left|\cov(X_{j,[uT]}^{(p)}, X_{[uT]+\tau,T}^{(q)}) - \tilde{c}_{j}^{(p,q)}(u,\tau) \right| &= \mathcal{O}(2^{-j/2}T^{-1}),\notag \mbox{ and }\\
    \left|\cov(X_{[uT],T}^{(p)}, X_{[uT]+\tau,T}^{(q)})-\tilde{c}^{(p,q)}(u,\tau) \right| &= \mathcal{O}(T^{-1}), \mbox{  for any channels }(p,q).\notag
\end{align}
\end{corollary} 

\noindent \textsc{Proof:} See Appendix \ref{app:cov}.

These results show that when channels $p=q$, the term $\tilde{c}^{(p,p)}(u,\tau)$ represents the localised autocovariance function corresponding to the process channel $p$, while $\tilde{c}_j^{(p,p)}(u,\tau)$ encompasses its local autocovariance with its scale-$j$ subprocess. When $p \neq q$, $\tilde{c}^{(p,q)}(u,\tau)$ represents the localised CS-MvLSW process cross-covariance, while  $\tilde{c}_j^{(p,q)}(u,\tau)$ encapsulates the localised cross-covariance of process channel $q$  with the scale-$j$ subprocess of the $p$ channel. 

We next connect the above covariance structures to their corresponding spectral quantities.

\noindent 
\begin{property} \label{property1}
(i) For a CS-MvLSW process $\{\mathbf{X}_{t;T}\}$ as in Definition 1, the local (sub)process cross-covariance structure $\{ \tilde{c}^{(p,q)}_j(u, \tau) \}_j$ at rescaled time $u \in (0,1)$ and lag $\tau \in \mathbb{Z}$ is uniquely associated to its dual-scale spectral representation $\{ S_{jj'}^{(p,q)}(u) \}_{j'}$, and the (invertible) equations linking them for any channels $(p,q)$ are,
$$
\sum_{\tau \in \mathbb{Z}}\tilde{c}^{(p,q)}_j(u, \tau)\boldsymbol{\Psi}_{jl}(\tau)= \sum_{j'=1}^{\infty}A_{jj;lj'}S^{(p,q)}_{jj'}(u), \mbox{ for any scales }j, \, l.$$
(ii) For a CS-MvLSW process $\{\mathbf{X}_{t;T}\}$ as in Definition 1 with non-zero cross-spectral activity only for scales that are at most $h<J$ steps away from one another, i.e. ${S}_{jj'}^{(p,q)}(u) =0$ iff $(j,j')\notin \mathcal{B}_h$ with $\mathcal{B}_h=\{(j,j')/ |j-j'| \leq h\}$ for any $p,\, q$, the local process cross-covariance structure $\tilde{c}^{(p,q)}(u, \tau)$ at rescaled time $u$ and lag $\tau \in \mathbb{Z}$ is uniquely associated to its dual-scale spectral representation $\{ S_{jj'}^{(p,q)}(u) \}_{j,j'}$, and the (invertible) equations linking them are 
$$
\sum_{\tau \in \mathbb{Z}}\tilde{c}^{(p,q)}(u, \tau)\boldsymbol{\Psi}_{ll'}(\tau)= \sum_{j=1}^{\infty}\sum_{j'=1}^{\infty}A_{jj';ll'}S^{(p,q)}_{jj'}(u), \mbox{ for any scales }l, \, l'.
$$
\end{property}

\noindent \textsc{Proof:} See Appendix \ref{app:corr}, which also derives the inverse connections and obtains the dual-scale spectra as a linear combination of the scale-specific cross-covariances. 

These are important relationships and their use will become clear when estimating the dual-scale spectra in Section \ref{sec: estimation}. We next quantify the cross-scale dependence between different channels by defining their associated cross-scale wavelet coherence. 

\noindent \begin{definition} \label{def9}
For any pair of scales $(j,j')$ and rescaled time $u\in (0,1)$, the \textit{local dual-scale wavelet coherency matrix}, $\boldsymbol{\rho}_{jj'}(u)$ is defined as,
\begin{align}\label{eq:rho}
    \boldsymbol{\rho}_{jj'}(u)&=\mathbf{D}_{j}(u)\mathbf{S}_{jj'}(u)\mathbf{D}_{j'}(u), \mbox{ or equivalently,}\\
    &=\tilde{\mathbf{V}}_{j}(u)\mathbf{Q}_{jj'}(u)\tilde{\mathbf{V}}_{j'}^\top(u), \nonumber
\end{align}
where the matrices $\mathbf{D}_j(u)$ are diagonal with $p$th diagonal entry $\big\{S_{j}^{(p,p)}(u)\big\}^{(-1/2)}$ 
and $\tilde{\mathbf{V}}_{j}(u)=\mathbf{D}_{j}(u)\mathbf{V}_{j}(u)$ is the corresponding normalized transfer function. When $j=j'$, the quantity in \eqref{eq:rho} coincides with the coherency defined by \cite{ombao2014}, namely $ \boldsymbol{\rho}_{jj}(u)= \boldsymbol{\rho}_{j}(u)$.

The $(p,q)$ element of the wavelet coherency matrix, denoted $\rho_{jj'}^{(p,q)}(u)$, is the dual-scale coherency between scale $j$- channel $p$ and scale $j'$- channel $q$, which can be expressed as,
 \begin{align*} 
     \rho_{jj'}^{(p,q)}(u) = \frac{S_{jj'}^{(p,q)}(u)}{\sqrt{S_{j}^{(p,p)}(u)}\sqrt{S_{j'}^{(q,q)}(u)}}.
 \end{align*} 
\end{definition}

\noindent \begin{remark}
Note that $\rho_{jj'}^{(p,q)}(u) \in (-1,1)$, as determined by the local cross-scale dependence structure of the multivariate process channels $p$ and $q$. A value close to $\pm 1$ indicates a strong cross-scale $(j,j')$ linear association between these channels at  rescaled time $u$.
\end{remark}

\section{Estimation theory} \label{sec: estimation}
This section is devoted to the estimation of the spectral quantities associated with the proposed CS-MvLSW process embedding the cross-scale dependence framework. 
For this task we take two distinct avenues: one derived from the subprocesses (Section~\ref{sec:subest}), and the other process-centred (Section~\ref{sec:procest}).

We introduce the \textit{empirical wavelet coefficient vector at scale $j$ and time $k$} of the CS-MvLSW process $\{\mathbf{X}_{t;T}\}$: $\mathbf{d}_{j,k}= [d_{j,k}^{(1)},\ldots, d_{j,k}^{(P)}]^{\top}$ with $\mathbf{d}_{j,k;T}= \sum\limits_{t=1}^{T} \mathbf{X}_{t;T} \psi_{j,k}(t)$, where for ease of notation, we dropped the $T$ subscript, akin to \cite{ombao2014}.

\subsection{Subprocess-based estimation}\label{sec:subest}
Let us also introduce a new quantity, the scale-$j$ \textit{subprocess empirical wavelet coefficient vector} ${\mathbf{d}}_{jj',k}= [{d}_{jj',k}^{(1)},\ldots, {d}_{jj',k}^{(P)}]^{\top}$ at scale $j'$ and location $k$, defined as ${\mathbf{d}}_{jj',k}= \sum\limits_{t=1}^{T} \mathbf{X}_{j,t} \psi_{j'k}(t)$.

We now define the subprocess-centred \textit{cross-scale wavelet periodogram matrix}, $\mathbf{I}^{\S}_{jj',kk'}$ as,
\begin{align*}
\mathbf{I}^{\S}_{jj',kk'}=\mathbf{d}_{jj,k}\mathbf{d}_{j',k'}^{\top}
\end{align*}
which connects the scale-$j$ and scale-$j'$ localised wavelet decompositions of the same scale-$j$ subprocess and CS-MvLSW process, at times $k$ and $k'$ respectively. The superscript $^\S$ refers to the \underline{s}ubprocess-based development, not to be confused with the process spectrum $S$. In the above, we denote by $I_{jj',kk'}^{\S;(p,q)}=d_{jj,k}^{(p)}d_{j',k'}^{(q)}$ the $(p,q)$ entry of the cross-scale periodogram matrix at given dual-scale $(j,j')$ and times $k$ and $k'$, where $p,q= 1,\ldots,P$.

This raw cross-scale wavelet periodogram matrix, connecting the wavelet decompositions of the subprocesses with that of the overall process, will serve as the first step in developing a subprocess-based estimator for the dual-scale spectra with desirable asymptotic properties.

\noindent \begin{proposition} \label{proposition2}
Let $\{\mathbf{X}_{t;T}\}$ be a CS-MvLSW time series with cross-scale dependence structure and underlying cross-scale LWS structure denoted $\{\mathbf{S}_{jj'}(u)\}_{j,j'}$ at rescaled time $u$, as in Proposition ~\ref{proposition1}. Then, asymptotically for any times $k, \, k'$ and dual-scale $(j,j')$, 
\begin{equation*}
\begin{aligned}
    \E[\mathbf{I}_{jj',kk'}^{\S}] 
        &= \sum\limits_{l=1}^{\infty} A_{jj;lj'}^{(k-k')}\mathbf{S}_{jl}(k/T)
        + \mathcal{O}(T^{-1}) \quad \mbox{and} \\[0.2cm]
    \var \!\left(I_{jj',kk'}^{\S;(p,q)}\right)
        &= \left(\sum\limits_{l=1}^{\infty} A_{jj;lj} S_{jl}^{(p,p)}(k/T)
           \right)\left(\sum\limits_{l=1}^{\infty} A_{j'j';lj'} S_{j'l}^{(q,q)}(k'/T)\right)  \\[0.2cm]
        &\quad + \left( \sum\limits_{l=1}^{\infty} A_{jj;lj'}^{(k-k')} S_{jl}^{(p,q)}(k/T) \right)^2 
        + \mathcal{O}(2^{j}T^{-1}) + \mathcal{O}(2^{j'}T^{-1}), \quad \forall (p,q).
\end{aligned}
\end{equation*}
\end{proposition} 
\noindent \textsc{Proof:} See Appendix~\ref{app:subprocest}.
The above results indicate that the raw cross-scale wavelet periodogram matrix $\mathbf{I}_{jj',kk'}^{\S}$ is both asymptotically biased and inconsistent for the true spectral structure. 
For times $k':=k$, we denote the \underline{s}ubprocess-rephrased spectral quantity 
\begin{equation}\label{eq:betaspec}
\boldsymbol{\beta}_{jj'}^{\S}(k/T)=\sum\limits_{l=1}^{\infty} A_{jj;lj'} \mathbf{S}_{jl}(k/T),
\end{equation}
and adopt a classical approach to estimate the original spectral structure $\{\mathbf{S}_{jj'}(k/T)\}_{j,j'}$, namely we first smooth the raw periodogram and then correct its bias \citep{nason2000wavelet}. In a similar vein to \cite{embleton2022multiscale}, we apply a rectangular kernel smoother with window of length $(2M+1)$ across time to smooth the raw periodogram, yielding
\begin{equation}\label{eq:subprocper}
        \Tilde{\mathbf{I}}_{jj',kk'}^{\S}=\frac{1}{2M+1}\sum\limits_{m=-M}^{M} \mathbf{I}_{jj',(k+m)(k'+m)}^{\S}.  
\end{equation}

\noindent \begin{proposition} \label{proposition3}
Let $\{\mathbf{X}_{t;T}\}$ be a CS-MvLSW time series as in Proposition~\ref{proposition1}, and let us further assume Gaussian innovations and uniform local process and subprocess (auto- and cross-) covariance absolute summability, specifically $\sup\limits_{u\in(0,1)}\sum\limits_{\tau}|\tilde{c}^{(p,q)}(u,\tau)|=\mathcal{O}(1)$ and $\sup\limits_{u\in(0,1)}\sum\limits_{\tau}|\tilde{c}_j^{(p,q)}(u,\tau)|=\mathcal{O}(1)$ for each scale $j$.\\ 
Then the estimator in equation~\eqref{eq:subprocper} for any channels $(p,q)$, dual-scale $(j,j')$ and time $k':=k$ converges in mean-square to its true counterpart, the rephrased spectral quantity $\beta_{jj'}^{\S;(p,q)}(k/T)$ defined in~\eqref{eq:betaspec} when $M, \, T \to \infty$ such that $M/T \to 0$, with the following asymptotic properties
\begin{equation}
\begin{aligned}
\E (\Tilde{I}_{jj',k}^{\S;(p,q)})&=\beta_{jj'}^{\S;(p,q)}(k/T) +\mathcal{O}(MT^{-1}) \quad \mbox{and} \notag\\ 
     \var (\Tilde{I}_{jj',k}^{\S;(p,q)}) &=
     \mathcal{O}(2^{2j}M^{-1}) + \mathcal{O}(2^{2j'}M^{-1}) +\mathcal{O}(MT^{-1}).\notag
\end{aligned} 
\end{equation}
\end{proposition} 

\noindent \textsc{Proof:} See Appendix~\ref{app:subprocest}.

\noindent \begin{remark} \label{remark8}
 In the limit, as $M, T \to \infty \mbox{ with } M/T \to 0$, the estimator in~\eqref{eq:subprocper} is consistent and asymptotically unbiased, with an asymptotically vanishing variance, $\var(\Tilde{I}_{jj',k}^{\S;(p,q)}) \to 0$ for fine enough scales $j, \, j'$ such that $2^j, \, 2^{j'}=o((2M+1)^{1/2})$. There is a trade-off between bias and variance: increasing $M$ reduces the variance but increases the bias. 
\end{remark}  

Note that the subprocesses $\{\mathbf{X}_{j,t}\}_t$ are conceptually introduced to define the cross-scale wavelet coefficients $\{\mathbf{d}_{jj',k}\}_{j',k}$, but themselves are not directly observable. Instead, we approximate them using an average-basis representation \citep{abramovich2000wavelet} by projecting the observed process onto each scale $j$. The approximation, denoted as $\{\Tilde{\mathbf{X}}_{j,t}\}_t$ still fulfills $\mathbf{X}_{t,T}=\sum_{j}\Tilde{\mathbf{X}}_{j,t}$ and provides a practical, computationally efficient surrogate for the latent subprocess coefficients $\{\tilde{\mathbf{d}}_{jj',k}\}$ involved in the estimation procedure above, while preserving the dominant scale-specific features required for estimating dependence.

Let us define an estimator akin to~\eqref{eq:subprocper} that uses the surrogate subprocess wavelet coefficients, 
\begin{equation}\label{eq:approxsubprocper}
        \Tilde{\Tilde{\mathbf{I}}}_{jj',kk'}^{\S}=\frac{1}{2M+1}\sum\limits_{m=-M}^{M} \tilde{\mathbf{d}}_{jj,k+m}{\mathbf{d}}_{j',k'+m}^{\top}.
\end{equation}

\begin{proposition}\label{prop:approx}
Let $\{\mathbf{X}_{t;T}\}$ be a CS-MvLSW time series as in Proposition~\ref{proposition3} that satisfies the active cross-scale condition in Property~\ref{property1}(ii). 
Then the estimator in equation~\eqref{eq:approxsubprocper} for any channels $(p,q)$, dual-scale $(j,j')$ and time $k':=k$  is consistent and asymptotically unbiased for its true counterpart,  $\beta_{jj'}^{\S;(p,q)}(k/T)$, when $M, \, T \to \infty$ such that $M/T \to 0$.
\end{proposition}

\noindent \textsc{Proof:}
See Appendix~\ref{app:subprocest}. These desirable results suggest that we can adopt a bias-correction approach for the smoothed cross-scale periodogram matrix constructed in~\eqref{eq:approxsubprocper},
\begin{align} \label{eq:est1}
    \hat{S}_{jj'}^{\S;(p,q)}(k/T) =\sum\limits_{l=1}^{J} \left[\boldsymbol{A}^{jj}\right]^{-1}_{j'l}
    \Tilde{\Tilde{I}}_{jl,k}^{\S;(p,q)} \mbox{ for all }(p,q),
\end{align}
where $\boldsymbol{A}^{jj}$ is the $J \times J$ matrix whose $(j', \, l)$ entry is $A_{jj;j'l}$. The inner product matrix $\boldsymbol{A}^{jj}$ is established to be invertible as a by-product of the proof of Property 1(i).

\begin{proposition}\label{prop:subprocconsist}
Let $\{\mathbf{X}_{t;T}\}$ be a CS-MvLSW process as in Proposition~\ref{prop:approx}. Then the estimator in~\eqref{eq:est1} for any channels $(p,q)$, dual-scale $(j,j')$ and time $k$  is consistent and asymptotically unbiased for its true counterpart, $S_{jj'}^{(p,q)}(k/T)$, when $M, \, T \to \infty$ with $M/T \to 0$.
\end{proposition}

\noindent \textsc{Proof:}
See Appendix~\ref{app:subprocest}. This result paves the way for the consistent estimation of cross-scale channel dependency, as we describe in Section~\ref{sec:cohest}.

\subsection{Process-based estimation}\label{sec:procest}
By means of the $P$-dimensional empirical wavelet coefficient vector $\mathbf{d}_{j,k;T}= \sum\limits_{t=1}^{T} \mathbf{X}_{t;T} \psi_{j,k}(t)$, we proceed to define the \textit{overall process cross-scale wavelet periodogram matrix}, $\mathbf{I}_{jj',kk'}^{\P}$ as,
\begin{align}\label{eq:proc_raw}    \mathbf{I}_{jj',kk'}^{\P}=\mathbf{d}_{j,k}\mathbf{d}_{j',k'}^{\top},
\end{align}
where again for ease of notation we dropped the $T$ subscript. The superscript $^\P$ refers to the \underline{p}rocess-based development, not to be confused with the process dimension $P$.

In the above, denote $I_{jj',kk'}^{\P;(p,q)}={d_{j,k}^{(p)}d_{j',k'}^{(q)}}$ to be the $(p,q)$ entry of the cross-scale periodogram matrix at a given pair of scales $(j,j')$, where $p,q = 1,\ldots,P$. 
With this raw cross-scale wavelet periodogram matrix obtained from the original process as the starting point, we next develop the process-based estimator with the desired asymptotic mean-squared consistency.

\noindent \begin{proposition} \label{proposition4}
Let $\{\mathbf{X}_{t;T}\}$ be a CS-MvLSW time series as in Proposition ~\ref{proposition3}. 
Then, asymptotically for any times $k, \, k'$ and cross-scale $(j,j')$,
\begin{equation}
    \begin{aligned}
\E[\mathbf{I}_{jj',kk'}^{\P}]&=\sum\limits_{l=1}^{\infty}\sum\limits_{l'=1}^{\infty} A_{ll';jj'}^{(k-k')}\mathbf{S}_{ll'}(k/T)+\mathcal{O}(T^{-1}) \quad \mbox{and} \notag \\
   \var(I_{jj',kk'}^{\P;(p,q)}) &= \left(\sum\limits_{l=1}^{\infty}\sum\limits_{l'=1}^{\infty} A_{ll';jj}S_{ll'}^{(p,p)}(k/T) \right)\left(\sum\limits_{l=1}^{\infty}\sum\limits_{l'=1}^{\infty}  
   A_{ll';j'j'}S_{ll'}^{(q,q)}(k'/T) \right)\notag \\
   &+ \left( \sum\limits_{l=1}^{\infty}\sum\limits_{l'=1}^{\infty} A_{ll';jj'}^{(k-k')}S_{ll'}^{(p,q)}(k/T) \right)^2 + \mathcal{O}(2^{2j}T^{-1}) + \mathcal{O}(2^{2j'}T^{-1}),  \, \forall (p,q). \notag 
    \end{aligned}
\end{equation}  
\end{proposition}
\noindent \textsc{Proof:} See  Appendix~\ref{app:procest}. 
The above results indicate that the raw wavelet periodogram matrix is both asymptotically biased and inconsistent, thus we proceed to smooth it and then investigate how to correct its bias. As before, we apply a rectangular kernel smoother with window length $(2M+1)$ across time, yielding the smoothed $P \times P$ periodogram matrix
\begin{equation}\label{eq:procest}
        \Tilde{\mathbf{I}}_{jj',kk'}^{\P}=\frac{1}{2M+1}\sum\limits_{m=-M}^{M} \mathbf{I}_{jj',(k+m)(k'+m)}^{\P}.
\end{equation}
\noindent \begin{proposition} \label{proposition5}
Let $\{\mathbf{X}_{t;T}\}$ be a CS-MvLSW time series as in Proposition \ref{proposition3}. 
The estimator in equation~\eqref{eq:procest} converges in mean-square to $\boldsymbol{\beta}^{\P}_{jj'}(k/T)=\sum\limits_{l=1}^{\infty}\sum\limits_{l'=1}^{\infty} A_{ll';jj'}^{(k-k')}\mathbf{S}_{ll'}(k/T)$ when $k':=k$ and $M, \, T \to \infty$ such that $M/T \to 0$, and has the following asymptotic properties,
\begin{align}
\E(\Tilde{I}_{jj',kk'}^{\P;(p,q)})&=\sum\limits_{l=1}^{\infty}\sum\limits_{l'=1}^{\infty} A_{ll';jj'}^{(k-k')}S_{ll'}^{(p,q)}(k/T)+\mathcal{O}(MT^{-1}) \quad \mbox{and} \notag\\ 
     \var(\Tilde{I}_{jj',kk'}^{\P;(p,q)}) &=\mathcal{O}(2^{2j}M^{-1})+ \mathcal{O}(2^{2j'}M^{-1}) +\mathcal{O}(M T^{-1}). \notag
\end{align}  
\end{proposition}  
\noindent \textsc{Proof:} See Appendix~\ref{app:procest}, which makes use of the \underline{p}rocess-based rephrased spectral quantity $\boldsymbol{\beta}^{\P}_{jj'}(\cdotp)$. In the event of no cross-scale activity, the estimator and its rates collapse to the ones postulated in the MvLSW environment \citep{ombao2014}. 

\begin{remark} \label{remark9}
As in the subprocess-based construction, we derive consistency for the process-based smoothed periodogram since for $M, \,T \to \infty, \mbox{ with } M/T \to 0$ and for fine enough scales $j, \, j'$ such that $2^j, \, 2^{j'}=o((2M+1)^{1/2})$, we have $\var(\Tilde{I}_{jj',kk'}^{\P;(p,q)}) \to 0$. The proofs of Propositions~\ref{prop:approx} and~\ref{proposition5} highlight that the subprocess- and process-based constructions yield estimators of the subprocess- or process- based rephrased spectrum respectively, $\beta^{\S;(p,q)}_{jj'}(\cdotp)$ or $\beta^{\P;(p,q)}_{jj'}(\cdotp)$, with different asymptotic biases and necessitate tighter assumptions $2^{(j'-j)/2}=o(T)$ on the cross-scale activity for the subprocess-based construction. 
\end{remark}  
In order to derive an estimator for the cross-scale spectra $\{S_{jj'}^{(p,q)}(\cdotp)\}_{j,j'}$ using its connection to the smoothed process-based periodogram in Proposition~\ref{proposition5}, we recall the assumption that ensures tractability by only allowing neighbouring scales to be connected (see Property~\ref{property1}(ii)). This assumption is aligned with the neurological data that motivates this work \citep{eegdata}. 
Namely, the $J \times J$ cross-scale spectral matrices, $\boldsymbol{S}^{(p,q)}(u)=\left({S}^{(p,q)}_{jj'}(u)\right)_{j,j'=1}^J$, share a {\em banded} diagonal structure across rescaled times $u$, with bandwidth $h<J$ for all channels $(p,q)$, i.e. ${S}^{(p,q)}_{jj'}(u)=0$ if and only if $(j,j')\notin \mathcal{B}_h$ where $\mathcal{B}_h=\{ (j,j')/ |j-j'| \leq h\}$. 



For each channel pair $(p,q)$ and time $k$, concatenating the $J^{\star}:=J(2h+1)-h(h+1)$ scale-dependent entries assumed to be active, we connect the smoothed periodogram and unknown spectral column vectors by using the asymptotic expectation result in Proposition~\ref{proposition5},
\begin{align*}     
\E \left[\Tilde{\Tilde{\mathbf I}}_{k}^{\P;(p,q)} \right]&=\tilde{\mathbf{A}}\tilde{\mathbf S}^{(p,q)}(k/T) +\boldsymbol{\mathcal{O}}(MT^{-1}),     
\end{align*}
where $\Tilde{\Tilde{\mathbf I}}_{k}^{\P;(p,q)}:=\left[
(\Tilde{I}_{j,j+\delta';k}^{\P;(p,q)})_{\delta'=0:h, j=1:J-\delta'} \, ,
(\Tilde{I}_{j+\delta',j;k}^{\P;(q,p)}
)_{\delta'=1:h, j=1:J-\delta'}
\right]^\top$ is the column vector of smoothed periodograms matching the scale $j'':=(j,j')$ ordering of the unknown cross-spectral vector $\tilde{\mathbf S}^{(p,q)}(k/T)$ across the dual-scales in $\mathcal{B}_h$. Here we defined $\tilde{\mathbf{A}}$ to be the Gram matrix of $\boldsymbol{\Psi}(\tau)$, the wavelet cross-correlation vector given by $\boldsymbol{\Psi}(\tau):=\left[\{\boldsymbol{\Psi}_{j, j+\delta}(\tau)\}_{{\delta=0:h},{j=1:J-\delta}}, \{\boldsymbol{\Psi}_{j, j+\delta}(-\tau)\}_{{\delta=1:h},{j=1:J-\delta}} \right]^\top$. 
From Property \ref{property1}(ii), the components of $\boldsymbol{\Psi}(\tau)$ are linearly independent for Haar wavelets and the matrix $\tilde{\mathbf{A}}$ is invertible, hence we obtain the process-based estimator of the cross-scale spectrum as 
\begin{align}\label{eq:procestS}
    \hat{S}^{\P;(p,q)}_{jj'}(k/T) =\sum_{l^{''}=(l,l')\in \mathcal{B}_h}\tilde{\mathbf{A}}^{-1}_{j''l''}
    \Tilde{I}_{l'';k}^{\P;(p,q)} \mbox{ for all }(p,q) \mbox{ and time }k.
\end{align}

\begin{proposition}\label{prop:procconsist}
Let $\{\mathbf{X}_{t;T}\}$ be a CS-MvLSW time series as in Proposition~\ref{prop:approx}. Then the estimator in~\eqref{eq:procestS} for any channels $(p,q)$, dual-scale $(j,j')$ and time $k$  is consistent and asymptotically unbiased for the true spectrum, $S_{jj'}^{(p,q)}(k/T)$, when $M, \, T \to \infty$ with $M/T \to 0$.
\end{proposition}
\noindent \textsc{Proof:} See Appendix~\ref{app:procest}. As an illustration, in Appendix~\ref{app:Aconstruct} we explicitly provide the construction for the case when the spectral matrix $\mathbf{S}^{(p,q)}(k/T)$ is tridiagonal, i.e., it has non-zero elements only on the main diagonal and on its upper and lower diagonals ($h=1$).

\subsection{Coherence estimation}\label{sec:cohest}
The $P\times P$ matrix estimator $\hat{\mathbf{S}}_{jj'}(k/T)$ whose $(p,q)$ entry is $\hat{S}_{jj'}^{(p,q)}(k/T)$ and may be obtained using the subprocess- or process-based approach, can be subsequently used to estimate the $(j,j')$ cross-scale wavelet coherence localised at rescaled time $u=k/T$. Letting $\hat{\mathbf{D}}_j(\cdotp)$, $\hat{\mathbf{D}}_{j'}(\cdotp)$ be diagonal matrices whose elements are $(\hat{S}_{j}^{(p,p)}(\cdotp))^{-(1/2)}$ and $(\hat{S}_{j'}^{(p,p)}(\cdotp))^{-(1/2)}$ respectively, we define the local cross-scale $(j,j')$ wavelet coherence matrix estimator to be,
\begin{align} \label{rho}
    \hat{\boldsymbol{\rho}}_{jj'}(u) = \hat{\mathbf{D}}_{j}(u) \hat{\mathbf{S}}_{jj'}(u) \hat{\mathbf{D}}_{j'}(u) \mbox{ at each rescaled time }u.
\end{align}
The $(p,q)$-th element of the $\hat{\boldsymbol{\rho}}_{jj'}(\cdotp)$ matrix is the estimated time-varying cross-scale wavelet coherence between channels $p, \,q$ at a given pair of scales $(j,j')$.

\begin{proposition}
Under the assumptions of Proposition~\ref{prop:approx}, the coherence estimator in~\eqref{rho} that uses the spectrum estimator that is subprocess-based in~\eqref{eq:est1} or process-based in~\eqref{eq:procestS} is consistent for the true cross-scale coherence ${\boldsymbol{\rho}}_{jj'}(\cdotp)$ at each fixed cross-scale pair $(j, \, j')$.
\end{proposition}

\noindent \textsc{Proof:}  The localised coherence definition for the true and estimated quantities~\eqref{eq:rho} and~\eqref{rho}, coupled with a direct application of the continuous mapping theorem \citep{billingsley1999convergence} for the consistent spectral estimators set out in Propositions~\ref{prop:subprocconsist} and~\ref{prop:procconsist} for the subprocess- and process-based approaches respectively, yield the desired consistency properties for their corresponding cross-scale coherence estimator.

\section{Simulation study}\label{simulation study}

The simulation experiments aim to assess whether the proposed CS-MvLSW framework (i) accurately estimates single-scale dependence (coherence) both in the absence of and in the presence of an imposed cross-scale dependence structure; (ii) recovers time-varying cross-scale dependence, and (iii) for single scale estimation, we compare our framework against the existing method of \cite{ombao2014} (referred to as `MvLSW' in the reported results), in scenarios without and with cross-scale structure, although we note here that the MvLSW formulation is tailored to settings without cross-scale coherence. This method is implemented using the \textsf{R} package \texttt{mvLSW} \citep{TaylorParkEckley2019mvLSW}. 

For reference, we also report results obtained using the subprocess-realisations according to Definition~\ref{def2}-- these appear labelled as `Subprocess (True)', although we note they still present an element of randomness, in the way in which the overall process realisation does. Since the bias-correction matrix properties are theoretically developed for Haar wavelets, all simulation analyses are conducted using the Haar wavelets.

A trivariate process $\mathbf{X}_{t;T}=[X_{t;T}^{(1)},X_{t;T}^{(2)},X_{t;T}^{(3)}]^\top$ is generated from a CS-MvLSW model with prescribed within-scale spectra $\{\mathbf{S}_j(u)\}$ and innovation cross-scale covariances $\{\boldsymbol{Q}_{j,j'}(u)\}$, where $u=t/T\in(0,1)$.
In all experiments, $R=1{,}000$ independent replicates are generated for $T\in\{512,1024,4096\}$, and estimation uses a smoothing window with $M=\lfloor\sqrt{T}\rfloor$.
Complete data-generating specifications are provided in Appendix~\ref{app:sim}. 

At cross-scale $(j,j')$, coherence estimation accuracy is summarized by the time-averaged squared errors and their mean (and similarly for the bias, see equation~\eqref{eq:msb} in Appendix~\ref{app:B}),
\begin{align*}
    MSE_{jj'}&=\frac{1}{RT}\sum_{r=1}^R\sum_{t=1}^{T}\big(\hat{\boldsymbol{\rho}}_{jj'}^{(r)}(t/T)-\boldsymbol{\rho}_{jj'}(t/T)\big)^2.
\end{align*}

\begin{figure}[htbp]
  \centering
    \includegraphics[width=.75\linewidth]{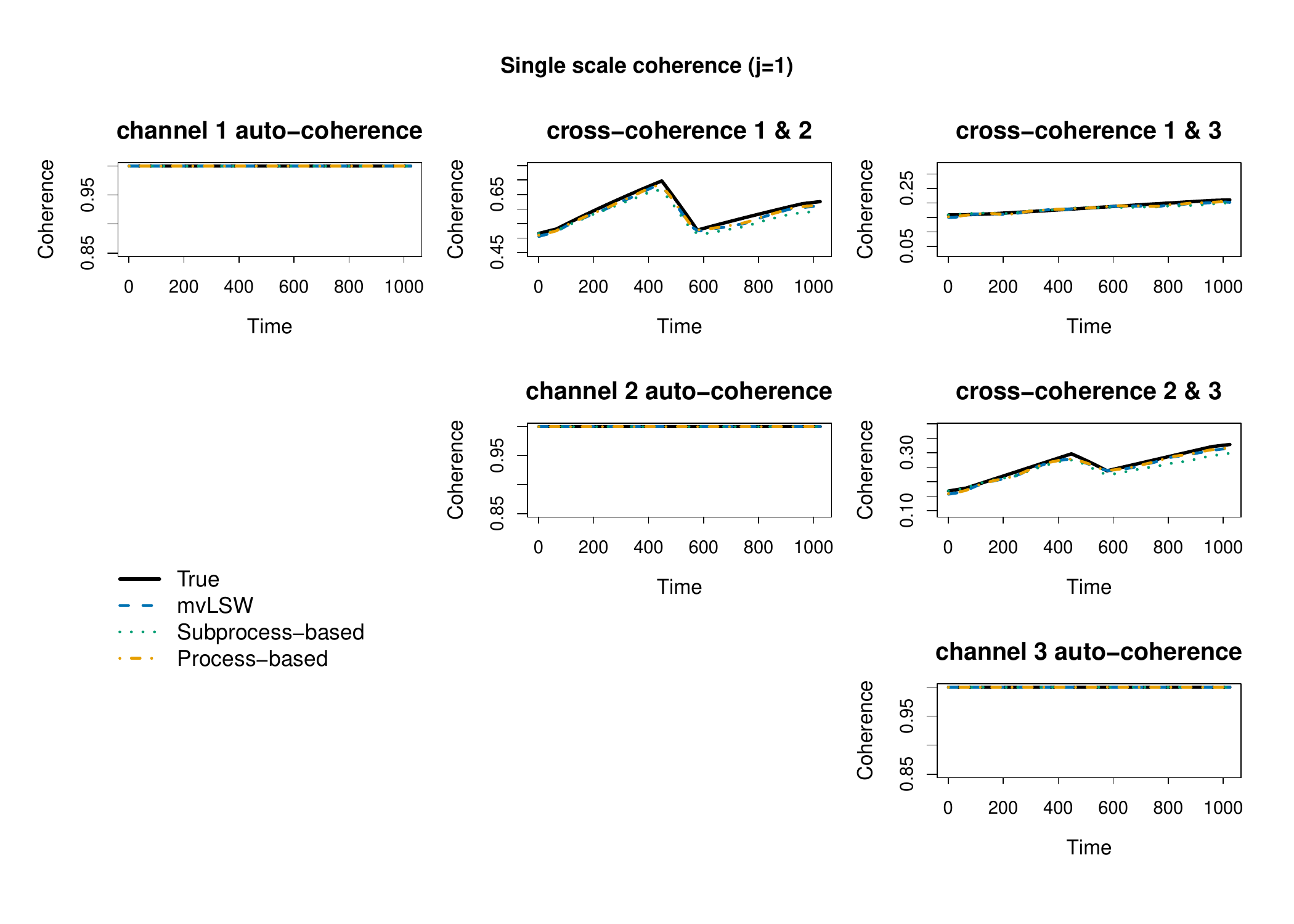}
    \hfill 
    \includegraphics[width=.75\linewidth]{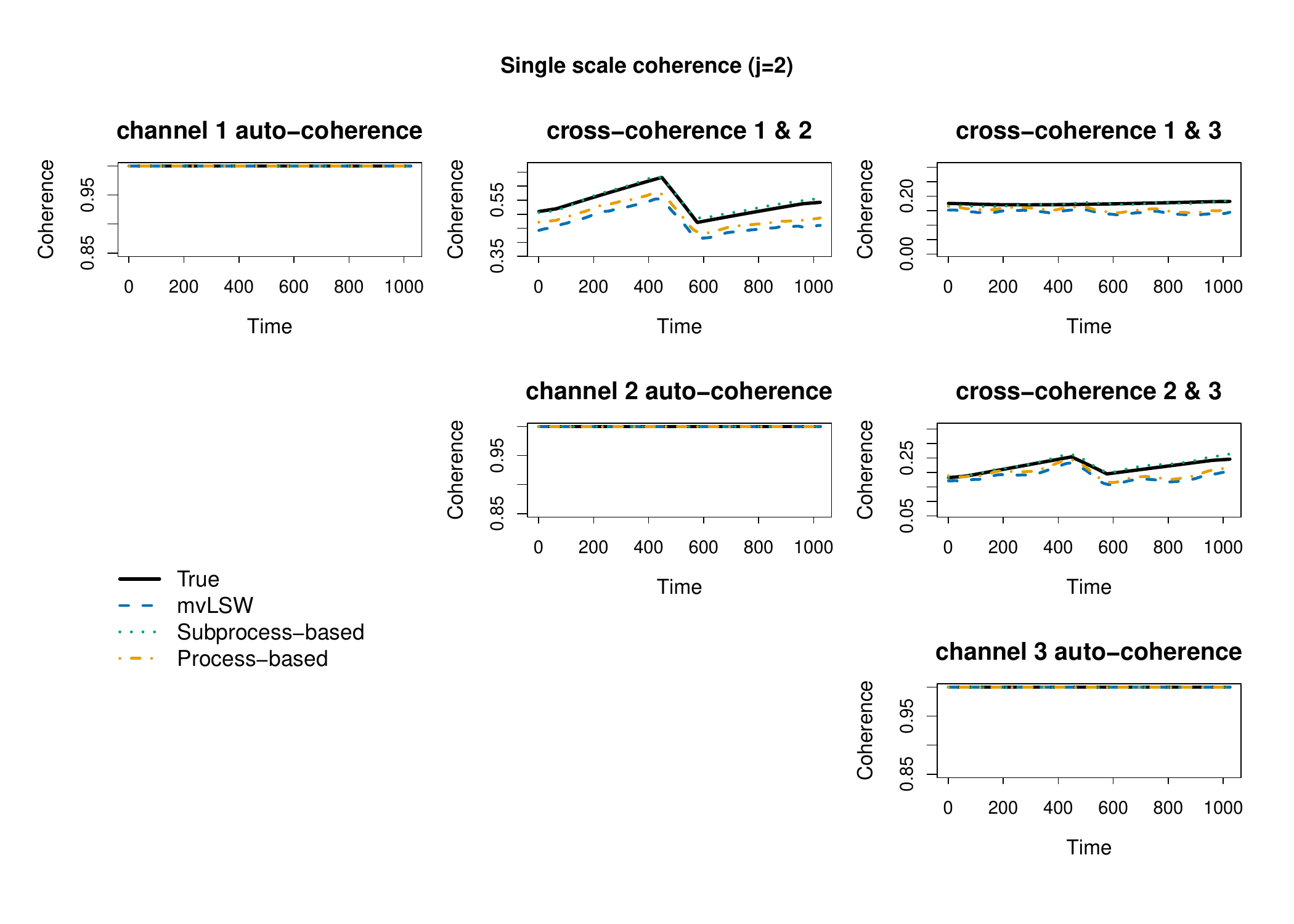}
  \caption{Single-scale coherence estimates at $j=1$ (top) and $j=2$ (bottom) for $T=1024$ (Scenario 1), averaged over 1{,}000 replicates. Black: truth; coloured: mvLSW, subprocess-based (approximated), and process-based estimates.}
  \label{fig:sin_coh}
\end{figure}

\textbf{Scenario 1: Single-scale dependence only.} Non-zero power is restricted to the two finest scales $j_1=1$ and $j_2=2$, and no cross-scale dependence is imposed.
Specifically, $\boldsymbol{Q}_{j,j}(u)=\boldsymbol{I}$ ($P \times P$ identity matrix) for all $j$ and $u$, while $\boldsymbol{Q}_{j,j'}(u)=\mathbf{0}$ for $j\neq j'$; innovations $\{\mathbf{z}_{jk}\}$ are therefore uncorrelated across scales.
The classical MvLSW estimator \citep{ombao2014} and the proposed subprocess (approximated)- and process-based estimators from Section \ref{sec: estimation} are applied to each process replicate, $\{X_{t;T}^{(r)}\}$.
Figure~\ref{fig:sin_coh} reports the mean single-scale coherence estimates at $j_1=1$ and $j_2=2$ for $T=1024$.
All three methods recover the main coherence patterns, with the proposed estimators (especially the subprocess-based method) showing closer agreement with the ground truth. 
Table \ref{tab:coh_msb_mse_s1} (Appendix \ref{appB:add results}) reports the mean squared bias (MSB) and mean squared error (MSE) of the proposed methods for channels $(p,q)=(1,2)$ at scales $j_1=1$ and $j_2=2$, respectively, as $T$ increases (and hence $M$ increases). In terms of bias, MvLSW and both proposed estimators are all small and broadly comparable. In contrast, both proposed estimators reduce the MSE relative to MvLSW, with the subprocess-based estimator achieving the largest improvements even in this no-cross-scale-dependence scenario. As $T$ increases, the subprocess- and process-based errors decrease markedly, consistent with the asymptotic regime described in Remarks \ref{remark8} and \ref{remark9}; by comparison, MvLSW errors improve only weakly with $T$. For the process-based estimator, gains are modest at $T=512$ but become pronounced for larger $T$, whereas the subprocess-based estimator delivers substantial reductions already at moderate sample sizes. Finally, we emphasize that cross-channel $(1,2)$ is representative rather than cherry-picked: Figures \ref{fig:sin_se_j1} and \ref{fig:sin_se_j2} (Appendix \ref{appB:add results}) summarize the SE distributions across replicates for all auto- and cross-channel pairs, providing a global view that aligns with Table \ref{tab:coh_msb_mse_s1}.

\textbf{Scenario 2: Time-varying cross-scale dependence.}
The same within-scale spectra as in Scenario~1 are used, while cross-scale dependence is introduced through non-zero off-diagonal blocks of $\boldsymbol{Q}_{j,j'}(u)$.
Diagonal blocks satisfy $\boldsymbol{Q}_{j,j}(u)=\boldsymbol{I}$, and cross-scale dependence between scales $(j_1,j_2)$ is specified by a time-varying $P\times P$ block
\begin{align*}
\boldsymbol{Q}_{j_1,j_2}(u)=
\left[\begin{array}{lll}
0.3 + 0.2u &  \quad0.1 + 0.2u  & \quad0.1 + 0.1u \\
0.1 + 0.2u & \quad0.4 + 0.1f(u-0.5) & \quad0.1 + 0.1f(u-0.5) \\
0.1 + 0.1u &\quad 0.1 + 0.1f(u-0.5) &\quad 0.3
\end{array}\right],
\end{align*}
with the block-symmetry condition $\boldsymbol{Q}_{j_2,j_1}(u)=[\boldsymbol{Q}_{j_1,j_2}(u)]^{\top}$, where $f(x)=0 $ for $x<0$ and $f(x)=1$ for $x \geq 0$. In this design, $\boldsymbol{Q}_{j_1,j_2}(u)$ is itself symmetric, hence $\boldsymbol{Q}_{j_2,j_1}(u)=\boldsymbol{Q}_{j_1,j_2}(u)$.

In this scenario, to assess the flexibility and robustness of the proposed methods, we consider three cases (Cases 1–3) with active scale pairs $(j_1,j_2)\in \{(1,2),(3,4),(1,4)\}$, as specified in our setup. For single-scale coherence, we continue to benchmark the proposed estimators against MvLSW, while the current MvLSW framework is not formulated to accommodate cross-scale coherence. Cross-scale coherence is estimated using both the proposed subprocess- and process-based estimators. Our empirical investigations indicate that the subprocess approach can introduce a noticeable (finite-sample) bias, particularly for smaller $T$. We therefore treat the process-based estimator as the main procedure in this setting, while still reporting the subprocess-based results (together with the corresponding `True' subprocess benchmark) to validate the theoretical developments. 

\begin{figure}[htbp] 
\centering \includegraphics[width=.55\linewidth]{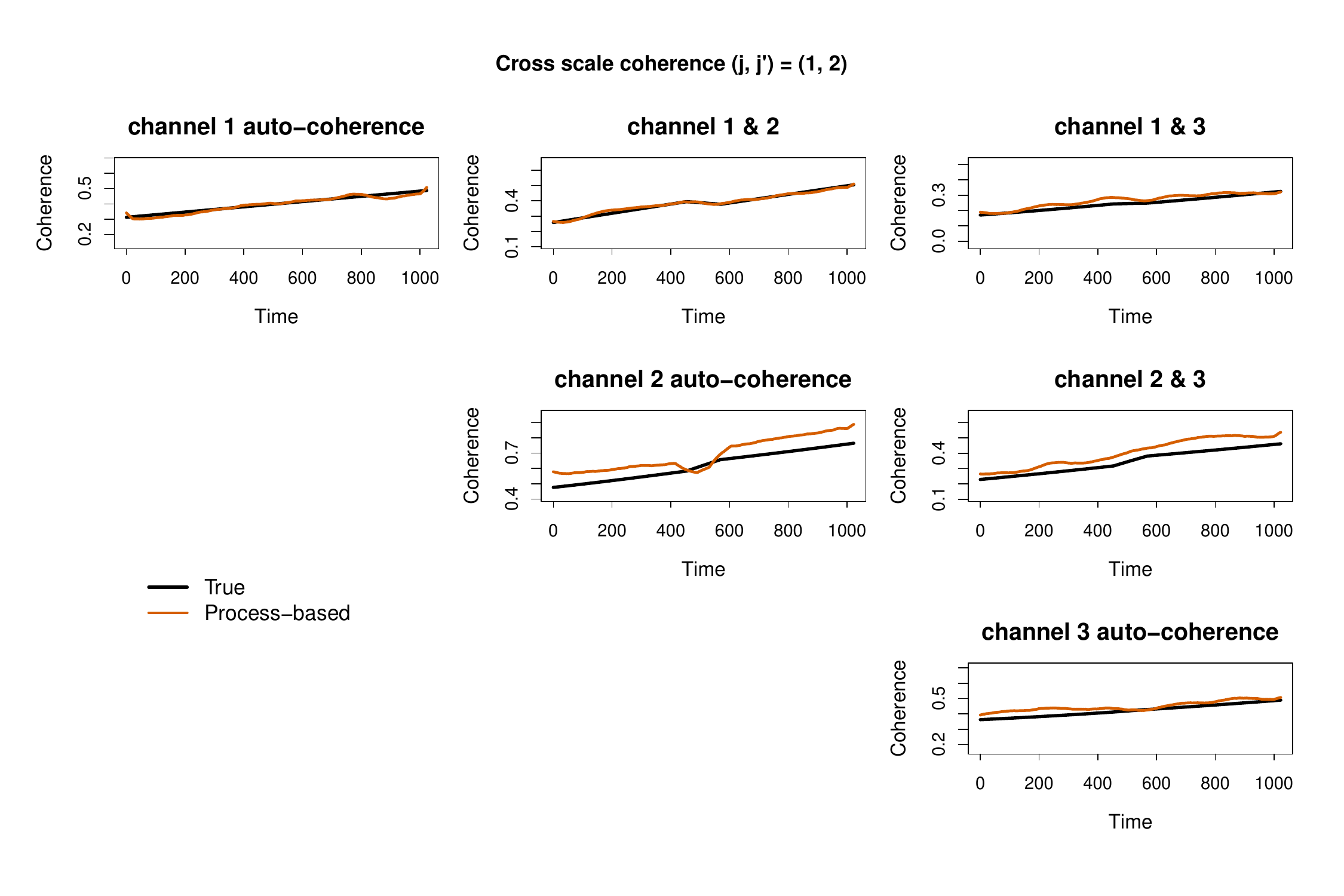} 
\hfill
\includegraphics[width=.55\linewidth]{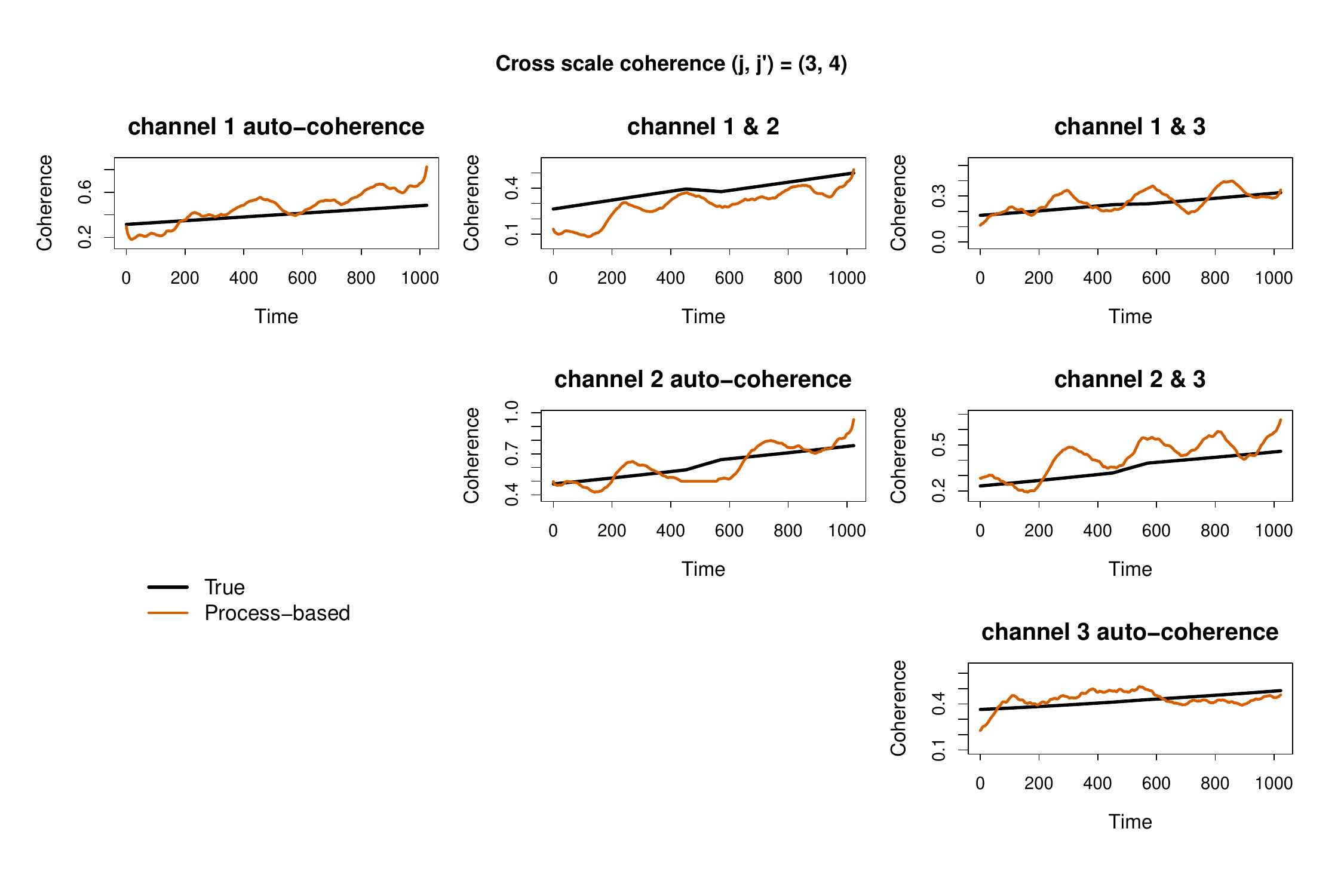} 
\hfill 
\includegraphics[width=.55\linewidth]{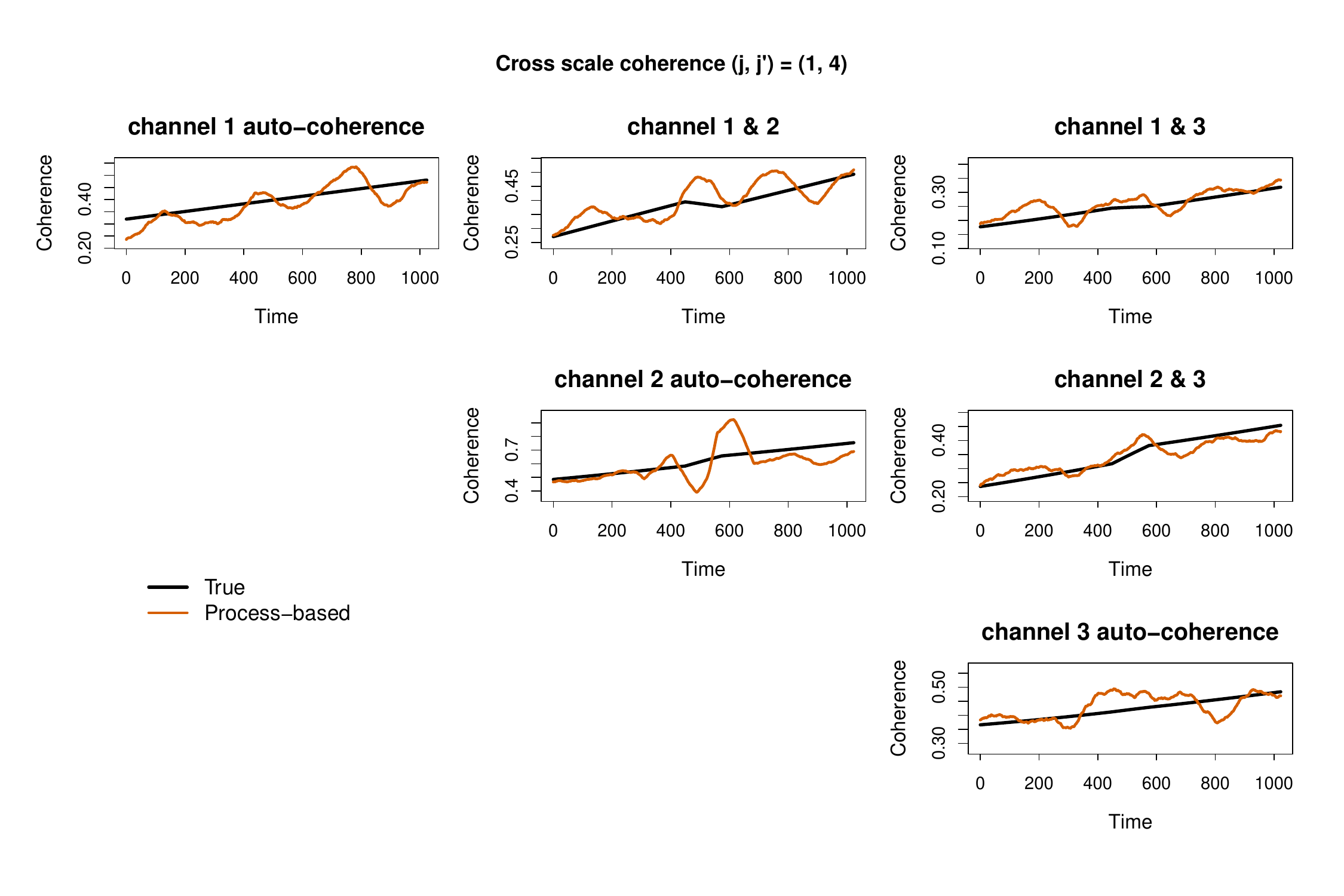} 
\caption{Estimated cross-scale coherence (Scenario 2) for $T=1024$ for three pairs of scales $(j_1, j_2)$, averaged over 1{,}000 replicates. Black: truth; coloured: process-based estimator.}
\label{fig:cross-scale-coherence}
\end{figure}

Figure \ref{fig:cross-scale-coherence} displays cross-scale coherence estimates for the above three cases using the process-based estimator, 
showing close agreement with the truth across channel pairs and over time.
Uncertainty and asymptotic behaviour are also assessed with MSB and MSE (see Tables~\ref{tab:coh_msb_mse_s2c1}--\ref{tab:coh_msb_mse_s2c3} in Appendix \ref{appB:add results}). The results indicate that for single-scale coherence estimation, MvLSW performs poorly once cross-scale dependence is present (see Figures \ref{fig:sin_coh_s2c1}, \ref{fig:sin_coh_s2c2} and \ref{fig:sin_coh_s2c3} in Appendix \ref{appB:add results}). In particular, both bias and variance can be inflated, and the error does not necessarily decrease with $T$ in a manner consistent with the usual asymptotic regime. This effect is most pronounced when the signal power is concentrated towards medium coarser scales, which constitutes a more challenging setting. In contrast, the process-based estimator shows some degradation when cross-scale structure is present but nevertheless, both its bias and MSE decrease substantially as $T$ increases. 

The subprocess-based estimator is largely robust to the presence of cross-scale dependence, and it consistently delivers the best performance for single-scale coherence estimation. On the other hand, this estimator may exhibit a shift in cross-scale coherence, dramatically counterbalanced by a much smaller variance that decreases furthermore with the increasing sample size $T$. In the idealised simulation benchmark where the true subprocesses are available, the same estimator recovers the cross-scale dependence almost perfectly. This indicates that the practical limitation arises primarily from the accuracy of the subprocess approximation, rather than from the estimator itself. By comparison, the proposed process-based estimator exhibits smaller bias but larger variation for smaller $T$ in this challenging cross-scale setting, both improving rapidly and converging to satisfactory levels as $T$ increases. Overall, these results suggest that the process-based estimator provides a reliable approach for capturing cross-scale coherence, while the subprocess-based estimator remains most competitive when the goal is single-scale coherence estimation, both in the presence of complex dependence structures. Although not the purpose of this work, in the absence of cross-scale activity, the subprocess-based estimator is turning out to strongly outperform the (only) equivalent variant in the current literature, namely the MvLSW estimator of \cite{ombao2014}.

Additional stress-tests under more complex multiscale settings are detailed in Appendix~\ref{app:sim} to further verify the effectiveness of process-based estimator, where non-zero spectra occur at three or four scales and multiple cross-scale links coexist. Corresponding estimation results are reported in Figure \ref{fig:multiple_scales}, Appendix~\ref{appB:add results}. These supplementary results further support stable recovery of cross-scale coherence beyond the two-scale configurations shown here.

\section{EEG data analysis}\label{EEG analysis}
To further assess the proposed CS-MvLSW methodology, we analyze EEG recorded from children with attention deficit hyperactivity disorder (ADHD) during cognitive tasks and compare the results with those from healthy controls (children without any registered psychiatric disorder). Our goal is to characterize group differences in cross-channel connectivity, with particular emphasis on cross-scale interactions. Based on the simulation results, we use the subprocess-based estimator for single-scale dependence and the process-based estimator for cross-scale dependence, as this combination provides stable and interpretable inference for noisy, finite-length EEG recordings. This enables us to quantify interactions between long-term dynamics in one channel and short-term dynamics in another.

\begin{figure}[htbp]
    \centering
    \includegraphics[width=0.65\textwidth]{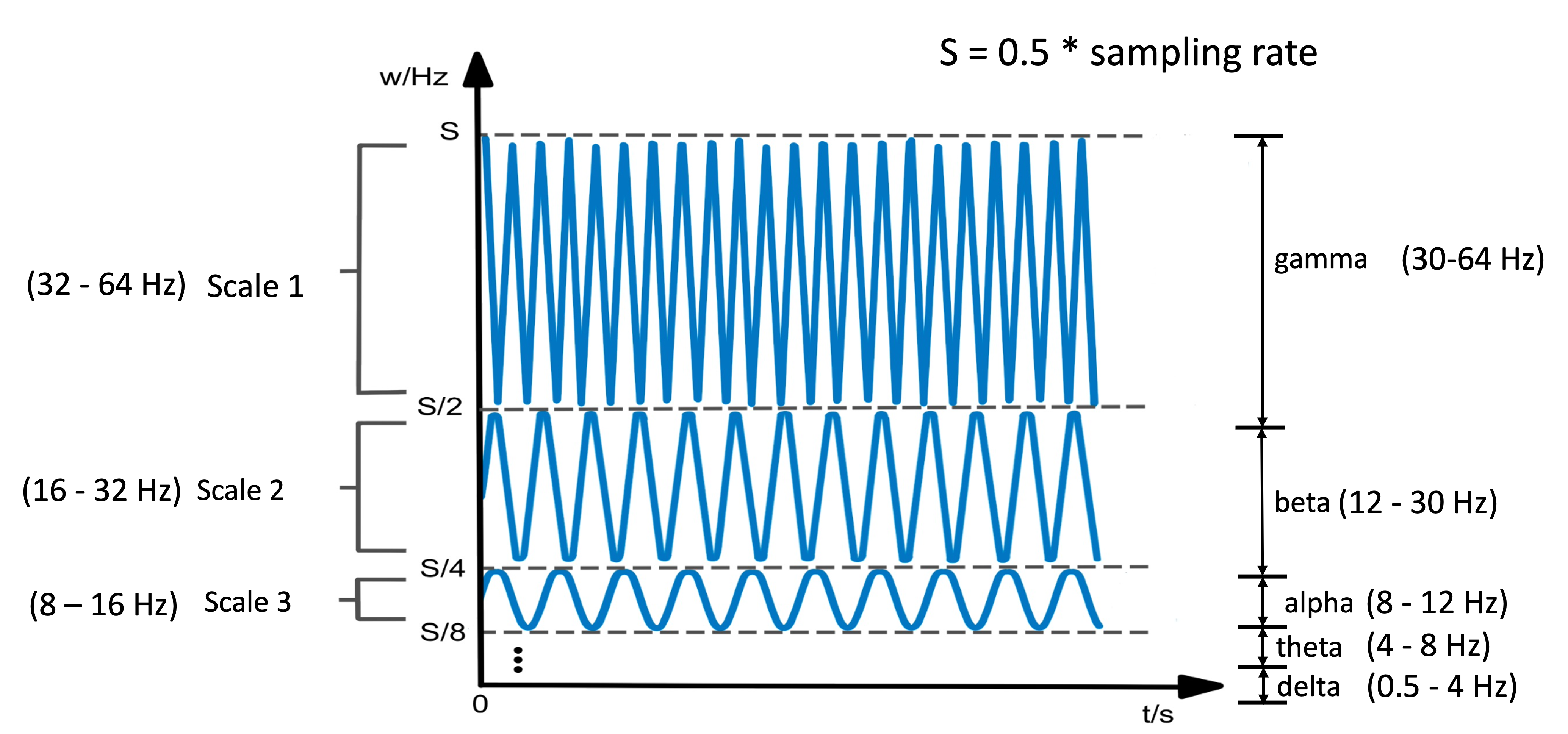}
    \caption{Relationships between the scale and frequency band of components at every scale in terms of the sampling rate of the original signal.}
    \label{fig:scale-freq}
\end{figure}

We analyzed an EEG dataset collected by \cite{eegdata}, consisting of 19-channel recordings sampled at 128 Hz from 50 subjects with ADHD and from 50 healthy controls. Preprocessing was performed using the PREP pipeline \citep{BigdelyShamlo2015PREP} to improve signal quality and remove artifacts arising from electrical interference and muscle activity, including eye blinks and ear movements. To capture brain regions most involved in the visual–cognitive task, we selected six EEG channels: Fp1 (left prefrontal), Fp2 (right prefrontal), T7 (left temporal), T8 (right temporal), O1 (left occipital), and O2 (right occipital) (see Figure \ref{fig:eegplot}), and for each subject we extracted an approximately 10-second segment corresponding to the core period of the task for subsequent analysis. Due to the low-pass and high-pass quadrature mirror filters used in the decomposition and reconstruction \citep{daubechies1992ten}, the wavelet spectra of the subprocesses at each scale have a specific correspondence with frequency bands, as illustrated in Figure \ref{fig:scale-freq}. 

Specifically, we aim to identify meaningful connectivity between components from different scales (frequency bands) across the selected channels in children with ADHD/ healthy control, motivated by the distinct functional roles of the corresponding brain regions: the frontal region is primarily associated with attention, the temporal region with speech and memory processing, and the occipital region with visual processing \citep{BjorgeEmaus2017}.

To study the dynamics of brain activity, we inspect the time-evolving single- and cross-scale dependencies among selected channels and subprocesses at multiple scales. The Haar wavelet was used as the analyzing wavelet. Figure \ref{fig:eeg_coh} (top) in Appendix \ref{app: add eeg results} shows the sample mean of single-scale $j=2$ coherence estimates among 50 ADHD subjects and 50 healthy control subjects, corresponding to the components from 16--32 Hz (roughly beta band) from the six channels.  The single-scale coherence analysis reveals clear frequency-dependent interactions across bilateral and inter-regional brain areas. In particular, the coherence between the prefrontal channels Fp1 (left) and Fp2 (right) at scale $j=2$ indicates pronounced linear dependence with distinct temporal patterns between the ADHD and control groups. This suggests altered functional coupling between the left and right prefrontal cortices, which are closely related to attentional control and executive functioning. In addition, the occipital channels O1 (left) and O2 (right) exhibit time-varying coherence at the same scale, reflecting dynamic inter-hemispheric interactions within the visual cortex. Notably, coherence between prefrontal and occipital regions (e.g., Fp1–O1, Fp1–O2) displays marked temporal variability, indicating nonstationary long-range functional connectivity between frontal and visual areas. These patterns differ systematically between the two groups, suggesting that cross-regional information integration across cognitive and sensory systems may be altered in children with ADHD.

We next utilize the proposed cross-scale coherence to investigate interactions between neural components operating at different frequency bands. Figure~\ref{fig:eeg_coh} (bottom) in Appendix \ref{app: add eeg results} displays the group-averaged time-varying cross-scale coherence between scales $(j,j')=(1,2)$, corresponding to the 32--64 Hz  (roughly gamma) and 16--32 Hz bands, respectively. Pronounced cross-scale dependencies are observed across multiple channel pairs, with clear temporal modulation. In particular, cross-scale coherence between bilateral prefrontal channels (Fp1–Fp2) and between prefrontal and occipital regions (e.g., Fp1–O1, Fp2–O2) exhibits distinct time-varying patterns that differ between the ADHD and control groups. Cross-scale interactions are also evident within the occipital region (O1–O2), indicating nonstationary coupling between higher- and lower-frequency components of visual cortical activity. These results suggest that cross-frequency coordination across both local and long-range brain networks differs between children with ADHD and healthy controls. Additional coherency results for other scale pairs are reported in Appendix \ref{app: add eeg results}, further indicating that dual-scale interactions are widespread across components and vary with the choice of scales.

The follow-up question is on the potential role of dual-scale dependencies to help discriminate between the ADHD and control groups. We address this question by evaluating cross-scale coherence among the above six channels for all 50 ADHD subjects and 50 control subjects. We employ the permutation test \citep{RazZhengOmbaoTuretsky2003} with 10,000 random permutations. This permutes the coherence between the given subprocesses across all subjects in ADHD and healthy control groups, and using the means of the estimated cross-scale coherence from two groups, determines whether the ADHD and control groups are statistically indistinguishable in terms of dual-scale coherence. To control for multiple comparisons across channel pairs, the permutation-based $p$-values were adjusted using the Benjamini–Hochberg false discovery rate (FDR) procedure \citep{BenjaminiHochberg1995FDR}. Full details are given in Appendix \ref{app: add eeg results}.

Denoting by `$X_{j}^{(p)}$' the scale-$j$ subprocess of signal from channel $p$, the results suggest that the dependence between $X_2^{(\text{Fp1})}$ and $X_1^{(\text{O2})}$ in the ADHD group is significantly weaker  than that in the control group ($p$-value $< 0.05$). Hence ADHD alters functional connectivity between these two brain regions at the corresponding frequency bands (16--32 Hz and 32--64 Hz). Statistically significant differences also exist in dependence between $X_2^{(\text{T7})}$ and $X_1^{(\text{O2})}$.

\begin{figure}[htbp]
    \centering
    \includegraphics[width=.9\textwidth]{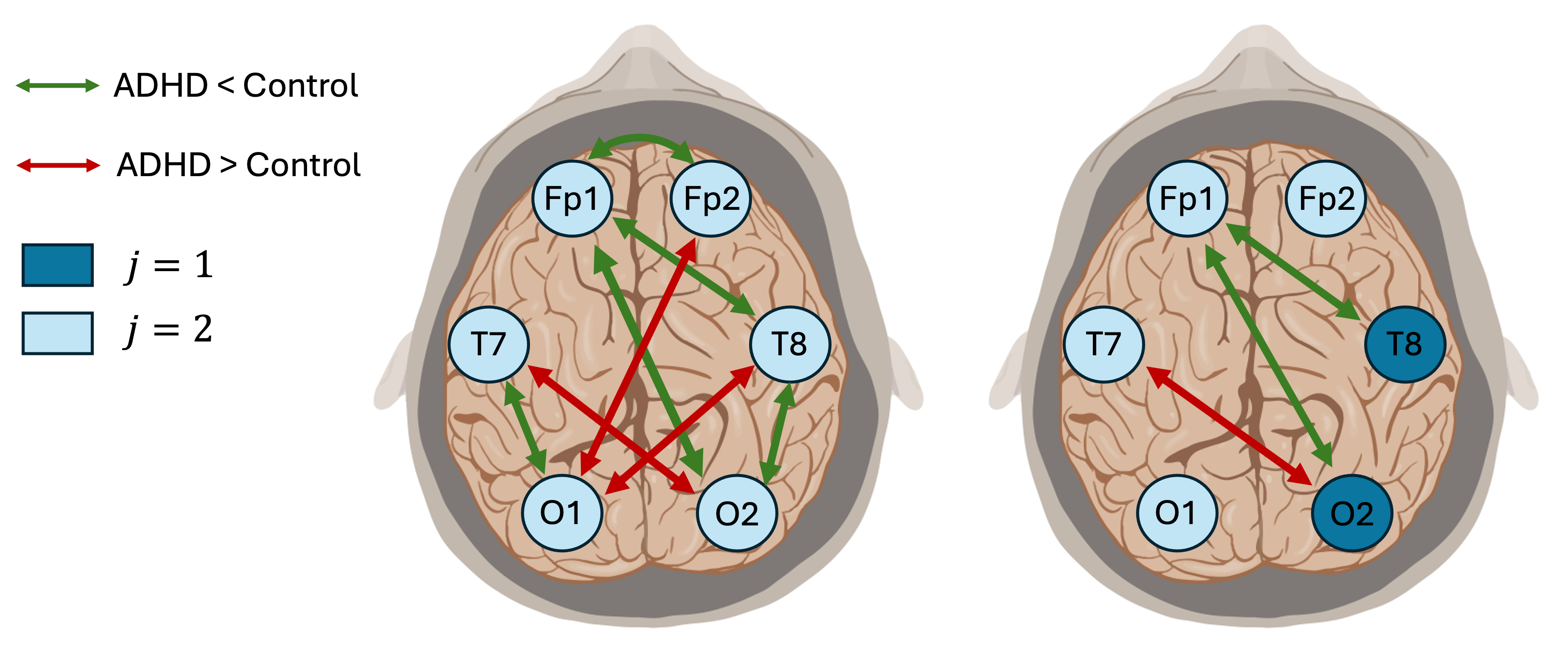}
    \caption{Schematic summary of group differences in functional connectivity between ADHD and control subjects. Left panel: single-scale connectivity at scale $j=2$ (approx. 16--32 Hz). Right panel: cross-scale connectivity between scales $j=1$ (32--64 Hz) and $j=2$ (16--32 Hz). Red (green) arrow: stronger (weaker) coherence in the ADHD group than in control.}
    \label{fig:dynamics}
\end{figure}

Pictorially, the brain dynamic patterns within and across the left and right hemispheres are depicted in Figure \ref{fig:dynamics}, and identify different functional single- and cross-scale (frequency) connectivity structures in children with ADHD compared to healthy controls.

Within the six selected channels, the analysis reveals clear ADHD-related alterations in both single-scale ($j=2$) and cross-scale ($(j,j')=(1,2)$) connectivity. Overall, several connections show stronger coherence in ADHD than in control subjects, but there are also prominent reductions. These findings suggest that ADHD is associated with both single-scale and cross-scale (frequency) disruptions in functional connectivity, especially involving occipital (O2) interactions with frontal (Fp1) and temporal (T7) activity.

Consistent with previous studies based on this dataset and related ADHD literature, our results confirm pronounced interactions among frontal, temporal, and occipital regions during cognitive tasks, particularly in the beta and gamma frequency bands, which are widely associated with attention and executive processing \citep{RedondoHuserOmbao2025}. 

Beyond these established findings, 
the proposed method reveals previously unreported cross-scale (frequency) interactions across the left and right brain hemispheres. Specifically, we identify cross-scale coherence between high-frequency (gamma/beta) and lower-frequency (alpha) components linking frontal–temporal and frontal–occipital networks. Notably, temporal-occipital cross-scale interactions are significantly enhanced in children with ADHD compared to controls, indicating altered cross-frequency coordination otherwise not captured by conventional single-scale or band-limited coherence analyses. 

\section{Discussion} \label{s:discuss}
We developed a rigorous multiscale wavelet based modeling framework which can capture time-evolutionary {\em cross-scale dependence} between components of multivariate {\em nonstationary} time series. This contribution has major practical significance  because, {\em for the first time}, it allows neuroscientists to investigate associations between long-term dynamics in one channel and short-term dynamics in another through the proposed {cross-scale} wavelet coherence. Theoretical contributions are the \underline{subprocess} introduction; formal definitions of novel cross-scale dependence measures, such as the \underline{local dual-scale covariance and coherence}; and the associated theory with their estimation procedure. The proposed construction concentrates on extracting the dependence structure of subprocesses at \underline{different} scales. The analysis of EEG data from ADHD versus healthy control children illustrates that cross-scale interactions exist and are time-evolving and informative for characterising group differences in brain networks. More broadly, the framework is applicable to other multiscale systems where short-term variability may relate to longer-term structure, including environmental, engineering, and economic time series. Simulation results  illustrate the dramatic impact that the presence of cross-scale activity has on the quality of MvLSW-based coherence estimation. A natural direction for future work could concentrate on improving the approximation of the latent subprocess components, especially for shorter series and challenging dependence regimes.

\section{Disclosure statement}\label{disclosure-statement}

The authors have  no conflicts of interest.

\section{Data Availability Statement}\label{data-availability-statement}

The real data used in Section \ref{EEG analysis} are publicly available \hyperlink{https://ieee-dataport.org/open-access/eeg-data-adhd-control-children}{\textbf{here}}.

\section{Acknowledgement}
MIK gratefully acknowledges support from  EPSRC NeST Programme Grant EP/X002195/1.


\phantomsection\label{supplementary-material}
\bigskip

\begin{center}

{\large\bf SUPPLEMENTARY MATERIAL}

\end{center}

The supplementary material includes appendices containing technical proofs, details of the simulation setup, additional results for simulation studies and real data analysis, as well as R code for implementing the proposed methods and reproducing the reported results.




\bibliography{references}

@article{abramovich2000wavelet,
  title={Wavelet analysis and its statistical applications},
  author={Abramovich, Felix and Bailey, Trevor C. and Sapatinas, Theofanis},
  journal={Journal of the Royal Statistical Society: Series D},
  volume={49},
  number={1},
  pages={1--29},
  year={2000},
  publisher={Wiley Online Library}
}

@article{cen2025inference,
author = {Zetai Cen and Yudong Chen and Clifford Lam},
title = {Inference on Dynamic Spatial Autoregressive Models with Change Point Detection},
journal = {Journal of Business \& Economic Statistics},
pages = {1-14},
year = {2026},
publisher = {Taylor \& Francis},
doi = {10.1080/07350015.2025.2572768},
}

@article{cribben2016estimating,
    author = {Cribben, Ivor and Yu, Yi},
    title = {Estimating Whole-Brain Dynamics by Using Spectral Clustering},
    journal = {Journal of the Royal Statistical Society: Series C (Applied Statistics)},
    volume = {66},
    number = {3},
    pages = {607-627},
    year = {2017},
    month = {09}
}

@article{palasciano2025continuous,
    author = {Palasciano, H A and Knight, M I and Nason, G P},
    title = {Continuous-time locally stationary wavelet processes},
    journal = {Biometrika},
    volume = {112},
    number = {2},
    pages = {asaf015},
    year = {2025},
    month = {03},
    issn = {1464-3510}
}

@book{daubechies1992ten,
 author={Daubechies, Ingrid},
  title={Ten lectures on wavelets},
  year={1992},
  publisher={SIAM}
}

@article{ombao2014, 
author={Park, Timothy and Eckley, Idris and Ombao, Hernando},
title={Estimating Time-Evolving Partial Coherence Between Signals via Multivariate Locally Stationary Wavelet Processes},  
year={2014},  
pages={5240--5250},
volume={62}, 
journal={IEEE Transactions on Signal Processing}, 
}

@article{nason2000wavelet,
author={Nason, Guy P. and Sachs, Rainer and Kroisandt, Gerald},  
title={Wavelet processes and adaptive estimation of the evolutionary wavelet spectrum},
journal={Journal of the Royal Statistical Society: Series B (Statistical Methodology)},
volume={62},
number={2},
pages={271--292},
year={2000},
publisher={Wiley Online Library}
}

@book{Priestley,
title = { Spectral analysis and time series},
author = { Priestley, M. B. },
publisher = { Academic Press London ; New York },
pages = { 2 v. (xvii, [45], 890 p.) : },
year = { 1981 },
type = { Book },
}

@article{Dahlaus,
  title = { Fitting Time Series Models to Nonstationary Processes},
  author = {Dahlhaus, Rainer},
 journal = {The Annals of Statistics},
 number = {1},
 pages = {1--37},
 publisher = {Institute of Mathematical Statistics},
 volume = {25},
 year = {1997}
}

@article{fiecas2016modeling,
  title={Modeling the evolution of dynamic brain processes during an associative learning experiment},
  author={Fiecas, Mark and Ombao, Hernando},
  journal={Journal of the American Statistical Association},
  volume={111},
  number={516},
  pages={1440--1453},
  year={2016},
  publisher={Taylor \& Francis}
}

@article{embleton2022multiscale,
title={Multiscale spectral modelling for nonstationary time series within an ordered multiple-trial experiment},
author={Embleton, Jonathan and Knight, Marina I. and Ombao, Hernando},
journal={Annals of Applied Statistics},
volume={16},
number={4},
pages={2774-2803},
year={2022},
publisher={York}
}

@article{killick2020,
title={The local partial autocorrelation function and some applications},
author={Killick, Rebecca and Knight, Marina I. and Nason, Guy P. and Eckley, Idris A.},
journal={Electronic Journal of Statistics}, 
volume={14}, 
pages={3268-3314},
year={2020}
}

@article{sanderson,
 title = {Estimating linear dependence between nonstationary time series using the locally stationary wavelet model},
 author = {J. Sanderson and P. Fryzlewicz and M. W. Jones},
 journal = {Biometrika},
 number = {2},
 pages = {435--446},
 volume = {97},
 year = {2010}
}

@article{koopmans1964multivariate,
  title={On the multivariate analysis of weakly stationary stochastic processes},
  author={Koopmans, LH},
  journal={The Annals of Mathematical Statistics},
  volume={35},
  number={4},
  pages={1765--1780},
  year={1964},
  publisher={JSTOR}
}

@article{ombao2005,
title = {{SLEX} Analysis of Multivariate Nonstationary Time Series},
author = {Ombao, Hernando and Sachs, Rainer and Guo, Wensheng },
journal = {Journal of the American Statistical Association},
number = {470},
pages = {519--531},
publisher = {[American Statistical Association, Taylor & Francis, Ltd.]},
volume = {100},
year = {2005}
}

@article{moment,
 author = {Isserlis, Leon},
  title = {On a Formula for the Product-Moment Coefficient of any Order of a Normal Frequency Distribution in any Number of Variables},
 journal = {Biometrika},
 number = {1/2},
 pages = {134--139},
 publisher = {[Oxford University Press, Biometrika Trust]},
 volume = {12},
 year = {1918}
}

@article{stock,
author = {Basu, Sumanta and Das, Sreyoshi and Michailidis, George and Purnanandam, Amiyatosh},
title = {A System-wide Approach to Measure Connectivity in the Financial Sector},
journal = {Social Science Research Network},
year = {2019}

}

@article{eegdata,
doi = {10.21227/rzfh-zn36},
url = {https://dx.doi.org/10.21227/rzfh-zn36},
author = {Ali Motie Nasrabadi and Armin Allahverdy and Mehdi Samavati and Mohammad Reza Mohammadi},
publisher = {IEEE Dataport},
title = {{EEG} data for {ADHD} / Control children},
year = {2020} }

@article{TCAM1070,
	author = {Brassarote, Gabriela and Souza, Eniuce  and Monico, João },
	title = {Non-decimated Wavelet Transform for a Shift-invariant Analysis},
	journal = {Trends in Computational and Applied Mathematics},
	volume = {19},
	number = {1},
	year = {2018},
	pages = {93}
	
}

@Book{billingsley1999convergence,
  Title                    = {Convergence of Probability Measures},
  Author                   = {Billingsley, P.},
  Publisher                = {Wiley},
  Year                     = {1999},
  Address                  = {New York}
}

@Article{fryz03:forecasting,
  Title                    = {Forecasting non-stationary time series by wavelet process modelling},
  Author                   = {Fryzlewicz, P. and Van~Bellegem, S. and von~Sachs, R.},
  journal = {Annals of the Institute of Statistical Mathematics},
  Year                     = {2003},
  Pages                    = {737--764},
  Volume                   = {55}
}

@Article{knight24:jcgs,
  Title                    = {Adaptive wavelet domain principal component analysis for nonstationary time series},
  Author                   = {Knight, Marina I. and Nunes, Matthew A. and Hargreaves, Jessica K.},
  Journal                  = {Journal of Computational and Graphical Statistics},
  Year                     = {2024},
  Pages                    = {1--14},
  Volume                   = {33}
}

@article{BjorgeEmaus2017,
  author    = {Bj{\o}rge, L.-E. N. and Emaus, T. H.},
  title     = {Identification of {EEG}-based Signature Produced by Visual Exposure to the Primary Colors {RGB}},
  school    = {Norwegian University of Science and Technology},
  year      = {2017},
  type      = {Master's thesis}
}

@article{RazZhengOmbaoTuretsky2003,
  author  = {Raz, Jonathan and Zheng, Hui and Ombao, Hernando and Turetsky, Bruce},
  title   = {Statistical tests for {fMRI} based on experimental randomization},
  journal = {NeuroImage},
  volume  = {19},
  number  = {2},
  pages   = {226--232},
  year    = {2003},
  month   = {6},
  publisher = {Elsevier}
}

@article{RedondoHuserOmbao2025,
  author  = {Redondo, Paolo Victor and Huser, Rapha{\"e}l and Ombao, Hernando},
  title   = {Measuring information transfer between nodes in a brain network through spectral transfer entropy},
  journal = {The Annals of Applied Statistics},
  volume  = {19},
  number  = {3},
  pages   = {2386--2411},
  year    = {2025},
  publisher = {Institute of Mathematical Statistics}
}

@article{BigdelyShamlo2015PREP,
  author  = {Bigdely-Shamlo, Nima and Mullen, Tim and Kothe, Christian and Su, Kyung-Min and Robbins, Kay A.},
  title   = {The {PREP} pipeline: standardized preprocessing for large-scale {EEG} analysis},
  journal = {Frontiers in Neuroinformatics},
  year    = {2015},
  volume  = {9},
  pages   = {16}
}

@article{BenjaminiHochberg1995FDR,
  author  = {Benjamini, Yoav and Hochberg, Yosef},
  title   = {Controlling the False Discovery Rate: A Practical and Powerful Approach to Multiple Testing},
  journal = {Journal of the Royal Statistical Society: Series B (Statistical Methodology)},
  year    = {1995},
  volume  = {57},
  number  = {1},
  pages   = {289--300}
}

@article{TaylorParkEckley2019mvLSW,
  author  = {Taylor, S. A. C. and Park, T. and Eckley, I. A.},
  title   = {Multivariate Locally Stationary Wavelet Analysis with the {mvLSW} {R} Package},
  journal = {Journal of Statistical Software},
  year    = {2019},
  volume  = {90},
  number  = {11},
  pages   = {1--19}
}

@article{OmbaoPinto2024,
  title   = {Spectral Dependence},
  author  = {Ombao, Hernando and Pinto, Marco},
  journal = {Econometrics and Statistics},
  volume  = {32},
  pages   = {122--159},
  year    = {2024}
}

@article{Canolty2010,
  title   = {The functional role of cross-frequency coupling},
  author  = {Canolty, Ryan T. and Knight, Robert T.},
  journal = {Trends in Cognitive Sciences},
  volume  = {14},
  number  = {11},
  pages   = {506--515},
  year    = {2010}
}

@article{Siebenhuhner2016,
  title   = {Cross-frequency synchronization connects networks of fast and slow oscillations during visual working memory maintenance},
  author  = {Siebenh{\"u}hner, Felix and Wang, Sheng H. and Palva, J. Matias and Palva, Satu},
  journal = {eLife},
  volume  = {5},
  pages   = {e13451},
  year    = {2016}
}

@article{deHemptinne2013,
  title   = {Exaggerated phase-amplitude coupling in the primary motor cortex in Parkinson disease},
  author  = {de Hemptinne, Coralie and Ryapolova-Webb, Elena S. and Air, Ellen L. and Garcia, Paul A. and Miller, Kai J. and Ojemann, Jeffrey G. and Ostrem, Jill L. and Galifianakis, Nicholas B. and Starr, Philip A.},
  journal = {Proceedings of the National Academy of Sciences of the United States of America},
  volume  = {110},
  number  = {12},
  pages   = {4780--4785},
  year    = {2013}
}

@article{Tanaka2022,
  title   = {Magnetoencephalography detects phase-amplitude coupling in {P}arkinson's disease},
  author  = {Tanaka, Masahito and Yanagisawa, Takufumi and Fukuma, Ryohei and others},
  journal = {Scientific Reports},
  volume  = {12},
  pages   = {1835},
  year    = {2022}
}

\clearpage

\begin{center}
{\Large\bfseries Supplementary Material for `Dynamic cross-scale wavelet coherence'\par}
\vspace{0.5em}
{\normalsize \textit{ Haibo Wu, Marina I. Knight and Hernando Ombao}\par}
\end{center}
\vspace{1em}

\appendix

\section{Appendix A (Technical proofs)}
\subsection{Covariance approximations}\label{app:cov}

\textsc{Proof of Proposition~\ref{proposition1}:}

Recalling the representation of a scale-$j$ component of the $P$-variate locally stationary wavelet process introduced in (3) yields
\begin{align*}
    &\cov(X_{j,[uT]}^{(p)},X_{j,[uT]+\tau}^{(q)}) = \E \left[ X_{j,[uT]}^{(p)} X_{j,[uT]+\tau}^{(q)}\right] \notag \\
    &=\E \left[  \left(\sum\limits_{k} \mathbf{V}_j^{(p)}(k/T) \psi_{j,k}([uT]) \mathbf{z}_{j,k}\right)
     \times  \left(\sum\limits_{k'} \mathbf{V}_{j}^{(q)}(k'/T)\psi_{j,k'}([uT]+\tau)\mathbf{z}_{j,k'} \right)^{\top}     \right] \notag \\
    & = \sum\limits_{k}\sum\limits_{k'} \mathbf{V}_j^{(p)}(k/T) \psi_{j,k}([uT]) E[\mathbf{z}_{j,k}\mathbf{z}^{\top}_{j,k'}]\psi_{j,k'}([uT]+\tau)\mathbf{V}_{j}^{(q)}(k'/T)^\top,
\end{align*}
where $\mathbf{V}_j^{(p)}(u)$ denotes the $p$th row of the $\mathbf{V}_j(u)$ transfer matrix of $\{ \mathbf{X}_{t;T}\}$.

From the random vector innovation construction, $\E[\mathbf{z}_{j,k}\mathbf{z}^{\top}_{j,k'}]=\cov(\mathbf{z}_{j,k},\mathbf{z}_{j,k'})=\boldsymbol{Q}_{j,j}(k/T)\delta_{k,k'}$, where $\boldsymbol{Q}_{j,j}(k/T)=\boldsymbol{I}$ and using the definition of the LWS matrix, $S_{j}^{(p,q)}(u)= \mathbf{V}_{j}^{(p)}(u)\boldsymbol{Q}_{j,j}(u)\mathbf{V}_{j}^{(q)}(u)^{\top}$, letting $m=k-[uT]$ we obtain
\begin{align*}
    \cov(X_{j,[uT]}^{(p)},X_{j,[uT]+\tau}^{(q)})= \sum\limits_{m}S^{(p,q)}_j\left(\frac{[uT]+m}{T}\right)\psi_{j,m}(0)\psi_{j,m}(\tau).
\end{align*} 
Analogous to the method proposed by \cite{koopmans1964multivariate}, using the Lipschitz continuity of $S_j^{(p,q)}(u)$, we consider the difference between this single-scale covariance and the subprocess-based version, 
namely\\ $\cov(X_{j,[uT]}^{(p)},X_{j,[uT]+\tau}^{(q)})-c_j^{(p,q)}(u,\tau)  = \sum\limits_{m}S^{(p,q)}_j\left(\frac{[uT]+m}{T}\right)\psi_{j,m}(0)\psi_{j,m}(\tau) - c_j^{(p,q)}(u,\tau)$. Thus,  
\begin{align*}
   & \left| \cov(X_{j,[uT]}^{(p)},X_{j,[uT]+\tau}^{(q)})-c_j^{(p,q)}(u,\tau) \right| \\
    &\leq T^{-1}\sum\limits_{m}|m|L^{(p,q)}_{jj}|\psi_{j,m}(0)\psi_{j,m}(\tau)|=\mathcal{O}(2^{-j}T^{-1}), \forall p, \, q
\end{align*}
where we used the scale-$j$ autocorrelation wavelet property of having a compact support of length $2^j$, coupled with $\boldsymbol{\Psi}_{j}(\tau)=\mathcal{O}(1)$ and the property of the Lipschitz constants $\sum\limits_{j}2^{2j}L_{jj}^{(p,q)}<\infty$. Similarly, 
\begin{align*}
    &\cov(X_{j,[uT]}^{(p)},X_{j',[uT]+\tau}^{(q)}) = \E \left[ X_{j,[uT]}^{(p)} X_{j',[uT]+\tau}^{(q)}\right] \notag \\
    &=\E \left[ \left( \sum\limits_{k} \mathbf{V}_j^{(p)}(k/T) \psi_{j,k}([uT]) \mathbf{z}_{j,k}\right)
     \times  \left(\sum\limits_{k'} \mathbf{V}_{j'}^{(q)}(k'/T)\psi_{j',k'}([uT]+\tau)\mathbf{z}_{j',k'} \right)^{\top}     \right] \notag \\
    & = \sum\limits_{k}\sum\limits_{k'} \mathbf{V}_j^{(p)}(k/T) \psi_{j,k}([uT]) \E[\mathbf{z}_{j,k}\mathbf{z}^{\top}_{j',k'}]\psi_{j',k'}([uT]+\tau)\mathbf{V}_{j'}^{(q)}(k'/T)^{\top},
\end{align*}
where $\E[\mathbf{z}_{j,k}\mathbf{z}^{\top}_{j',k'}]=\cov(\mathbf{z}_{j,k},\mathbf{z}_{j',k'})=\boldsymbol{Q}_{j,j'}(k/T)\delta_{k,k'}$. From the definition of the cross-scale LWS matrix, $S_{jj'}^{(p,q)}(u)= \mathbf{V}_{j}^{(p)}(u)\boldsymbol{Q}_{j,j'}(u)\mathbf{V}_{j'}^{(q)}(u)^{\top}$, taking $m=k-[uT]$, the above becomes 
\begin{align*}
    \cov(X_{j,[uT]}^{(p)},X_{j',[uT]+\tau}^{(q)})= \sum\limits_{m}S^{(p,q)}_{jj'}\left(\frac{[uT]+m}{T}\right)\psi_{j,m}(0)\psi_{j',m}(\tau).
\end{align*}
 Using the Lipschitz continuity of each $S_{jj'}^{(p,q)}(u)$, we consider the difference 
\begin{align*}
   & \left| \cov(X_{j,[uT]}^{(p)},X_{j',[uT]+\tau}^{(q)})-c_{jj'}^{(p,q)}(u,\tau) \right| \\
    &=\left| \sum\limits_{m}S^{(p,q)}_{jj'}\left(\frac{[uT]+m}{T}\right)\psi_{j,m}(0)\psi_{j',m}(\tau) - c_{jj'}^{(p,q)}(u,\tau)\right| \\
    &\leq T^{-1}\sum\limits_{m}|m|L_{jj'}^{(p,q)}|\psi_{j,m}(0)\psi_{j',m}(\tau)|=\mathcal{O}({2^{-(j+j')/2}}T^{-1}),
\end{align*}
where we used the Cauchy-Schwarz inequality and the scales $j$, $j'$ autocorrelation wavelet properties above, coupled with the Lipschitz constants property $\sum\limits_{j,j'}2^{j+j'}L_{jj'}^{(p,q)}<\infty, \forall p, \, q$.
\\
\noindent \textsc{Proof of Corollary~\ref{corllary1}:}

Using the process representation $X_{t;T}^{(p)}=\sum_{j}X_{j,t}^{(p)}$ for any channel $p$ and the spectral Lipschitz properties, let us first take the process cross-covariance with its scale-$j$ subprocess, 
\begin{align*}
   & \left| \cov(X_{j,[uT]}^{(p)},X_{[uT]+\tau;T}^{(q)})-\tilde{c}_{j}^{(p,q)}(u,\tau) \right| \\
   &= \left| \sum_{j'} \left(\cov(X_{j,[uT]}^{(p)},X_{j';[uT]+\tau}^{(q)})-\tilde{c}_{jj'}^{(p,q)}(u,\tau) \right)\right| \\
    &\leq \sum_{j'} \left| \sum\limits_{m}S^{(p,q)}_{jj'}\left(\frac{[uT]+m}{T}\right)\psi_{j,m}(0)\psi_{j',m}(\tau) - c_{jj'}^{(p,q)}(u,\tau)\right| \\
    &\leq T^{-1}\sum_{j'}\sum\limits_{m}|m|L_{jj'}^{(p,q)}|\psi_{j,m}(0)\psi_{j',m}(\tau)|=\mathcal{O}(2^{-j/2}T^{-1}).
\end{align*}
Similarly, we obtain the overall process cross-covariance approximation.
\\

\subsection{Covariance-spectrum characterisation} \label{app:corr}
\noindent \textsc{Proof of Property~\ref{property1}:}

(i) As with any process representation, we need to establish its uniqueness in the sense that if we assume two different subprocess spectral representations $\{\mathbf{S}_{jl}(\cdotp)\}_l$ and $\{\tilde{\mathbf{S}}_{jl}(\cdotp)\}_l$ exist corresponding to a unique subprocess cross-covariance structure $\{\tilde{c}_{j}(u,\tau)\}_j$, then the spectral representations should coincide. Mathematically, the common scale $j$-subprocess cross-covariance structure $\{\tilde{c}_{j}(u,\tau)\}_j$ at rescaled time $u$ and lag $\tau$ can be re-written as 
$$
\sum_{l=1}^J\left(S_{jl}^{(p,q)}(u)-\tilde{S}_{jl}^{(p,q)}(u) \right)
\boldsymbol{\Psi}_{jl}(\tau)=0, \forall \tau, u,
$$
and we want to show that this implies $e_{jl}^{(p,q)}(u)=0$, for all scales $(j,l)\in \mathcal{B}_h$, channel pairs $(p,q)$ and rescaled time $u$, where we used the notation $e_{jl}^{(p,q)}(u)=S_{jl}^{(p,q)}(u)-\tilde{S}_{jl}^{(p,q)}(u)$. This is equivalent to showing that for each scale $j$, the family $\{ \boldsymbol{\Psi}_{jl}(\tau) \}_{l=1}^J$ is linearly independent.

As in \cite{nason2000wavelet}, we take the expression above in the Fourier domain via Parseval's identity, here for fixed channels $p, \, q$, which in this particular context become
\begin{align*}
\sum_{l=1}^J{e}_{jl}^{(p,q)}(u) \boldsymbol{\Psi}_{jl}(\tau)&=0, \forall \tau, u,\\
\sum_{l}\sum_{l'}{e}_{jl}^{(p,q)}(u){e}_{jl'}^{(p,q)}(u) \sum_{\tau}\boldsymbol{\Psi}_{jl}(\tau)\boldsymbol{\Psi}_{jl'}(\tau)&=0, \forall u,\\
\sum_{l}\sum_{l'}{e}_{jl}^{(p,q)}(u){e}_{jl'}^{(p,q)}(u) \frac{1}{2\pi}\int \widehat{\boldsymbol{\Psi}}_{jl}(\omega)\overline{\widehat{\boldsymbol{\Psi}}}_{jl'}(\omega)d \omega&=0, \forall u,\\
\int \left| \sum_{l} {e}_{jl}^{(p,q)}(u)\widehat{\boldsymbol{\Psi}}_{jl}(\omega)\right|^2 d \omega&=0, \forall u,
\end{align*}
where here $\widehat \cdotp$ in the above denotes the Fourier transform and $\left| \cdotp \right|$ denotes the complex modulus. We thus have for each scale $j$,
$$
\sum_{l} {e}_{jl}^{(p,q)}(u)\widehat{\boldsymbol{\Psi}}_{jl}(\omega)=0, \forall \omega, u,
$$
where it is easily shown that the Fourier transform of the cross-scale autocorrelation wavelet is $\widehat{\boldsymbol{\Psi}}_{jl}(\omega)= \hat{\psi}_{j}(\omega)\overline{\hat{\psi}}_{l}(\omega)$.

We will next show that the above equations imply that $\tilde{e}_{jl}^{(p,q)}(u)=0, \forall j,\,l,\,u$ for each $p,\,q$ for the Haar wavelets, where $\tilde{e}_{jl}^{(p,q)}(u)=2^{-(j+l)/2}{e}_{jl}^{(p,q)}(u)$, and the proof follows in a similar manner for Daubechies wavelets with compact support (by exploiting the periodic nature of their zeroes in the Fourier domain).

\cite{killick2020} have shown the Fourier transform of the discrete Haar wavelet at scale $j$ is $\hat{\psi}_{j}(\omega)=2^{-j/2} \frac{(1-\mbox{exp}{(-i \omega 2^{j-1}}))^2}{1-\mbox{exp}{(-i \omega)}}$. The Fourier transform of the cross-correlation wavelet is 
$
\widehat{\boldsymbol{\Psi}}_{jl}(\omega)= 2^{-(j+l)/2} (1-cos \omega 2^{j-1})(1-cos \omega 2^{l-1})\mbox{exp}{(-i \omega(2^{j-1}-2^{l-1}))}/ (1-cos \omega),
$
and
$$
\sum_{l=1}^J \tilde{e}_{jl}^{(p,q)}(u)(1-cos \omega 2^{l-1})\mbox{exp}{(i \omega2^{l-1})}=0, \forall u, \omega.
$$
Dropping for ease the channel pair superscript $(p,q)$ and taking $\omega:=\pi$ in the above, the only non-zero term is the one corresponding to $l=1$ which yields $\tilde e_{j1}(u)=0$ for all $u$, therefore $e_{j1}(u)=0$ for all $u$. In a similar vein, taking next $\omega:=\pi/2$ and using the fact that $\tilde e_{j1}(u)=0$ for all $u$, the only non-zero term corresponds to $l=2$, thus $e_{j2}(u)=0$ for all $u$. For a general scale $l$, taking $\omega:=\pi/2^{l-1}$ we obtain $e_{jl}(u)=0$ for all rescaled times $u$ and scales $j,\, l$, therefore the dual-scale spectral representation is unique ($\mathbf{S}=\tilde{\mathbf{S}}$). 

An additional implication is that for fixed $j$, the system $\{ \boldsymbol{\Psi}_{jl}(\tau) \}_{l=1}^J$ is linearly independent. Hence, since the matrices $\mathbf{A}^{jj}$ can be viewed as Gram matrices corresponding to the vectors $\Psi_{j1}, \ldots, \Psi_{jJ}$, they are invertible for each scale $j$.

Let us now establish the subprocess covariance connection to the cross-scale spectra. Recalling that $\tilde{c}^{(p,q)}_j(u,\tau)=\sum\limits_{j'}S^{(p,q)}_{jj'}(u)\boldsymbol{\Psi}_{jj'}(\tau)$, we obtain for any scales $j, \,l$, time $u$,
\begin{align*}
\sum_{\tau}\tilde{c}^{(p,q)}_j(u,\tau) \boldsymbol{\Psi}_{jl}(\tau)&=\sum_{j'}S^{(p,q)}_{jj'}(u)\sum_{\tau}\boldsymbol{\Psi}_{jj'}(\tau)\boldsymbol{\Psi}_{jl}(\tau),\\
&=\sum_{j'}A_{jj';jl}S^{(p,q)}_{jj'}(u)
=\sum_{j'}A_{jj;j'l}S^{(p,q)}_{jj'}(u).
\end{align*}
Using the symmetry of the $\mathbf{A}^{jj}$ matrix, the equation in Property 1 (i) is obtained. Furthermore, the invertibility of the (symmetrical) matrix $\mathbf{A}^{jj}$ ensures the invertibility of the above to yield all cross-scale $(j,j')$ spectra as,
$$
S_{jj'}^{(p,q)}(u)=\sum_{l} [(\mathbf{A}^{jj})^{-1}]_{j'l}\sum_{\tau}\tilde{c}^{(p,q)}_j(u,\tau) \boldsymbol{\Psi}_{jl}(\tau).
$$
(ii) Equation~\eqref{eq:proccovspec} in Section~\ref{sec:procest} shows that the process covariance structure may be written for any channel pairs $(p,q)$ in terms of the cross-spectral terms, as  
\begin{align*}  
\tilde{c}^{(p,q)}(u,\tau)
&= \sum\limits_{(j,j') \in \mathcal{B}_h} S_{jj'}^{(p,q)}(u) \boldsymbol{\Psi}_{j j'}(\tau),\\
&=\sum\limits_{\delta=0}^h\sum\limits_{j=1}^{J-\delta} S_{j,j+\delta}^{(p,q)}(u) \boldsymbol{\Psi}_{j, j+\delta}(\tau)
+ \sum\limits_{\delta=1}^h\sum\limits_{j=1}^{J-\delta} S_{j,j+\delta}^{(q,p)}(u) \boldsymbol{\Psi}_{j, j+\delta}(-\tau),
\, \forall u, \tau. 
\end{align*}
We aim to show that should another spectral representation $\{ \tilde{S}_{jj'}^{(p,q)}(u) \}_{j,j'}$ correspond to the process covariance structure $\tilde{c}^{(p,q)}(u,\tau)$, then we must have
$$
e_{jj'}^{(p,q)}(u):=S_{jj'}^{(p,q)}(u)-\tilde{S}_{jj'}^{(p,q)}(u) \equiv 0,
$$
for all scales $(j,j')\in \mathcal{B}_h$, channel pairs $(p,q)$ and rescaled time $u$.

Equivalently, we start with 
$
\sum\limits_{\delta=0}^h\sum\limits_{j=1}^{J-\delta} e_{j,j+\delta}^{(p,q)}(u) \boldsymbol{\Psi}_{j, j+\delta}(\tau)
+ \sum\limits_{\delta=1}^h\sum\limits_{j=1}^{J-\delta} e_{j,j+\delta}^{(q,p)}(u) \boldsymbol{\Psi}_{j, j+\delta}(-\tau) \equiv 0, \, \forall u, \, \tau,
$
which akin to the derivation at point (i), may be translated into the Fourier domain as 
\begin{equation}\label{eq:specFourier}
\sum\limits_{\delta=0}^h\sum\limits_{j=1}^{J-\delta} e_{j,j+\delta}^{(p,q)}(u) \widehat{\boldsymbol{\Psi}}_{j, j+\delta}(\omega)
+ \sum\limits_{\delta=1}^h\sum\limits_{j=1}^{J-\delta} e_{j,j+\delta}^{(q,p)}(u) \widehat{\boldsymbol{\Psi}}_{j, j+\delta}(-\omega) \equiv 0, \, \forall u, \, \omega.
\end{equation}

Showing that equation~\eqref{eq:specFourier} implies $\{ {e}_{jj'}^{(p,q)}(u) \equiv 0 \}_{(j,j')\in\mathcal{B}_h}$ is equivalent to showing that the family $\{ \{\boldsymbol{\Psi}_{j, j+\delta}(\tau)\}_{{\delta=0:h},{j=1:J-\delta}}, \{\boldsymbol{\Psi}_{j, j+\delta}(-\tau)\}_{{\delta=1:h},{j=1:J-\delta}} \}$ is linearly independent.

We will proceed by first showing that the sub-family $\{ 
\{\boldsymbol{\Psi}_{j, j}(\tau)\}_{j=1}^{J}, \{\boldsymbol{\Psi}_{j, j+h}(\tau)\}_{j=1}^{J-h}, \{\boldsymbol{\Psi}_{j, j+h}(-\tau)\}_{j=1}^{J-h} \}$ is linearly independent for any finite, fixed $h$ such that $h<J$, or equivalently, that
\begin{equation}\label{eq:specFourierh}
\sum\limits_{j=1}^{J} e_{j,j}^{(p,q)}(u) \widehat{\boldsymbol{\Psi}}_{j, j}(\omega)
+ \sum\limits_{j=1}^{J-h} \left( e_{j,j+h}^{(p,q)}(u) \widehat{\boldsymbol{\Psi}}_{j, j+h}(\omega)
+ e_{j,j+h}^{(q,p)}(u) \widehat{\boldsymbol{\Psi}}_{j, j+h}(-\omega) \right) \equiv 0, \, \forall u, \, \omega,
\end{equation}
implies $e_{j,j}^{(p,q)}(u)=e_{j,j+h}^{(p,q)}(u)=e_{j,j+h}^{(q,p)}(u) \equiv 0$. 

Writing equation~\eqref{eq:specFourierh} for both $\omega$ and $(-\omega)$, and subtracting gives
$$
\sum\limits_{j=1}^{J-h} \left( e_{j,j+h}^{(p,q)}(u) - e_{j,j+h}^{(q,p)}(u)\right) \left(\widehat{\boldsymbol{\Psi}}_{j, j+h}(\omega)
- \widehat{\boldsymbol{\Psi}}_{j, j+h}(-\omega) \right) \equiv 0, \, \forall u, \, \omega.
$$
Recalling from (i) that the Fourier transform of the cross-correlation Haar wavelet is 
$
\widehat{\boldsymbol{\Psi}}_{jl}(\omega)= 2^{-(j+l)/2} (1-cos \omega 2^{j-1})(1-cos \omega 2^{l-1})\mbox{exp}{(-i \omega(2^{j-1}-2^{l-1}))}/ (1-cos \omega),
$
we shall next recursively use the fact that $\widehat{\boldsymbol{\Psi}}_{j, j+h}(\omega)
- \widehat{\boldsymbol{\Psi}}_{j, j+h}(-\omega)=2 \mbox{Im}(\widehat{\boldsymbol{\Psi}}_{j, j+h}(\omega))$, namely
$$
\widehat{\boldsymbol{\Psi}}_{j, j+h}(\omega)
- \widehat{\boldsymbol{\Psi}}_{j, j+h}(-\omega)= 2 i \times 2^{-(2j+h)/2} (1-cos \omega 2^{j-1})(1-cos \omega 2^{j+h-1}) sin \omega 2^{j-1}(1-2^h)/ (1-cos \omega).
$$
Taking $\omega:=\pi/2^h$, we note that $1-cos \omega 2^{j+h-1}=0$ iff $j \geq 2$, hence $e_{1,1+h}^{(p,q)}(u) = e_{1,1+h}^{(q,p)}(u), \, \forall u$ as $cos(\pi/2^h) \neq 1$ and $sin \pi(1-2^{h})/2^h \neq 0$.

Similarly, take $\omega:=\pi/2^{h+1}$ and obtain that $1-cos \omega 2^{j+h-1}=0$ iff $j \geq 3$, hence using the same arguments we obtain $e_{2,2+h}^{(p,q)}(u) = e_{2,2+h}^{(q,p)}(u), \, \forall u$. In general, we obtain for any channel pair $(p,q)$,
$$
e_{j,j+h}^{(p,q)}(u) = e_{j,j+h}^{(q,p)}(u), \, \forall u, \, j\in \overline{1, J-h}.
$$
Replacing these into~\eqref{eq:specFourierh} and using $\widehat{\boldsymbol{\Psi}}_{j, j+h}(\omega)
+ \widehat{\boldsymbol{\Psi}}_{j, j+h}(-\omega)=2 \mbox{Re}(\widehat{\boldsymbol{\Psi}}_{j, j+h}(\omega))$, we have 
\begin{equation*}
\sum\limits_{j=1}^{J} e_{j,j}^{(p,q)}(u) \widehat{\boldsymbol{\Psi}}_{j, j}(\omega)
+ \sum\limits_{j=1}^{J-h} \left( e_{j,j+h}^{(p,q)}(u) \times 2 \mbox{Re}(\widehat{\boldsymbol{\Psi}}_{j, j+h}(\omega))\right)
\equiv 0, \, \forall u, \, \omega,
\end{equation*}
or equivalently,
\begin{equation}\label{eq:specFourierh2}
\sum\limits_{j=1}^{J-h} \left( e_{j,j}^{(p,q)}(u) \widehat{\boldsymbol{\Psi}}_{j, j}(\omega)
+ e_{j,j+h}^{(p,q)}(u) \times 2 \mbox{Re}(\widehat{\boldsymbol{\Psi}}_{j, j+h}(\omega))\right) +
\sum\limits_{j=J-h+1}^{J} e_{j,j}^{(p,q)}(u) \widehat{\boldsymbol{\Psi}}_{j, j}(\omega) \equiv 0.
\end{equation}
The first term in the equation above has $(1-cos\omega 2^{j-1})$ as a factor, which is zero when taking $\omega:=\pi$ iff $j \geq 2$. Hence $e_{1,1}^{(p,q)}(u) \equiv 0$. Next considering $\omega:=\pi/2$ and noting that $(1-cos\omega 2^{j-1})$ iff $j \geq 3$, and recursively using these arguments we obtain $$e_{j,j}^{(p,q)}(u) = 0, \, \forall u, \, j\in \overline{1, J-h}.$$

Using these results into equation~\eqref{eq:specFourierh2}, we obtain 
\begin{equation*}
\sum\limits_{j=1}^{J-h} \left( e_{j,j+h}^{(p,q)}(u) \times 2 \mbox{Re}(\widehat{\boldsymbol{\Psi}}_{j, j+h}(\omega))\right) +
\sum\limits_{j=J-h+1}^{J} e_{j,j}^{(p,q)}(u) \widehat{\boldsymbol{\Psi}}_{j, j}(\omega) \equiv 0, \, \forall u, \, \omega.
\end{equation*}

Now using $\omega:=\pi/2^h$, we note that $1-cos \omega 2^{j+h-1}=0$ iff $j \geq 2$, from which we obtain $e_{1,1+h}^{(p,q)}(u)=0, \, \forall u$, since the terms in the first sum are all zero as their scales yield $\omega 2^{j-1}$ is a multiple of $2\pi$ as $j-1 \geq J-h \geq 1$. Similarly, taking $\omega:=\pi/2^{h+1}$ is associated to non-zero terms only for $j=2$, and re-iterating this process we obtain 
$$
e_{j,j+h}^{(p,q)}(u) = 0, \, \forall u, \, j\in \overline{1, J-h}.
$$

Equation~\eqref{eq:specFourierh2} then becomes 
\begin{equation*}
\sum\limits_{j=J-h+1}^{J} e_{j,j}^{(p,q)}(u) \widehat{\boldsymbol{\Psi}}_{j, j}(\omega) \equiv 0, \, \forall u, \, \omega,
\end{equation*}
from which by replacing $\omega:=\pi/2^{J-h}$, we have the terms $1-cos \omega 2^{j-1}=0$ iff $j \geq J-h+1$, yielding the associated coefficients $e_{j,j}^{(p,q)}(u)$ to be zero when $j:=J-h+1$. Using the same principle, we obtain 
$$
e_{j,j}^{(p,q)}(u) = 0, \, \forall u, \, j\in \overline{J-h+1, J}, \, \forall (p,q).
$$

Hence for any finite fixed $h$ such that $h<J$ and channel pair $(p,q)$, we have at each rescaled time $u$ that the coefficients $e_{j,j}^{(p,q)}(u) = 0, \, j\in \overline{1, J}$ and $e_{j,j+h}^{(p,q)}(u)=e_{j,j+h}^{(q,p)}(u) = 0, \, j\in \overline{1, J-h}$, which concludes the first part of the proof that ensures that the sub-family $\{ 
\{\boldsymbol{\Psi}_{j, j}(\tau)\}_{j=1}^{J}, \{\boldsymbol{\Psi}_{j, j+h}(\tau)\}_{j=1}^{J-h}, \{\boldsymbol{\Psi}_{j, j+h}(-\tau)\}_{j=1}^{J-h} \}$ is linearly independent.

Secondly, we will show that the sub-family of neighbouring cross-scale correlation wavelets $ \{\{\boldsymbol{\Psi}_{j, j+1}(\tau)\}_{j=1}^{J-1}, \{\boldsymbol{\Psi}_{j, j+2}(-\tau)\}_{j=1}^{J-2} \}$ is also linearly independent, hence we show that 
\begin{equation*}
\sum\limits_{j=1}^{J-1} e_{j,j+1}^{(p,q)}(u) \widehat{\boldsymbol{\Psi}}_{j, j+1}(\omega)
+ \sum\limits_{j=1}^{J-2} \left( e_{j,j+2}^{(q,p)}(u) \widehat{\boldsymbol{\Psi}}_{j, j+2}(-\omega) \right) \equiv 0, \, \forall u, \, \omega,
\end{equation*}
implies $e_{j,j+1}^{(p,q)}(u)=e_{j,j+2}^{(q,p)}(u) \equiv 0$ for any channel pairs $(p,q)$. 

We follow the same strategy as in the first part of the proof. Noting that when $\omega:=\pi/2$, the terms contributing to the first sum are such that $1-cos \omega 2^j=0$ when $j \geq 2$, in turn yielding terms $1-cos \omega 2^{j+1}=0$ in the second sum, we obtain  $e_{1,2}^{(p,q)}(u)=0, \, \forall u$. 

Let us now take $\omega:=\pi/2^2$, which generates zero terms when $j \geq 3$ in the first sum, $e_{2,3}^{(p,q)}(u)$ and non-zero contributions in the second sum for $e_{1,3}^{(q,p)}(u)$. The term $e_{1,3}^{(q,p)}(u)$ is the only one associated to the real part of the overall sum, hence $e_{1,3}^{(q,p)}(u)=0, \, \forall u$, which in turn gives $e_{2,3}^{(p,q)}(u)=0, \, \forall u$. Re-iteration using $\omega:=\pi/2^3$ generates the same arguments and yields $e_{2,4}^{(q,p)}(u)=e_{3,4}^{(p,q)}(u)=0, \, \forall u$, and so on to get
$$e_{j,j+1}^{(p,q)}(u)=0, \, j\in\overline{1,J-1} \mbox{ and } e_{j,j+2}^{(q,p)}(u)=0, \, j\in\overline{1,J-2}, \, \forall u, \, \forall (p,q),$$
ensuring the second part of the proof, namely the sub-family $\{\{\boldsymbol{\Psi}_{j, j+1}(\tau)\}_{j=1}^{J-1}, \{\boldsymbol{\Psi}_{j, j+2}(-\tau)\}_{j=1}^{J-2} \}$ is linearly independent. Note that using cross-correlation wavelet properties for $\tau: =- \tau$, this also implies that the sub-family $\{\{\boldsymbol{\Psi}_{j+1, j}(\tau)\}_{j=1}^{J-1}, \{\boldsymbol{\Psi}_{j, j+2}(\tau)\}_{j=1}^{J-2} \}$ is linearly independent.

Lastly, we shall show that the sub-family $\{\{\boldsymbol{\Psi}_{j,j+1}(\tau)\}_{j=1}^{J-1}, \{\boldsymbol{\Psi}_{j, j+2}(\tau)\}_{j=1}^{J-2} \}$ is linearly independent, which equates to showing that 
\begin{equation}
\sum\limits_{j=1}^{J-1} e_{j,j+1}^{(p,q)}(u) \widehat{\boldsymbol{\Psi}}_{j, j+1}(\omega)
+ \sum\limits_{j=1}^{J-2} \left( e_{j,j+2}^{(p,q)}(u) \widehat{\boldsymbol{\Psi}}_{j, j+2}(\omega) \right) \equiv 0, \, \forall u, \, \omega, \notag
\end{equation}
implies $e_{j,j+1}^{(p,q)}(u)=e_{j,j+2}^{(p,q)}(u) \equiv 0$ for any channel pairs $(p,q)$. We follow the same arguments as for the second part and reach the desired conclusion. Again using the properties of cross-correlation wavelets for $\tau: =- \tau$, we have that the sub-family $\{\{\boldsymbol{\Psi}_{j+1, j}(\tau)\}_{j=1}^{J-1}, \{\boldsymbol{\Psi}_{j+2, j}(\tau)\}_{j=1}^{J-2} \}$ is also linearly independent.

Recursively using these results, namely the first independent sub-family with $h:=1$ and combining it with the second and third sub-families, shows the sub-family of cross-correlation wavelets that include steps $|\delta| \leq h=2$ is linearly independent, which again combined with the second and third independent sub-families results in the linear independence including steps up to $h=3$, and so on until we include all steps up to $h<J$, thus showing the family $\{ \{\boldsymbol{\Psi}_{j, j+\delta}(\tau)\}_{{\delta=0:h},{j=1:J-\delta}}, \{\boldsymbol{\Psi}_{j, j+\delta}(-\tau)\}_{{\delta=1:h},{j=1:J-\delta}} \}$ is linearly independent. Recall we denoted its associated Gram matrix as $\tilde{\mathbf{A}}$, hence this is an invertible matrix.

The overall process covariance--spectrum connection follows from a direct application of the definition of the process covariance, $\tilde{c}^{(p,q)}(u,\tau)=\sum\limits_{j,j'}S^{(p,q)}_{jj'}(u)\boldsymbol{\Psi}_{jj'}(\tau)$ for all $u$ and $\tau$, hence for any scales $l, \, l'$,
\begin{align*}
\sum_{\tau}\tilde{c}^{(p,q)}(u,\tau) \boldsymbol{\Psi}_{ll'}(\tau)&=\sum_{j,j'}S^{(p,q)}_{jj'}(u)\sum_{\tau}\boldsymbol{\Psi}_{jj'}(\tau)\boldsymbol{\Psi}_{ll'}(\tau),\\
&=\sum_{j,j'}A_{jj';ll'}S^{(p,q)}_{jj'}(u).
\end{align*}
The invertibility of the matrix $\tilde{\mathbf{A}}$ ensures we can collate the above and obtain the cross-scale spectra as
$$
S_{jj'}^{(p,q)}(u)=\sum_{(l,l')\in \mathcal{B}_h} [\tilde{\mathbf{A}}^{-1}]_{jj';ll'}\sum_{\tau}\tilde{c}^{(p,q)}(u,\tau) \boldsymbol{\Psi}_{ll'}(\tau), \, \forall (j,j')\in \mathcal{B}_h.
$$
\\

\subsection{Subprocess-based estimation}\label{app:subprocest}
For simplicity, in the remainder of this subsection we drop the superscript $^\S$ from the \underline{s}ubprocess-based periodogram $I_{jj',kk'}^{\S;(p,q)}$ and related quantities, unless needed for clarity.

\noindent \textsc{Proof of Proposition~\ref{proposition2}:}

Recall that ${d}_{jj,k}^{(p)} =\sum_t {X}_{j,t}^{(p)} \psi_{j,k}(t)$ and ${d}_{j',k'}^{(q)} =\sum_t {X}_{t;T}^{(q)} \psi_{j',k'}(t)$ where ${X}_{j,t}^{(p)}=\sum_m \mathbf{V}_{j}^{(p)}(m/T)\psi_{j,m}(t)\mathbf{z}_{j,m}$ and ${X}_{t;T}^{(q)}=\sum_l\sum_m \mathbf{V}_l^{(q)}(m/T)\psi_{l,m}(t)\mathbf{z}_{l,m}$, where we recall that $\mathbf{V}_j^{(p)}(u)$ denotes the $p$th row of the $\mathbf{V}_j(u)$ transfer matrix of $\{ \mathbf{X}_{t;T}\}$. Thus, using the covariance structure of the random innovations,
\begin{align*}
    \E(I_{jj',kk'}^{(p,q)}) &= \E \left( \left\{ \sum\limits_{t}{X}_{j,t}^{(p)}\psi_{j,k}(t)\right\} \left\{ \sum\limits_{t'}X_{t';T}^{(q)}\psi_{j',k'}(t') \right\} \right) \notag \\
    &= \sum\limits_{l} \sum\limits_{m} \mathbf{V}_{j}^{(p)}(m/T) \boldsymbol{Q}_{jl}(m/T) \mathbf{V}_{l}^{(q)}(m/T)^\top \notag \\
    & \times \left\{ \sum\limits_{t} \psi_{j,m}(t) \psi_{j,k}(t) \sum\limits_{t'} \psi_{l,m}(t') \psi_{j',k'}(t') \right\}.
\end{align*}
Letting $m=n+k$ in the equation above, we obtain
\begin{align}
    \E(I_{jj',kk'}^{(p,q)}) = \sum\limits_{l} \sum\limits_{n} \left\{S_{jl}^{(p,q)}\left(\frac{n+k}{T}\right) \right\} \left\{\sum\limits_{t}\psi_{j,n+k-t}\psi_{j,k-t}\sum\limits_{t'}\psi_{l,n+k-t'}\psi_{j',k'-t'}\right\}, \notag
\end{align}
and since for any scales $j,\,l$, $S_{jl}^{(p,q)}(u)$ is Lipschitz continuous with finite Lipschitz constant $L_{jl}^{(p,q)}$ satisfying the conditions in Proposition 1, for some fixed $n$, $\left|S_{jl}^{(p,q)}((n+k)/T)- S_{jl}^{(p,q)}(k/T) \right| \leq |n|L_{jl}^{(p,q)}/T$ and we obtain
\begin{align}
    \E(I_{jj',kk'}^{(p,q)}) = \sum\limits_{l} S_{jl}^{(p,q)}(k/T) \sum\limits_{n} \left\{\sum\limits_{t} \psi_{j,n+k-t}\psi_{j,k-t}\right\} \times \left\{\sum\limits_{t'}\psi_{l,n+k-t'}\psi_{j',k'-t'}\right\}+\mathcal{O}(T^{-1}). \notag
\end{align}
In establishing the order above for the approximation term $\sum\limits_{l}L_{jl}^{(p,q)}/T\sum\limits_{n} |n|\boldsymbol{\Psi}_{jj}(n)\boldsymbol{\Psi}_{lj'}(n+k-k')$ we used the fact that the number of wavelet cross-correlation product terms $\boldsymbol{\Psi}_{jj}\boldsymbol{\Psi}_{lj'}$ is finite and bounded as a function of $n$ due to their compact support whose length is bounded by $\mbox{min}\{2^j,2^{j'}+2^l\}\leq (2^j+2^{j'}+2^l)/2$, coupled with the Cauchy-Schwarz inequality to yield $\sum\limits_n | \boldsymbol{\Psi}_{jj}(n)\boldsymbol{\Psi}_{lj'}(n+k-k')| \leq A_{jj}^{1/2} A_{lj'}^{1/2}$ and with the property of the Lipschitz constants $\sum\limits_{j,l}2^{j+l}L_{jl}^{(p,q)}<\infty$ and $\sum\limits_{l}2^{-l}A_{lj'}=1$ \citep{fryz03:forecasting}.

From the definition of the cross-scale autocorrelation wavelets, we re-write the above as
\begin{align}
    \E(I_{jj',kk'}^{(p,q)}) &= \sum\limits_{l}  S_{jl}^{(p,q)}(k/T) \sum\limits_{n} \boldsymbol{\Psi}_{jj}(n)\boldsymbol{\Psi}_{lj'}(n+k-k') + \mathcal{O}(T^{-1}) \notag \\
    &= \sum\limits_{l} A_{jj;lj'}^{(k-k')}S_{jl}^{(p,q)}(k/T) + \mathcal{O}(T^{-1}), \mbox{ where }A_{jj;lj'}^{(k-k')}=A_{jl;jj'}^{(k-k')}.\label{eq:subperexp}
\end{align}

For the variance, we start by considering $\E\left((I_{jj',kk'}^{(p,q)})^2 \right)= \E\left( (d_{jj',k}^{(p)})^2 (d_{j,k'}^{(q)})^2 \right)$. Using the result derived by \cite{moment} and the entry-wise definition of the periodogram, we have
\begin{align*}
    \E\left((I_{jj',kk'}^{(p,q)})^2 \right)& = \E\left(I_{jj,kk}^{(p,p)} \right) \E\left(I_{j'j',k'k'}^{(q,q)} \right) + 2 \E \left( I_{jj',kk'}^{(p,q)} \right)^2 , \mbox{ hence using ~\eqref{eq:subperexp} we obtain} \\
\var(I_{jj',kk'}^{(p,q)}) &=\bigg( \sum\limits_{l} A_{jj;lj}^{(0)}S_{jl}^{(p,p)}(k/T) + \mathcal{O}(T^{-1}) \bigg) \\
    &\times \bigg( \sum\limits_{l} A_{j'j';lj'}^{(0)}S_{j'l}^{(q,q)}(k'/T) + \mathcal{O}(T^{-1}) \bigg) \\
    &+ \bigg( \sum\limits_{l} A_{jj;lj'}^{(k-k')}S_{jl}^{(p,q)}(k/T) + \mathcal{O}(T^{-1}) \bigg)^2.
\end{align*}

For fixed $\delta$, using $A_{jj;lj'}^{(\delta)}=A_{jl;jj'}^{(\delta)}$ and its definition, terms of the form 
\begin{align*}
\sum\limits_{l} A_{jj;lj'}^{(\delta)}S_{jl}^{(p,q)}(k/T) & =  
\sum\limits_{l} \sum\limits_{n} \boldsymbol{\Psi}_{jl}(n)\boldsymbol{\Psi}_{jj'}(n+\delta) S_{jl}^{(p,q)}(k/T), \\
& = \sum\limits_{n} \left(\sum\limits_{l} S_{jl}^{(p,q)}(k/T) \boldsymbol{\Psi}_{jl}(n)\right) \boldsymbol{\Psi}_{jj'}(n+\delta),\\
& \leq  \bigg| \sum\limits_{n} \tilde{c}^{(p,q)}_j(k/T,n)\boldsymbol{\Psi}_{jj'}(n+\delta)\bigg|,\\
&\leq \sum\limits_{n} \bigg|\boldsymbol{\Psi}_{jj'}(n+\delta)\bigg|= \mathcal{O}(2^j)+\mathcal{O}(2^{j'}),
\end{align*}
due to the process having finite covariance and the cross-scale correlation  wavelet having compact support with length of order $2^j+2^{j'}$.
     
Thus, it is easily verified that,
\begin{align*}
     \var(I_{jj',kk'}^{(p,q)}) &= \left(\sum\limits_{l} A_{jj;lj}S_{jl}^{(p,p)}(k/T)\right)\left(\sum\limits_{l} A_{j'j';lj'}S_{j'l}^{(q,q)}(k'/T) \right)\notag \\
     &+ \left(\sum\limits_{l} A_{jj;lj'}^{(k-k')}S_{jl}^{(p,q)}(k/T)\right)^2 + \mathcal{O}(2^{j}T^{-1})+ \mathcal{O}(2^{j'}T^{-1}).
\end{align*}
\\
\noindent \textsc{Proof of Proposition~\ref{proposition3}:}

Recall that the smoothed periodogram is given by $    \Tilde{\mathbf{I}}_{jj',kk'}=\frac{1}{2M+1} \sum\limits_{m=-M}^{M} \mathbf{I}_{jj',(k+m)(k'+m)}$, where $(2M+1)$ is the length of the rectangular smoothing window.

For any channels $(p,q)$, the expected value of this estimator can be derived to be
\begin{align*}
      \E \Big( \Tilde{I}_{jj',kk'}^{(p,q)}\Big)&=\frac{1}{2M+1}\sum\limits_{m=-M}^{M} \E \Big(I_{jj',(k+m)(k'+m)}^{(p,q)}\Big)\\
      &= \frac{1}{2M+1}\sum\limits_{m=-M}^{M}  \sum\limits_{l}
      \bigg\{A_{jj;lj'}^{(k-k')}S_{jl}^{(p,q)}\left(\frac{k+m}{T}\right) + \mathcal{O}(T^{-1}) \bigg\},
\end{align*}
where for the last equality we made use of the expectation result~\eqref{eq:subperexp} in Proposition~\ref{proposition2}. 

\cite{nason2000wavelet} proved that $\sum\limits_{n} |\boldsymbol{\Psi}_{j}(n)| = \mathcal{O}(2^j)$, hence $A_{jj;lj'}^{(k-k')} = \sum\limits_{n} \boldsymbol{\Psi}_{j}(n) \boldsymbol{\Psi}_{lj'}(n+k-k') \leq \left(\sum\limits_{n} |\boldsymbol{\Psi}_{jj}(n|^2 \right)^{1/2} \left(\sum\limits_{n} |\boldsymbol{\Psi}_{lj'}(n+k-k')|^2 \right)^{1/2}= A_{jj}^{1/2}A_{lj'}^{1/2}=\mathcal{O}(2^{j+j'})+\mathcal{O}(2^{j+l})$. 
Coupling this with the Lipschitz continuity of the spectral components and with the property of the Lipschitz constants, we  obtain
\begin{align*}
     \E \Big( \Tilde{I}_{jj',kk'}^{(p,q)}\Big) = \sum\limits_{l} A_{jj;lj'}^{(k-k')}S_{jl}^{(p,q)}\left(k/T\right) + \mathcal{O}(MT^{-1}).
\end{align*}


For the variance part, we begin by considering
$\E \Big( (\Tilde{I}_{jj',kk'}^{(p,q)})^2 \Big)$.

Replacing the definition of the cross-scale (smoothed) periodogram, we have
\begin{align*}
    \E \Big( (\Tilde{I}_{jj',kk'}^{(p,q)})^2 \Big) &= \frac{1}{(2M+1)^2}\sum\limits_{m=-M}^{M}\sum\limits_{m'=-M}^{M} \E \Big(I_{jj',(k+m)(k'+m)}^{(p,q)}I_{jj',(k+m')(k'+m')}^{(p,q)}\Big), \\
    &= \frac{1}{(2M+1)^2} \sum\limits_{m=-M}^{M} \sum\limits_{\tau=-M-m}^{M-m} \E \Big(d_{jj,k+m}^{(p)} d_{j',k'+m}^{(q)} d_{jj,k+m+\tau}^{(p)} d_{j',k'+m+\tau}^{(q)}\Big), \mbox{ with }\tau=m'-m.
\end{align*}
With an application of \cite{moment}, the variance of the smoothed periodogram can be shown to be
\begin{align*}
    \var \Big( \Tilde{I}_{jj',kk'}^{(p,q)} \Big) &= \frac{1}{(2M+1)^2} \bigg\{  \sum\limits_{m=-M}^{M} \sum\limits_{\tau}
    \E\Big(d_{jj,k+m}^{(p)}d_{jj,k+m+\tau}^{(p)} \Big) \times \E \Big(d_{j',k'+m}^{(q)}d_{j',k'+m+\tau}^{(q)} \Big)\\& +\sum\limits_{m=-M}^{M} \sum\limits_{\tau} \E \Big(d_{jj,k+m}^{(p)}d_{j',k'+m+\tau}^{(q)} \Big) \times\ \E \Big(d_{jj,k+m+\tau}^{(p)} d_{j',k'+m}^{(q)} \Big) \bigg\}, \\
    &= \frac{1}{(2M+1)^2} \sum\limits_{m=-M}^M \bigg\{\sum\limits_{\tau}\Big( A_{jj}^{(\tau)} S_{jj}^{(p,p)}(k/T) \Big) \times \Big( \sum\limits_{l,l'} A_{ll';j'j'}^{(\tau)} S_{ll'}^{(q,q)}(k'/T) \Big) \\ &+ \sum\limits_{\tau}\Big(\sum\limits_{l} A_{jj;lj'}^{(k-k'-\tau)} S_{jl}^{(p,q)}(k/T) \Big) \times \Big(\sum\limits_{l'} A_{jj;l'j'}^{(k-k'+\tau)} S_{jl'}^{(p,q)}(k'/T) \Big) \\ &+ \sum\limits_{\tau}(|m|+1)\mathcal{O}(T^{-1}) + \sum\limits_{\tau} (|m|+1)^2 \mathcal{O}(T^{-2}) \bigg\}.
\end{align*}
In the above, the term $\E\Big(d_{jj,k+m}^{(p)}d_{jj,k+m+\tau}^{(p)} \Big)$ was obtained through a straightforward application of the definition ${d}_{jj,k}^{(p)} =\sum_t {X}_{j,t}^{(p)} \psi_{j,k}(t)$ where $\mathbf{X}_{j,t}^{(p)}=\sum_m \mathbf{V}_j^{(p)}(m/T)\psi_{j,m}(t)\mathbf{z}_{j,m}$, coupled with the covariance structure assumed for the random innovations $\{\mathbf{z}_{j,m}\}_{l,m}$ of a CS-MvLSW process as in Definition~\ref{def1}.

We derived the next term, $\E \Big(d_{j',k'+m}^{(q)}d_{j',k'+m+\tau}^{(q)} \Big)$, as follows. In general, ${d}_{j,k}^{(p)} =\sum_t {X}_{t;T}^{(p)} \psi_{j,k}(t)$ where $\mathbf{X}_{t;T}^{(p)}=\sum_l\sum_m \mathbf{V}_l^{(p)}(m/T)\psi_{l,m}(t)\mathbf{z}_{l,m}$, and similarly for the $q$th channel, thus
\begin{align*}
    \E({d}_{j,k}^{(p)}{d}_{j',k'}^{(q)}) &= \E \left( \left\{ \sum\limits_{t}{X}_{t;T}^{(p)}\psi_{j,k}(t)\right\} \left\{ \sum\limits_{t'}X_{t';T}^{(q)}\psi_{j',k'}(t') \right\}^\top \right) \notag \\
    &= \sum\limits_{l} \sum\limits_{l'} \sum\limits_{m} \mathbf{V}_l^{(p)}(m/T) \boldsymbol{Q}_{ll'}(m/T) \mathbf{V}_{l'}^{(q)}(m/T)^{\top} \notag \\
    & \times \left\{ \sum\limits_{t} \psi_{l,m}(t) \psi_{j,k}(t) \sum\limits_{t'} \psi_{l',m}(t') \psi_{j',k'}(t') \right\}.
\end{align*}
Letting $m=n+k$ into the above, we have
\begin{align}
    \E({d}_{j,k}^{(p)}{d}_{j',k'}^{(q)}) = \sum\limits_{l}\sum\limits_{l'} \sum\limits_{n} \left\{S_{ll'}^{(p,q)}\left(\frac{n+k}{T}\right) \right\} \left\{\sum\limits_{t}\psi_{l,n+k-t}\psi_{j,k-t}\sum\limits_{t'}\psi_{l',n+k-t'}\psi_{j',k'-t'}\right\}, \notag
\end{align}
and using the Lipschitz continuity of the dual-scale spectrum, we obtain
\begin{align*}
    \E({d}_{j,k}^{(p)}{d}_{j',k'}^{(q)}) = \sum\limits_{l}\sum\limits_{l'}  S_{ll'}^{(p,q)}(k/T) \sum\limits_{n} \left\{\sum\limits_{t} \psi_{l,n+k-t}\psi_{j,k-t}\right\} \times \left\{\sum\limits_{t'}\psi_{l',n+k-t'}\psi_{j',k'-t'}\right\}+\mathcal{O}(T^{-1}). \notag
\end{align*}
Replacing the definition of the cross-scale autocorrelation wavelets subsequently yields,
\begin{align}
    \E({d}_{j,k}^{(p)}{d}_{j',k'}^{(q)}) &= \sum\limits_{l}\sum\limits_{l'}  S_{ll'}^{(p,q)}(k/T) \sum\limits_{n} \boldsymbol{\Psi}_{lj}(n)\boldsymbol{\Psi}_{l'j'}(n+k-k') + \mathcal{O}(T^{-1}) \notag \\
    &= \sum\limits_{l}\sum\limits_{l'} A_{ll';jj'}^{(k-k')}S_{ll'}^{(p,q)}(k/T) + \mathcal{O}(T^{-1}),\label{eq:procI}
\end{align}
which we use for $p=q$, $j:=j'$, $k:=k'+m$ and $k':=k'+m+\tau$.
The final $(p,q)$ cross-terms in the variance of the smoothed periodogram are obtained as in the proof corresponding to the expectation part of Proposition~\ref{proposition2}.

In order to bound these terms, recall that in general $A_{ll';jj'}^{(\tau)} = \sum\limits_{n} \boldsymbol{\Psi}_{ll'}(n)\boldsymbol{\Psi}_{jj'}(n+\tau)$ and such terms can be expressed as 
\begin{align*}
    \sum\limits_{\tau}\bigg|\sum\limits_{l,l'} A_{ll';jj'}^{(\tau)} S_{ll'}^{(p,q)}(k/T)\bigg|
    &=\sum\limits_{\tau}\bigg|\sum\limits_{l,l'} \sum_n \boldsymbol{\Psi}_{ll'}(n)\boldsymbol{\Psi}_{jj'}(n+\tau) S_{ll'}^{(p,q)}(k/T) \bigg|\\
    &= \sum\limits_{\tau}\bigg| \sum_n \boldsymbol{\Psi}_{jj'}(n+\tau) \sum\limits_{l,l'} \boldsymbol{\Psi}_{ll'}(n) S_{ll'}^{(p,q)}(k/T) \bigg|\\
    &= \sum\limits_{\tau}\bigg| \sum_n \tilde{c}^{(p,q)}(k/T,n) \boldsymbol{\Psi}_{jj'}(n+\tau) \bigg|\\
    &\leq \sum_n |\tilde{c}^{(p,q)}(k/T,n)| \sum\limits_{\tau} |\boldsymbol{\Psi}_{jj'}(n+\tau)|,
\end{align*}
where we recall the assumption of summable process covariance and the property of the cross-scale autocorrelation wavelets $\boldsymbol{\Psi}_{jj'}$ to have compact support with length of order $2^j+2^{j'}$. Hence we can bound each of the terms in the variance formula as follows. 

For the first term, coupling the above result with $\sum\limits_{\tau}\bigg|A_{jj}^{(\tau)} S_{jj}^{(p,p)}(k/T)\bigg|\leq \sum\limits_n |{c}_{j}^{(p,p)}(k/T,n)| \sum\limits_{\tau} |\boldsymbol{\Psi}_{j}(\tau)|=\mathcal{O}(2^{j})$, we obtain
\begin{align}
    &\sum\limits_{\tau}A_{jj}^{(\tau)} S_{jj}^{(p,p)}(k/T) \sum\limits_{l,l'} A_{ll';j'j'}^{(\tau)} S_{ll'}^{(q,q)}(k'/T) \nonumber \\
    & \leq \bigg(\sum\limits_{\tau}\bigg| A_{jj}^{(\tau)} S_{jj}^{(p,p)}(k/T)\bigg|\bigg) \bigg(\sum\limits_{\tau} \bigg| \sum\limits_{l,l'} A_{ll';j'j'}^{(\tau)} S_{ll'}^{(q,q)}(k'/T) \bigg| \bigg) \nonumber \\ 
    &= 
    \mathcal{O}(2^{j+j'}).\label{eq:term1}
\end{align}
Similarly for the second term, 
\begin{align}
    &\sum\limits_{\tau}\sum\limits_{l} A_{jj;lj'}^{(k-k'-\tau)} S_{jl}^{(p,q)}(k/T) \sum\limits_{l} A_{jj;lj'}^{(k-k'+\tau)} S_{jl}^{(p,q)}(k'/T)  \nonumber\\
    & \leq \bigg(\sum\limits_{\tau}\bigg|\sum\limits_{l} A_{jl;jj'}^{(k-k'-\tau)} S_{jl}^{(p,q)}(k/T)\bigg|\bigg) \bigg(\sum\limits_{\tau} \bigg| \sum\limits_{l} A_{jl;jj'}^{(k-k'+\tau)} S_{jl}^{(p,q)}(k'/T) \bigg| \bigg), \mbox{ as }A^{(\delta)}_{jj;lj'}=A^{(\delta)}_{jl;jj'}  \nonumber\\ 
    &\leq \sum_n |\tilde{c}_j^{(p,q)}(k/T,n)| 
    \sum_{n'} |\tilde{c}_{j'}^{(p,q)}(k'/T,n')|
    \left(\sum\limits_{\tau'} |\boldsymbol{\Psi}_{jj'}(\tau')|\right)^2,\nonumber\\
    &= \mathcal{O}(2^{2j})+
    \mathcal{O}(2^{2j'}), \label{eq:term2} \mbox{ where we also used the assumption }\mbox{sup}_{u}\sum_{n}|\tilde{c}^{(p,q)}_{j}(u,n)|<\infty, \, \forall j. 
\end{align}
Then by replacing~\eqref{eq:term1} and~\eqref{eq:term2} into the variance of the smoothed periodogram and retaining the highest order terms, we obtain
\begin{align*}
    \var \Big( \Tilde{I}_{jj',kk'}^{(p,q)} \Big) = 
    \mathcal{O}(2^{2j}M^{-1}) + \mathcal{O}(2^{2j'}M^{-1}) +\mathcal{O}(MT^{-1}).
\end{align*}
Hence, the proposed smoothed wavelet periodogram is asymptotically mean-squared consistent for the true spectrum as $T \to \infty, M \to \infty, {M}/{T} \to 0$ and for fine enough scales, $2^{j}=o((2M+1)^{1/2}), \, 2^{j'}=o((2M+1)^{1/2})$.

\noindent \textsc{Proof of Proposition~\ref{prop:approx}:}

We denote by $\{\boldsymbol{\v}_{t;T}\}$ the error process obtained in practice, $\boldsymbol{\v}_{t;T}=\sum_{j=1}^J \boldsymbol{\v}_{j,t}$, whose subprocesses $\{\v_{j,t}\}$ quantify the approximation error, namely $\boldsymbol{\v}_{j,t}=\Tilde{\mathbf{X}}_{j,t}-\mathbf{X}_{j,t}$. 
Let us denote the (zero-mean) non-decimated wavelet coefficients at scale $j'$ and time $k$ associated to the scale-$j$ error subprocess as ${\mathbf{d}}_{jj',k}^{\v}$. Due to the linearity of the wavelet transform, we have $\mathbf{d}_{jj',k}^{\v}=\tilde{\mathbf{d}}_{jj',k}-\mathbf{d}_{jj',k}$ across all channels $p=1, \ldots, P$.

\noindent We decompose the error process  covariance function using 
\begin{equation}\label{eq:residcov}
\cov(\boldsymbol{\v}_{j,t},\boldsymbol{\v}_{j',t+\tau})= \cov(\Tilde{\mathbf{X}}_{j,t},\Tilde{\mathbf{X}}_{j',t+\tau})
+ \cov({\mathbf{X}}_{j,t},{\mathbf{X}}_{j',t+\tau})
- \cov(\Tilde{\mathbf{X}}_{j,t},{\mathbf{X}}_{j',t+\tau})
- \cov({\mathbf{X}}_{j,t},\Tilde{\mathbf{X}}_{j',t+\tau}),    
\end{equation}
%
%
and observe that with both $\{\mathbf{X}_{j,t}\}_{j,t}$ and $\{\Tilde{\mathbf{X}}_{j,t}\}_{j,t}$ representing the information in the observed CS-MvLSW process $\{\mathbf{X}_{t,T}\}$, the  uniqueness of the covariance--spectral representation established in Property~\ref{property1}(ii) ensures that the spectral structures $\{\mathbf{S}^{X}_{jj'}(\cdotp)\}_{j,j'}$ and $\{\mathbf{S}^{\Tilde X}_{jj'}(\cdotp)\}_{j,j'}$ coincide as their covariance structures coincide. Hence, further using the properties established in Proposition~\ref{proposition1} which connect $\cov({X}^{(p)}_{j,t},{X}^{(q)}_{j',t+\tau})=c_{jj'}^{X;(p,q)}(t/T,\tau)+\mathcal{O}(2^{-(j+j')/2}T^{-1})$, we obtain from~\eqref{eq:residcov} that 
$
\cov({\v}^{(p)}_{j,t},{\v}^{(q)}_{j',t+\tau})=c_{jj'}^{\v;(p,q)}(t/T,\tau)+\mathcal{O}(2^{-(j+j')/2}T^{-1})=\mathcal{O}(2^{-(j+j')/2}T^{-1})$ for any scales $j, \, j'$, time $t$, lag $\tau$, and channels $p, \,q$.

Dropping the channel superscript $p$ for simplicity, the non-decimated wavelet coefficients used for the basis averaging \citep{abramovich2000wavelet} yield mean squared errors
\begin{align*}
\E |{{d}}_{jj',k}^{\v}|^2&=
\E\left(\sum_{t}\v_{j,t}\psi_{j',k}(t)\right)^2,\\
&=\sum_{t}\sum_{t'} \E\left(\v_{j,t}\v_{j,t'}\right)\psi_{j',k}(t)\psi_{j',k}(t'),\\
&=\sum_{t}\sum_{\delta} \E\left(\v_{j,t}\v_{j,t+\delta}\right)\psi_{j',k}(t)\psi_{j',k}(t+\delta),\\
&=\sum_{t''}\sum_{\delta} \cov\left(\v_{j,t''+k},\v_{j,t''+k+\delta}\right)\psi_{j',0}(t'')\psi_{j',0}(t''+\delta),\\ 
&=\sum_{\delta} \left(c_{jj}^{\v}(k/T,\delta) +\mathcal{O}(2^{-(j+j')/2}T^{-1}) \right) \boldsymbol{\Psi}_{j'}(\delta) =\mathcal{O}(2^{(j'-j)/2}T^{-1}),
\end{align*}
where for the last equality we used similar arguments to those in the proof of Proposition~\ref{proposition1} and the property  $\sum_{\delta} |\boldsymbol{\Psi}_{j'}(\delta)| = \mathcal{O}(2^{j'})$. Hence, since the active dual scales $(j,j')$ must be `close enough' from Property~\ref{property1}(ii), we have $2^{(j'-j)/2}=o(T)$ and $\E|{{d}}_{jj',k}^{\v}|^2 =\mathcal{O}(2^{(j'-j)/2}T^{-1})$.   

The spectral estimator at $u=k/T$ can then be expressed as
\begin{align*}
  \Tilde{\Tilde{{I}}}_{jj',k}^{\S; (p,q)}&=\frac{1}{2M+1}\sum\limits_{m=-M}^{M} \tilde{{d}}_{jj,k+m}^{(p)}{d}_{j',k+m}^{(q)}, \\
  &=  \frac{1}{(2M+1)} \sum_{m=-M}^M (d_{jj,k+m}^{\v;(p)}+d_{jj,k+m}^{(p)}) {d}_{j',k+m}^{(q)}, \\
  &=  \frac{1}{(2M+1)} \sum_{m=-M}^M (d_{jj,k+m}^{\v;(p)}{d}_{j',k+m}^{(q)}+d_{jj,k+m}^{(p)} {d}_{j',k+m}^{(q)}), \\
  &=B1+B2.
\end{align*}

We first consider
\begin{align}\label{eq:prodbound}
  |\E(d_{jj,k}^{\v;(p)}{d}_{j',k}^{(q)})|^2 &\leq \E(d_{jj,k}^{\v;(p)})^2  \E({d}_{j',k}^{(q)})^2 \mbox{ from the Cauchy-Schwarz inequality} \nonumber\\
  &\leq K T^{-1} \E({d}_{j',k}^{(q)})^2, \mbox{ since for }  j=j', \, \E|d_{jj,k}^{\v;(p)}|^2=\mathcal{O}(T^{-1}), \nonumber\\
  &=K T^{-1} \left(\sum\limits_{l}\sum\limits_{l'} A_{ll';j'j'}S_{ll'}^{(q,q)}(k/T)+\mathcal{O}(T^{-1})\right),\nonumber\\
  &=\mathcal{O}(2^{j'}T^{-1}),
\end{align}
where in the penultimate expression we used equation~\eqref{eq:procI} with $j=j', \, k=k'$ and $p=q$, and for the final entry we used the bound below  and collected the maximum order terms. The bound follows as
\begin{align*}
\left|\sum\limits_{l}\sum\limits_{l'} A_{ll';j'j'}S_{ll'}^{(q,q)}(k/T)\right| &= \left|\sum_{l,l'}\sum _{\uptau}\Psi_{j'}(\uptau)\Psi_{ll'}(\uptau) S_{ll'}^{(q,q)}\left(k/T\right)\right| \mbox{ from the definition of }A \\
&= \left|\sum _{\uptau}\left(\sum_{l,l'} S_{ll'}^{(q,q)}\left(k/T\right)\Psi_{ll'}(\uptau)\right)\Psi_{j'}(\uptau)\right|\\
&=  \left|\sum _{\uptau}\tilde{c}^{(q,q)}\left(k/T,\uptau \right)\Psi_{j'}(\uptau)\right|   \mbox { from the local process autocovariance definition}\\
& \leq \sum _{\uptau}\left|\tilde{c}^{(q,q)}\left(k/T,\uptau \right)\right| \left|\Psi_{j'}(\uptau)\right| \mbox { using the triangle inequality}\\
&= \mathcal{O}(2^{j'}),
\end{align*}
since $\left|\tilde{c}^{(q,q)}\left(k/T,\uptau \right)\right| <\infty$ for all $k,\,\uptau$ and $\sum _{\uptau}\left|\Psi_{j'}(\uptau)\right| = \mathcal{O}(2^{j'})$ (see \cite{embleton2022multiscale}).

Thus $d_{jj,k}^{\v;(p)}{d}_{j',k}^{(q)}=o_P(1)$, from  which the term $B1=o_P(1)$. 


Next observe that the second term term is in fact $B2={\Tilde{{I}}}_{jj',k}^{\S; (p,q)}$, hence by Proposition~\ref{proposition3} we have $B2=\beta_{jj'}^{(p,q)}(k/T)+o_P(1)$ as $T \rightarrow \infty$.

Combining these results, we obtain $\Tilde{\Tilde{{I}}}_{jj',k}^{\S; (p,q)}=\beta_{jj'}^{(p,q)}(k/T)+o_P(1)$ as $T \rightarrow \infty$.

Let us now investigate 
\begin{equation}\label{eq:expapprox}
\E(\Tilde{\Tilde{{I}}}_{jj',k}^{\S; (p,q)})=\E(B1)+\E(B2).
\end{equation}

For the first term in~\eqref{eq:expapprox}, using the result in~\eqref{eq:prodbound} we have $\E(d_{jj,k}^{\v;(p)}{d}_{j',k}^{(q)})=\mathcal{O}(2^{j'/2}T^{-1/2})$ for all channels $p, \, q$. 

Since $B2={\Tilde{{I}}}_{jj',k}^{\S; (p,q)}$, the second term is $\E(B2)=\beta_{jj'}^{(p,q)}(k/T)+\mathcal{O}(MT^{-1})$  from Proposition~\ref{proposition3}. Hence $\E(\Tilde{\Tilde{{I}}}_{jj',k}^{\S; (p,q)})=\beta_{jj'}^{(p,q)}(k/T)+\mathcal{O}(M'_T)$, with $M'_T=\max\{2^{j'/2}T^{-1/2},MT^{-1}\} \to 0$ as $M, \, T \to \infty$ with $M/T \to 0$ and scales $j, \,j'$ such that $2^{j'}=o(T)$ and $2^{(j'-j)/2}=o(T)$.
\\


\noindent \textsc{Proof of Proposition~\ref{prop:subprocconsist}:}

As $M, T \rightarrow \infty$, for each $j$, $u$, the consistency result $\hat{S}_{jj'}^{\S;(p,q)}(u)\stackrel{P}{\rightarrow} {S}^{(p,q)}_{jj'}(u)$ follows from the consistency results
$\Tilde{\Tilde{{I}}}_{jj',k}^{\S; (p,q)} \stackrel{P}{\rightarrow} \beta_{jj'}^{(p,q)}(k/T)$  for all fine enough scales $j,\, j'$ (as shown in Proposition~\ref{prop:subprocconsist}) and then using the continuous mapping theorem \citep{billingsley1999convergence} for the continuous function $g(x_1, \ldots, x_J)= \sum_{l=1}^J (A^{jj})^{-1}_{j'l} x_l$ that defines their linear combination with coefficients given by the matrix $(A^{jj})^{-1}$ entries.

Additionally, using the properties of the matrix $A^{jj}$, we obtain the estimator asymptotic unbiasedness from the linearity of the expectation operator and from the asymptotic unbiasedness of the corrected periodogram, as follows
\begin{eqnarray*}
\E(\hat{S}_{jj'}^{\S;(p,q)}(u))&=&\E\left(\sum_{l=1}^J (A^{jj})^{-1}_{j'l}
\Tilde{\Tilde{{I}}}_{jl,k}^{\S; (p,q)}\right),\\&=& \sum_{l=1}^J (A^{jj})^{-1}_{j'l}
\E\left(\Tilde{\Tilde{{I}}}_{jl,k}^{\S; (p,q)}\right),\mbox{ then from the expectation part of Proposition~\ref{prop:subprocconsist}}\\
&=&\sum_{l=1}^J (A^{jj})^{-1}_{j'l} \left(\beta_{jl}^{(p,q)}(k/T) + \mathcal{O}(M'_T)\right), \mbox{ and using the definition of }\beta\\
&=& \sum_{l=1}^J (A^{jj})^{-1}_{j'l}\sum_{l'} A_{jj;j'l'}S_{jl'}^{(p,q)}(k/T)+ \mathcal{O}(M'_T),\\
&=& \sum_{l'} \left(\sum_{l=1}^J (A^{jj})^{-1}_{j'l}A^{jj}_{l,l'}\right)S_{jl'}^{(p,q)}(k/T)+ \mathcal{O}(M'_T),\\
&=& \sum_{l'} \left( (A^{jj})^{-1} (A^{jj}) \right)_{j',l'}S_{jl'}^{(p,q)}(k/T)+ \mathcal{O}(M'_T),\\
&=&S_{jj'}^{(p,q)}(k/T)+\mathcal{O}(\max\{2^{j'/2}T^{-1/2},MT^{-1}\}).
\end{eqnarray*}

\subsection{Process-based estimation}\label{app:procest}
For simplicity, in the remainder of this subsection we drop the superscript $^\P$ from the \underline{p}rocess-based periodogram $I_{jj',kk'}^{\P;(p,q)}$ and related quantities, unless needed for clarity.

\noindent \textsc{Proof of Proposition~\ref{proposition4}:}

Recall that ${d}_{j,k}^{(p)} =\sum_t {X}_{t;T}^{(p)} \psi_{j,k}(t)$ where $\mathbf{X}_{t;T}^{(p)}=\sum_l\sum_m \mathbf{V}_l^{(p)}(m/T)\psi_{l,m}(t)\mathbf{z}_{l,m}$, and similarly for the $q$th channel decomposition at scale $j'$ and time $k'$. In the proof of Proposition~\ref{proposition3} we obtained equation~\eqref{eq:procI}, which is the first desired result, namely for any channels $(p,q)$,
\begin{align*}
    \E(I_{jj',kk'}^{(p,q)}) 
    &= \sum\limits_{l}\sum\limits_{l'} A_{ll';jj'}^{(k-k')}S_{ll'}^{(p,q)}(k/T) + \mathcal{O}(T^{-1}).
\end{align*}

For the variance, we consider $\E\left((I_{jj',kk'}^{(p,q)})^2 \right)= E\left( (d_{j,k}^{(p)})^2 (d_{j',k'}^{(q)})^2 \right)$ and again using \cite{moment} and the entry-wise definition of the process-based periodogram, we have
\begin{align*}
    \E\left((I_{jj',kk'}^{(p,q)})^2 \right)& = \E\left(I_{jj,kk}^{(p,p)} \right) \E\left(I_{j'j',k'k'}^{(q,q)} \right) + 2 \E \left( I_{jj',kk'}^{(p,q)} \right)^2 , \mbox{ hence } \\
\var(I_{jj',kk'}^{(p,q)}) &=\bigg( \sum\limits_{l}\sum\limits_{l'} A_{ll';jj}^{(0)}S_{ll'}^{(p,p)}(k/T) + \mathcal{O}(T^{-1}) \bigg) \\
    &\times \bigg( \sum\limits_{l}\sum\limits_{l'} A_{ll';j'j'}^{(0)}S_{ll'}^{(q,q)}(k'/T) + \mathcal{O}(T^{-1}) \bigg) \\
    &+ \bigg( \sum\limits_{l}\sum\limits_{l'} A_{ll';jj'}^{(k-k')}S_{ll'}^{(p,q)}(k/T) + \mathcal{O}(T^{-1}) \bigg)^2.
\end{align*}
Using similar bounding arguments as for the variance part of Proposition~\ref{proposition3} results in 
\begin{align*}
     \var(I_{jj',kk'}^{(p,q)}) &= \sum\limits_{l}\sum\limits_{l'} A_{ll';jj}S_{ll'}^{(p,p)}(k/T)\sum\limits_{l}\sum\limits_{l'} A_{ll';j'j'}S_{ll'}^{(q,q)}(k'/T) \notag \\
     &+ \sum\limits_{l}\sum\limits_{l'} A_{ll';jj'}^{(k-k')}S_{ll'}^{(p,q)}(k/T) + \mathcal{O}(2^{2j}T^{-1})+ \mathcal{O}(2^{2j'}T^{-1}). 
\end{align*}
\\
\noindent \textsc{Proof of Proposition~\ref{proposition5}:}

Recall that the smoothed periodogram is given by $    \Tilde{\mathbf{I}}_{jj',kk'}=\frac{1}{2M+1} \sum\limits_{m=-M}^{M} \mathbf{I}_{jj',(k+m)(k'+m)}$, where $(2M+1)$ is the length of the rectangular smoothing window. 

For any channels $(p,q)$, the expected value of this estimator can be derived to be
\begin{align*}
      \E \Big( \Tilde{I}_{jj',kk'}^{(p,q)}\Big)&=\frac{1}{2M+1}\sum\limits_{m=-M}^{M} \E \Big(I_{jj',(k+m)(k'+m)}^{(p,q)}\Big)\\
      &= \frac{1}{2M+1}\sum\limits_{m=-M}^{M}  \bigg\{\sum\limits_{l}\sum\limits_{l'} A_{ll';jj'}^{(k-k')}S_{ll'}^{(p,q)}\left(\frac{k+m}{T}\right) + \mathcal{O}(T^{-1}) \bigg\}.
\end{align*}
Using the Lipschitz continuity of the spectral components, we then obtain
\begin{align*}
     \E \Big( \Tilde{I}_{jj',kk'}^{(p,q)}\Big) = \sum\limits_{l}\sum\limits_{l'} A_{ll';jj'}^{(k-k')}S_{ll'}^{(p,q)}\left(k/T\right) + \mathcal{O}(MT^{-1}). 
\end{align*}

For the variance part, we begin by considering  
$\E \Big( (\Tilde{I}_{jj',kk'}^{(p,q)})^2 \Big)$ and by replacing the definition of the cross-scale (smoothed) periodogram, we have
\begin{align*}
    \E \Big( (\Tilde{I}_{jj',kk'}^{(p,q)})^2 \Big) &= \frac{1}{(2M+1)^2}\sum\limits_{m=-M}^{M}\sum\limits_{m'=-M}^{M} \E \Big(I_{jj',(k+m)(k'+m)}^{(p,q)}I_{jj',(k+m')(k'+m')}^{(p,q)}\Big), \\
    &= \frac{1}{(2M+1)^2} \sum\limits_{m=-M}^{M} \sum\limits_{\tau=-M-m}^{M-m} \E \Big(d_{j,k+m}^{(p)} d_{j',k'+m}^{(q)} d_{j,k+m+\tau}^{(p)} d_{j',k'+m+\tau}^{(q)}\Big), \mbox{ with }\tau=m'-m.
\end{align*}

With an application of \cite{moment}, the variance of the smoothed periodogram can be shown to be
\begin{align*}
    \var \Big( \Tilde{I}_{jj',kk'}^{(p,q)} \Big) &= \frac{1}{(2M+1)^2} \bigg\{  \sum\limits_{m=-M}^{M} \sum\limits_{\tau} 
    \E\Big(d_{j,k+m}^{(p)}d_{j,k+m+\tau}^{(p)} \Big) \times \E \Big(d_{j',k'+m}^{(q)}d_{j',k'+m+\tau}^{(q)} \Big)\\& +\sum\limits_{m=-M}^{M} \sum\limits_{\tau} \E \Big(d_{j,k+m}^{(p)}d_{j',k'+m+\tau}^{(q)} \Big) \times \E \Big(d_{j',k'+m}^{(q)}d_{j,k+m+\tau}^{(p)} \Big) \bigg\}, \\
    &= \frac{1}{(2M+1)^2} \sum\limits_{m=-M}^M \bigg\{\sum\limits_{\tau}\Big(\sum\limits_{l,l'} A_{ll';jj}^{(\tau)} S_{ll'}^{(p,p)}(k/T) \Big) \times \Big( \sum\limits_{l,l'} A_{ll';j'j'}^{(\tau)} S_{ll'}^{(q,q)}(k'/T) \Big) \\ &+ \sum\limits_{\tau}\Big(\sum\limits_{l,l'} A_{ll';jj'}^{(k-k'-\tau)} S_{ll'}^{(p,q)}(k/T) \Big) \times \Big(\sum\limits_{l,l'} A_{ll';jj'}^{(k-k'+\tau)} S_{ll'}^{(q,p)}(k'/T) \Big) \\ &+ \sum\limits_{\tau}(|m|+1)\mathcal{O}(T^{-1}) + \sum\limits_{\tau} (|m|+1)^2 \mathcal{O}(T^{-2}) \bigg\}. 
\end{align*}

Recalling that $A_{ll';jj'}^{(\delta)} = \sum_{n} \boldsymbol{\Psi}_{l,l'}(n)\boldsymbol{\Psi}_{j,j'}(n+\delta)$ for a general $\delta$, we have 
\begin{align*}
    \sum\limits_{\tau}\bigg|\sum\limits_{l,l'} A_{ll';jj'}^{(\tau)} S_{ll'}^{(p,q)}(k/T)\bigg| 
    &=\sum\limits_{\tau}\bigg|\sum\limits_{l,l'} \sum_n \boldsymbol{\Psi}_{l,l'}(n)\boldsymbol{\Psi}_{j,j'}(n+\tau) S_{ll'}^{(p,q)}(k/T) \bigg|\\
    &\leq \sum\limits_{\tau}\bigg| \sum_n \boldsymbol{\Psi}_{j,j'}(n+\tau) \sum\limits_{l,l'} \boldsymbol{\Psi}_{l,l'}(n) S_{ll'}^{(p,q)}(k/T) \bigg|\\
    &= \sum\limits_{\tau}\bigg| \sum_n \tilde{c}^{(p,q)}(k/T,n) \boldsymbol{\Psi}_{j,j'}(n+\tau) \bigg|\\
    &\leq \sum_n |\tilde{c}^{(p,q)}(k/T,n)| \sum\limits_{\tau} |\boldsymbol{\Psi}_{j,j'}(n+\tau)|,
\end{align*}
where we recall the assumption of summable process covariance and the property of the cross-scale autocorrelation wavelets $\boldsymbol{\Psi}_{j,j'}$ to have compact support with length of order $2^j+2^{j'}$. Hence we can bound each of the terms in variance formula, as follows.

For the first term,
\begin{align*}
    &\sum\limits_{\tau}\sum\limits_{l,l'} A_{ll';jj}^{(\tau)} S_{ll'}^{(p,p)}(k/T) \sum\limits_{l,l'} A_{ll';j'j'}^{(\tau)} S_{ll'}^{(q,q)}(k'/T)  \\
    & \leq \bigg(\sum\limits_{\tau}\bigg|\sum\limits_{l,l'} A_{ll';jj}^{(\tau)} S_{ll'}^{(p,p)}(k/T)\bigg|\bigg) \bigg(\sum\limits_{\tau} \bigg| \sum\limits_{l,l'} A_{ll';j'j'}^{(\tau)} S_{ll'}^{(q,q)}(k'/T) \bigg| \bigg)  \\ &= \mathcal{O}(2^{j+j'}), 
\end{align*}
and similarly for the second term
\begin{align*}
    &\sum\limits_{\tau}\sum\limits_{l,l'} A_{ll';jj'}^{(k-k'-\tau)} S_{ll'}^{(p,q)}(k/T) \sum\limits_{l,l'} A_{ll';jj'}^{(k-k'+\tau)} S_{ll'}^{(q,p)}(k'/T)  \\
    & \leq \bigg(\sum\limits_{\tau}\bigg|\sum\limits_{l,l'} A_{ll';jj'}^{(k-k'-\tau)} S_{ll'}^{(p,q)}(k/T)\bigg|\bigg) \bigg(\sum\limits_{\tau} \bigg| \sum\limits_{l,l'} A_{ll';jj'}^{(k-k'+\tau)} S_{ll'}^{(q,p)}(k'/T) \bigg| \bigg)  \\ &= \mathcal{O}(2^{2j})+\mathcal{O}(2^{2j'}). 
\end{align*}
Thus, retaining the highest order terms,
\begin{align*}
    \var \Big( \Tilde{I}_{jj',kk'}^{(p,q)} \Big) 
    = \mathcal{O}(2^{2j}M^{-1}) + \mathcal{O}(2^{2j'}M^{-1}) +\mathcal{O}(MT^{-1}).
\end{align*}
\\

\noindent \textsc{Proof of Proposition~\ref{prop:procconsist}:}

For times $k$ and $k':=k$, scales $j$ and $j':=j \pm \delta'$ with $\delta'=0, \ldots, h$, we re-write the expectation part of Proposition~\ref{proposition5} by exploiting the properties  $\mathbf{A}^{jj'}=[\mathbf{A}^{j'j}]^\top$, where $A^{jj'}_{ll'}:=A_{ll';jj'}=A_{l'l;j'j}$ and $\mathbf{\tilde{I}}_{jj';k}^{\P}=[\mathbf{\tilde{I}}_{j'j;k}^{\P}]^\top$ (from the process-based raw periodogram construction in \eqref{eq:proc_raw}), and obtain
\begin{align*}     \E(\Tilde{I}_{jj',k}^{\P;(p,q)})&=\sum\limits_{(l,l') \in \mathcal{B}_h} A_{ll'}^{jj'}S_{ll'}^{(p,q)}(k/T)+\mathcal{O}(MT^{-1})  \mbox{ for all channel pairs } (p,q),\\
&=\sum\limits_{\delta=-h}^h\sum\limits_{l=max\{1,1-\delta\}}^{min\{J-\delta,J\}} A_{l, l+\delta}^{jj'}S_{l,l+\delta}^{(p,q)}(k/T)+\mathcal{O}(MT^{-1}),\\
&=\sum\limits_{\delta=0}^h\sum\limits_{l=1}^{J-\delta} A_{l, l+\delta}^{jj'}S_{l,l+\delta}^{(p,q)}(k/T)+
\sum\limits_{\delta=1}^h\sum\limits_{l=1}^{J-\delta} A_{l, l+\delta}^{j'j}S_{l,l+\delta}^{(q,p)}(k/T)+\mathcal{O}(MT^{-1}).
\end{align*}

As $M, T \rightarrow \infty$, for each $j$, $u$, the consistency result $\hat{S}_{jj'}^{\P;(p,q)}(u)\stackrel{P}{\rightarrow} {S}^{(p,q)}_{jj'}(u)$ then follows from the consistency results
$\Tilde{\Tilde{{I}}}_{jj',k}^{\P; (p,q)} \stackrel{P}{\rightarrow} \beta_{jj'}^{(p,q)}(k/T)$  for all fine enough scales $j,\, j'$ (as shown in Proposition~\ref{proposition5}) and then using the continuous mapping theorem \citep{billingsley1999convergence} for the continuous function $g(x_1, \ldots, x_{J^{\star}})= \sum_{l''=1}^{J^{\star}} (\tilde{A})^{-1}_{j''l''} x_{l''}$ that defines their linear combination with coefficients given by the matrix $\tilde{\mathbf{A}}^{-1}$ entries. Each $j''\, ,l''$  corresponds to one of the $(l,l')$ scale combinations out of the total $J^{\star}:=J+2\sum\limits_{\delta=1}^h(J-\delta)=J(2h+1)-h(h+1)$ ones, as governed by the number of active scale combinations in $\mathcal{B}_h$ (see also the construction in Appendix~\ref{app:Aconstruct} below).

Additionally, using the properties of the matrix $\tilde{\mathbf{A}}$, we obtain the estimator asymptotic unbiasedness from the linearity of the expectation operator and from the asymptotic unbiasedness of the corrected periodogram (see the expectation part of Proposition~\ref{proposition5} proof). 

\subsection{Construction of the proposed process-based estimator} \label{app:Aconstruct}

In this section, we illustrate the implementation of our process-based estimator in Section~\ref{sec:procest}, as well as its theoretical underpinnings. Again, to avoid notational clutter, we drop the superscript $^\P$ from the \underline{p}rocess-based periodogram $I_{jj',kk'}^{\P;(p,q)}$ and related quantities.

A key relationship is that the process covariance structure may be written for any channels,
\begin{align*}    
\tilde{c}^{(p,q)}(u,\tau)&= \sum\limits_{(j,j') \in \mathcal{B}_h} S_{jj'}^{(p,q)}(u) \boldsymbol{\Psi}_{j j'}(\tau),\\
&=\sum\limits_{\delta=-h}^h\sum\limits_{j=max\{1,1-\delta\}}^{min\{J-\delta,J\}} S_{j,j+\delta}^{(p,q)}(u) \boldsymbol{\Psi}_{j, j+\delta}(\tau), \, \forall u, \tau.
\end{align*}
As $\boldsymbol{\Psi}_{j j'}(\tau)=\boldsymbol{\Psi}_{j' j}(-\tau)$ and $\mathbf{S}_{jj'}(u)=\mathbf{S}^\top_{j'j}(u)$ from their respective  definitions, we re-write
\begin{align}
\tilde{c}^{(p,q)}(u,\tau)&=\sum\limits_{\delta=0}^h\sum\limits_{j=1}^{J-\delta} S_{j,j+\delta}^{(p,q)}(u) \boldsymbol{\Psi}_{j, j+\delta}(\tau)
+ \sum\limits_{\delta=1}^h\sum\limits_{j=1}^{J-\delta} S_{j,j+\delta}^{(q,p)}(u) \boldsymbol{\Psi}_{j, j+\delta}(-\tau),\nonumber\\
&= \boldsymbol{\Psi}^{\top}(\tau) \tilde{\mathbf S}^{(p,q)}(u),
\, \forall u, \tau, \label{eq:proccovspec}
\end{align}
where for the last equality we denoted the wavelet cross-correlation vector by $\boldsymbol{\Psi}(\tau):=\left[\{\boldsymbol{\Psi}_{j, j+\delta}(\tau)\}_{{\delta=0:h},{j=1:J-\delta}}, \{\boldsymbol{\Psi}_{j, j+\delta}(-\tau)\}_{{\delta=1:h},{j=1:J-\delta}} \right]^\top$ and its corresponding localised cross-spectral vector $\tilde{\mathbf S}^{(p,q)}(u):=
[
(
{S}_{j,j+\delta}^{(p,q)}(u)
)_{\delta=0:h, j=1:J-\delta}\,  ,
(
{S}_{j,j+\delta}^{(q,p)}(u))_{\delta=1:h, j=1:J-\delta}
]^\top
$, each with $J^{\star}:=J+2\sum\limits_{\delta=1}^h(J-\delta)(=J(2h+1)-h(h+1))$ entries.

Using the working assumption that stipulates non-zero cross-spectral activity only for scales that are at most $h$ steps away from one another, i.e., ${S}_{jj'}^{(p,q)}(u) \neq 0$ iff $(j,j')\in \mathcal{B}_h$ with $\mathcal{B}_h=\{(j,j')/ |j-j'| \leq h\}$ and since $A_{ll';jj'}=A_{l'l;j'j}$, ${S}_{jj'}^{(p,q)}(u)={S}_{j'j}^{(q,p)}(u)$, we can re-write the expectation part of Proposition~\ref{proposition5} by using only the contributing scale pairs in $\mathcal{B}_h$, 
\begin{align*}     \E(\Tilde{I}_{jj',k}^{(p,q)})&=\sum\limits_{(l,l') \in \mathcal{B}_h} A_{ll'}^{jj'}S_{ll'}^{(p,q)}(k/T)+\mathcal{O}(MT^{-1}),\\
&=\sum\limits_{\delta=-h}^h\sum\limits_{l=max\{1,1-\delta\}}^{min\{J-\delta,J\}} A_{l, l+\delta}^{jj'}S_{l,l+\delta}^{(p,q)}(k/T)+\mathcal{O}(MT^{-1}).
\end{align*}
For example, when only neighbouring scales are connected ($h=1$), the above becomes 
\begin{align*}     
\E(\Tilde{I}_{jj',k}^{(p,q)})&=
\underbrace{\sum\limits_{l=1}^{J} A^{jj'}_{l, l}S_{l,l}^{(p,q)}(k/T)}_{\delta=0}+
\underbrace{\sum\limits_{l=1}^{J-1} A^{jj'}_{l, l+1}S_{l,l+1}^{(p,q)}(k/T)}_{\delta=1}+ \underbrace{\sum\limits_{l=2}^{J} A^{jj'}_{l, l-1}S_{l,l-1}^{(p,q)}(k/T)}_{\delta=-1}+
\mathcal{O}(MT^{-1})\\
&=
\underbrace{\sum\limits_{l=1}^{J} A^{jj'}_{l, l}S_{l,l}^{(p,q)}(k/T)}_{\delta=0}+
\underbrace{\sum\limits_{l=1}^{J-1} A^{jj'}_{l, l+1}S_{l,l+1}^{(p,q)}(k/T)}_{\delta=1}+\underbrace{\sum\limits_{l=1}^{J-1} A^{j'j}_{l,l+1}S_{l,l+1}^{(q,p)}(k/T)}_{\delta=-1}+
\mathcal{O}(MT^{-1}).
\end{align*}
Hence, for the case when the $J \times J$ matrix $\mathbf{S}^{(p,q)}(k/T)$ is tridiagonal, i.e., it has non-zero elements only on the main diagonal and on its upper and lower diagonals, the above is 
\begin{align}
\E(\Tilde{I}_{jj',k}^{(p,q)})&=
\underbrace{\left(diag(A^{jj'})^\top, upperdiag(A^{jj'})^\top, upperdiag(A^{j'j})^\top\right)}_{:=(\tilde{\mathbf{A}}^{j,j'})^\top \,is\,a\,row\, vector\, with \, \left(J+(J-1)+(J-1)\right) \, entries}\times \nonumber\\
&\times \underbrace{\left(diag(\mathbf{S}^{(p,q)}(k/T))^\top, upperdiag(\mathbf{S}^{(p,q)}(k/T))^\top, upperdiag(\mathbf{S}^{(q,p)}(k/T))^\top\right)^\top }_{:=\tilde{\mathbf{S}}^{(p,q)} (k/T)\,is\,a\,column\, vector\, with \, \left(J+(J-1)+(J-1)\right) \, entries}+\mathcal{O}(MT^{-1}), \mbox{ or}, \nonumber\\
\E(\Tilde{I}_{jj',k}^{(p,q)})&=
(\tilde{\mathbf{A}}^{j,j'})^\top\tilde{\mathbf{S}}^{(p,q)}(k/T) +\mathcal{O}(MT^{-1}), \forall (j,j')\in \mathcal{B}_1.\label{eq:crossscale}
\end{align}
For clarity, the notation $diag(C)=(c_{l,l})_{l=1}^{n}$ and $upperdiag(C)=(c_{l,l+1})_{l=1}^{n-1}$ extracts in column format the main diagonal ($n \times 1$ entries) and the upper diagonal ($(n-1) \times 1$ entries) of a general $n \times n$ matrix $C$, respectively.\\
Since the set $\mathcal{B}_1$ has $(J+2(J-1))$ elements, we can concatenate the periodograms associated to the pairs $(j,j') \in \mathcal{B}_1$, hence
\begin{align}
&\left(diag(\tilde{\mathbf{I}}^{(p,q)}_k)^\top, upperdiag(\tilde{\mathbf{I}}^{(p,q)}_k)^\top,upperdiag(\tilde{\mathbf{I}}^{(q,p)}_k)^\top\right)^\top= \nonumber\\
&=\underbrace{\left ( \left( \Tilde{I}_{11,k}^{(p,q)}, \Tilde{I}_{22,k}^{(p,q)}, \ldots, \Tilde{I}_{JJ,k}^{(p,q)}\right), 
\left( \Tilde{I}_{12,k}^{(p,q)}, \Tilde{I}_{23,k}^{(p,q)}, \ldots, \Tilde{I}_{J-1,J,k}^{(p,q)}\right),\left( \Tilde{I}_{12,k}^{(q,p)}, \Tilde{I}_{23,k}^{(q,p)}, \ldots, \Tilde{I}_{J-1,J,k}^{(q,p)}\right)\right)^\top}_{:=\tilde{\Tilde{\mathbf I}}_{k}^{(p,q)} \,is\,a\,column \, vector\, with \, \left(J+(J-1)+(J-1)\right) \, entries},\label{eq:concper} 
\end{align}
where we used the property $\mathbf{\tilde{I}}_{jj',k}=\mathbf{\tilde{I}}_{j'j,k}^\top$ (from the raw periodogram construction in equation \eqref{eq:proc_raw}).
Equations~\eqref{eq:crossscale} and~\eqref{eq:concper} thus yield
\begin{align*}
\E\left[\tilde{\Tilde{\mathbf I}}_{k}^{(p,q)} \right]=\tilde{\mathbf{A}} \tilde{\mathbf{S}}^{(p,q)}(k/T) +\mathcal{O}(MT^{-1}),
\end{align*}
where the $(J+2(J-1)) \times (J+2(J-1))$ matrix $\tilde{\mathbf{A}}$ matches the ordering in equation~\eqref{eq:concper} $$\tilde{\mathbf{A}}=\left(\left( \tilde{\mathbf{A}}^{11} | \ldots |\tilde{\mathbf{A}}^{JJ}\right)| \left( \tilde{\mathbf{A}}^{12} | \ldots |\tilde{\mathbf{A}}^{J-1,J}\right)| \left( \tilde{\mathbf{A}}^{21} | \ldots |\tilde{\mathbf{A}}^{J,J-1}\right) \right)^\top$$ and $|$ denotes the column concatenation for clarity.


\section{Appendix B (Simulation study)}\label{app:B}

For completeness, we evaluate the mean squared bias (MSB) associated to the coherence estimators as
\begin{align}\label{eq:msb}
    MSB_{jj'} = \frac{1}{T} \sum\limits_{t=1}^T \left( \frac{1}{R}\sum\limits_{r=1}^R \hat{\boldsymbol{\rho}}_{jj'}^{(r)} (t/T) -  \boldsymbol{\rho}_{jj'} (t/T) \right) ^2.
\end{align}

\subsection{Details for simulation settings}\label{app:sim}
\noindent \textbf{Two-scale setting (main text, Scenario 1).}
In this part we focus on the single-scale wavelet coherence, by structuring non-zero power at scale $j_1$ and $j_2$. Specifically, we have, 
\begin{align*}
\mathbf{S}_{j_1}(u) &=\left[\begin{array}{ccc}
4+16 u & 2+6 u & 1+2u \\
2+6u & 4+4f(u-0.5) &   1+2u\\
1+2u & 1+2u & 10
\end{array}\right]  \\  \mathbf{S}_{j_2}(u) &= \left[\begin{array}{ccc}
4+11 u & 2+4 u & 1+u \\
2+4u & 4+4f(u-0.5) &   1+u\\
1+u & 1+u & 8 
\end{array}\right]
\end{align*}
where $u \in(0,1)$ and $f(x)=0 $ for $x<0$ and $f(x)=1$ for $x \geq 0$. In this scenario, we specify $j_1 = 1$ and $j_2=2$, which means we have non-zero spectrum at only two finest level. On the other hand, 
 $\boldsymbol{Q}(u)$ is structured as a block matrix of $J \times J$ blocks, each block being a $P \times P$ matrix, e.g., $P=3$. Each block $\boldsymbol{Q}_{j_1, j_2}(u) \in \mathbb{R}^{3 \times 3}$ represents the localised covariance between scale $j_1$ and scale $j_2$ innovations, with
\begin{align*}
    \boldsymbol{Q}(u)=\left[\begin{array}{cccc}
\boldsymbol{Q}_{1, 1}(u) &\boldsymbol{Q}_{1, 2}(u) & \cdots & \boldsymbol{Q}_{1, J}(u) \\
\boldsymbol{Q}_{2, 1}(u) & \boldsymbol{Q}_{2, 2}(u) & \cdots & \boldsymbol{Q}_{2, J}(u) \\
\vdots & \vdots & \ddots & \vdots \\
\boldsymbol{Q}_{J, 1}(u) & \boldsymbol{Q}_{J, 2}(u)& \cdots & \boldsymbol{Q}_{J, J}(u)
\end{array}\right].
\end{align*}
For clarity, each submatrix $\boldsymbol{Q}_{j_1, j_2}(\cdotp)$ is a $P \times P$ matrix  
\begin{align*}
    \boldsymbol{Q}_{j_1, j_2}(k/T)=\left[\begin{array}{lll}
\operatorname{Cov}\left(z_{j_1,k}^{(1)},z_{j_2,k}^{(1)} \right) &  \operatorname{Cov}\left(z_{j_1,k}^{(1)},z_{j_2,k}^{(2)}  \right)  & \operatorname{Cov}\left(z_{j_1,k}^{(1)},z_{j_2,k}^{(3)} \right) \\
\operatorname{Cov}\left(z_{j_1,k}^{(2)},z_{j_2,k}^{(1)}\right) & \operatorname{Cov}\left(z_{j_1,k}^{(2)},z_{j_2,k}^{(2)}\right) & \operatorname{Cov}\left(z_{j_1,k}^{(2)},z_{j_2,k}^{(3)}\right) \\
\operatorname{Cov}\left(z_{j_1,k}^{(3)},z_{j_2,k}^{(1)}\right) & \operatorname{Cov}\left(z_{j_1,k}^{(3)},z_{j_2,k}^{(2)}\right) & \operatorname{Cov}\left(z_{j_1,k}^{(3)},z_{j_2,k}^{(3)}\right)
\end{array}\right]
\end{align*}
where $z_{j,k}^{(p)}$ denotes the random innovation of channel $p$ and scale-time $(j,k)$, e.g., $\operatorname{Cov}\left(z_{j_1,k}^{(2)},z_{j_2,k}^{(1)}\right)$ incorporates the dependence between scale $j_1$ subprocess of channel 2 and scale $j_2$ subprocess of channel 1 localised at $u=k/T$. For scenario 1 we have
\begin{align*}
\boldsymbol{Q}_{j_1, j_2}(u)=\left\{\begin{array}{ll}
\boldsymbol{I}_P, & j_1=j_2, \\
\boldsymbol{0}, & j_1 \neq j_2,
\end{array} \quad \forall u \in(0,1)\right.
\end{align*}

\noindent \textbf{Scenario 3: Multiple scales ($j=1,2,3$).}
We next consider a multiscale setting with non-zero spectral power across three scales $j\in\{1,2,3\}$. We specify the within-scale spectral matrices $\mathbf{S}_j(u)\in\mathbb{R}^{3\times 3}$ as follows
\begin{align*}
\mathbf{S}_{1}(u) &=
\left[\begin{array}{ccc}
4+16u &\quad 2+6u &\quad 2+2u\\
2+6u &\quad s_{1,22}(u) &\quad 2+2u\\
2+2u &\quad 2+2u &\quad 8+12u
\end{array}\right],\qquad
s_{1,22}(u)=
\begin{cases}
4+28u, & 0<u<0.5,\\
18, & 0.5\le u<1,
\end{cases}
\\[3pt]
\mathbf{S}_{2}(u) &=
\left[\begin{array}{ccc}
6+9u &\quad 2+4u &\quad 2+u\\
2+4u &\quad s_{2,22}(u) &\quad 1+4u\\
2+u &\quad 1+4u &\quad 12+4u
\end{array}\right],\qquad
s_{2,22}(u)=
\begin{cases}
4+24u, & 0<u<0.5,\\
16, & 0.5\le u<1,
\end{cases}
\\[3pt]
\mathbf{S}_{3}(u) &=
\left[\begin{array}{ccc}
8+6u &\quad 1+2u &\quad 2+2u\\
1+2u &\quad 6+6u &\quad 1+u\\
2+2u &\quad 1+u &\quad s_{3,33}(u)
\end{array}\right],\qquad
s_{3,33}(u)=
\begin{cases}
5+14u, & 0<u<0.5,\\
12, & 0.5\le u<1.
\end{cases}
\end{align*}

We impose cross-scale dependence via the innovation covariance block matrix $\boldsymbol{Q}(u)$, whose diagonal blocks satisfy
\begin{align*}
\boldsymbol{Q}_{j, j}(u) = \boldsymbol{I}_P,\qquad \forall j, \, u,
\end{align*}
and with non-zero off-diagonal blocks $\boldsymbol{Q}_{1,2}(u)$, $\boldsymbol{Q}_{2,1}(u)$, $\boldsymbol{Q}_{1,3}(u)$ and $\boldsymbol{Q}_{3,1}(u)$ given by  
\begin{align*}
\boldsymbol{Q}_{1,2}(u)&= 
\left[\begin{array}{ccc}
0.1+0.1u & \quad 0.1+0.15u &\quad 0.1+0.05u\\
0.1+0.15u &\quad 0.1+0.2u &\quad 0.05+0.1u\\
0.1+0.05u &\quad 0.05+0.1u &\quad 0.05+0.15u
\end{array}\right],\qquad
\boldsymbol{Q}_{2,1}(u)=\boldsymbol{Q}_{1,2}(u),
\\[3pt]
\boldsymbol{Q}_{1,3}(u) &= 
\left[\begin{array}{ccc}
0.2+0.1u &\quad 0.05+0.25u &\quad 0.1+0.05u\\
0.05+0.25u &\quad 0.2+0.1u &\quad 0.05+0.1u\\
0.1+0.05u &\quad 0.05+0.1u &\quad 0.1+0.2u
\end{array}\right],\qquad
\boldsymbol{Q}_{3,1}(u)=\boldsymbol{Q}_{1,3}(u),
\end{align*}
and all other off-diagonal blocks are set to zero.
This design introduces two concurrent cross-scale links (between scales $(j_1,j_2) = (1,2)$ and $(j_1,j_2) = (1,3)$), allowing us to assess performance when multiple cross-scale interactions coexist.

\noindent \textbf{Scenario 4: Multiple scales ($j=1,2,3,4$).} We further consider a setting with non-zero spectral power across four scales $j\in\{1,2,3,4\}$.
The within-scale spectra are specified as
\begin{align*}
\mathbf{S}_{1}(u) &=
\left[\begin{array}{ccc}
20-16u &\quad 10-8u &\quad 5-4u\\
10-8u &\quad 18-12u &\quad 5-2u\\
5-4u &\quad 5-2u &\quad 16-8u
\end{array}\right],\\[3pt]
\mathbf{S}_{2}(u) &=
\left[\begin{array}{ccc}
15-11u &\quad 6-4u &\quad 4-u\\
6-4u &\quad 12-6u &\quad 4-3u\\
4-u &\quad 4-3u &\quad 14-6u
\end{array}\right],\\[3pt]
\mathbf{S}_{3}(u) &=
\left[\begin{array}{ccc}
14-6u &\quad 4-2u &\quad 5-2u\\
4-2u &\quad 14-8u &\quad 4-2u\\
5-2u &\quad 4-2u &\quad 10-6u
\end{array}\right],\\[3pt]
\mathbf{S}_{4}(u) &=
\left[\begin{array}{ccc}
20-12u &\quad 2-u &\quad 3-2u\\
2-u &\quad 8-6u &\quad 2-u\\
3-2u &\quad 2-u &\quad 16-10u
\end{array}\right].
\end{align*}

For the innovation covariance blocks, we again set $\boldsymbol{Q}_{j, j}(u)=\boldsymbol{I}_3$ for all $j, \, u$ and impose two separate cross-scale links: one between scales $(j_1,j_2) = (1,2)$ and the other between scales $(j_1,j_2) = (3,4)$.
Specifically,
\begin{align*}
\boldsymbol{Q}_{1,2}(u) &= 
\left[\begin{array}{ccc}
0.4-0.1u &\quad 0.25-0.15u &\quad 0.15-0.05u\\
0.25-0.15u &\quad 0.3-0.2u &\quad 0.3-0.1u\\
0.15-0.05u &\quad 0.3-0.1u &\quad 0.2-0.1u
\end{array}\right],\qquad
\boldsymbol{Q}_{2,1}(u)=\boldsymbol{Q}_{1,2}(u),\\[3pt]
\boldsymbol{Q}_{3,4}(u) &= 
\left[\begin{array}{ccc}
0.3-0.2u &\quad 0.15-0.05u &\quad 0.3-0.2u\\
0.15-0.05u &\quad 0.4-0.2u &\quad 0.2-0.1u\\
0.3-0.2u &\quad 0.2-0.1u &\quad 0.3-0.1u
\end{array}\right],\qquad
\boldsymbol{Q}_{4,3}(u)=\boldsymbol{Q}_{3,4}(u),
\end{align*}
with all other off-diagonal blocks set to zero.
This configuration tests whether the estimator can recover cross-scale coherence when multiple scales are active but cross-scale dependence occurs only in specific scale pairs.

\subsection{Additional simulation results} \label{appB:add results}
This subsection reports supplementary results for the simulation studies in the main text.

\begin{figure}[htbp]
    \centering
    \includegraphics[width=0.9\linewidth,height=3.5in]{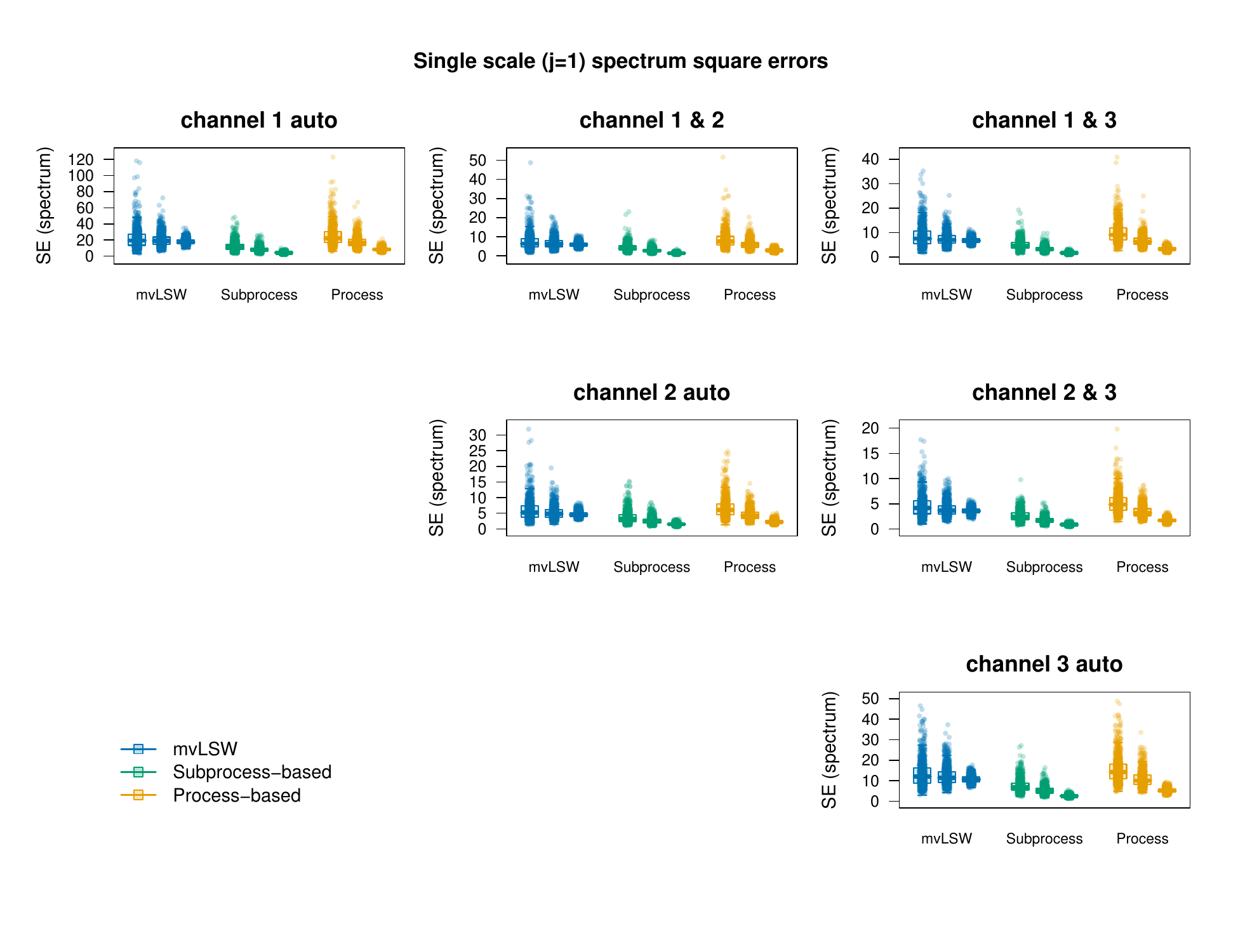}
  \hfill    \includegraphics[width=0.9\linewidth,height=3.5in]{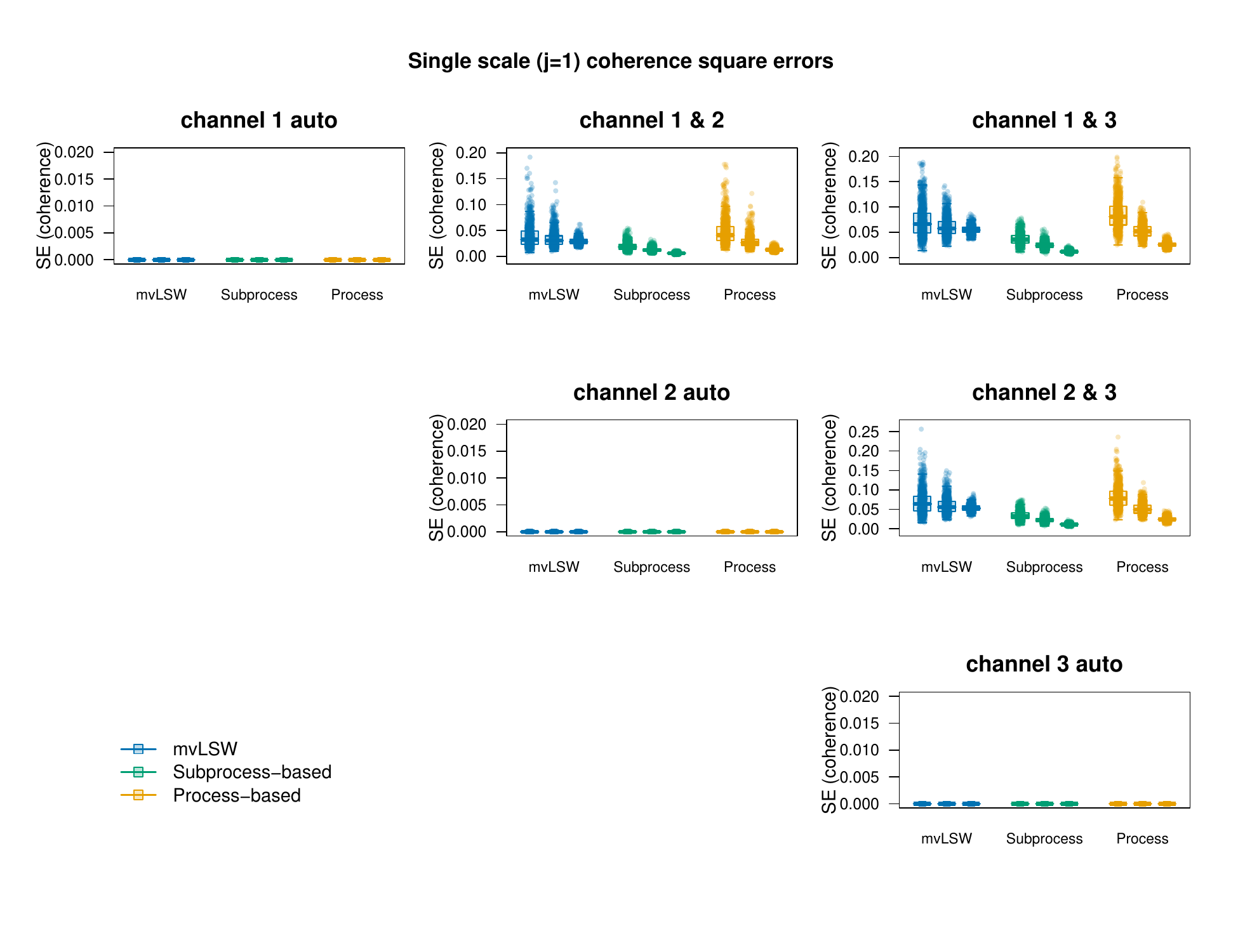}
  \caption{Time-averaged squared errors (SE) of the single-scale spectrum (top) and coherence (bottom) estimators at scale $j=1$ (Scenario 1). Each panel corresponds to an auto- or cross-channel pair. Boxplots summarize $R=1,000$ Monte Carlo replicates for mvLSW, subprocess-based (approximated), and process-based estimators. Within each estimator group, the three adjacent boxplots (left to right) correspond to $T=512$, $1024$, and $4096$.}
  \label{fig:sin_se_j1}
\end{figure}

\begin{figure}[htbp]
  \centering    \includegraphics[width=.9\linewidth,height=3.5in]{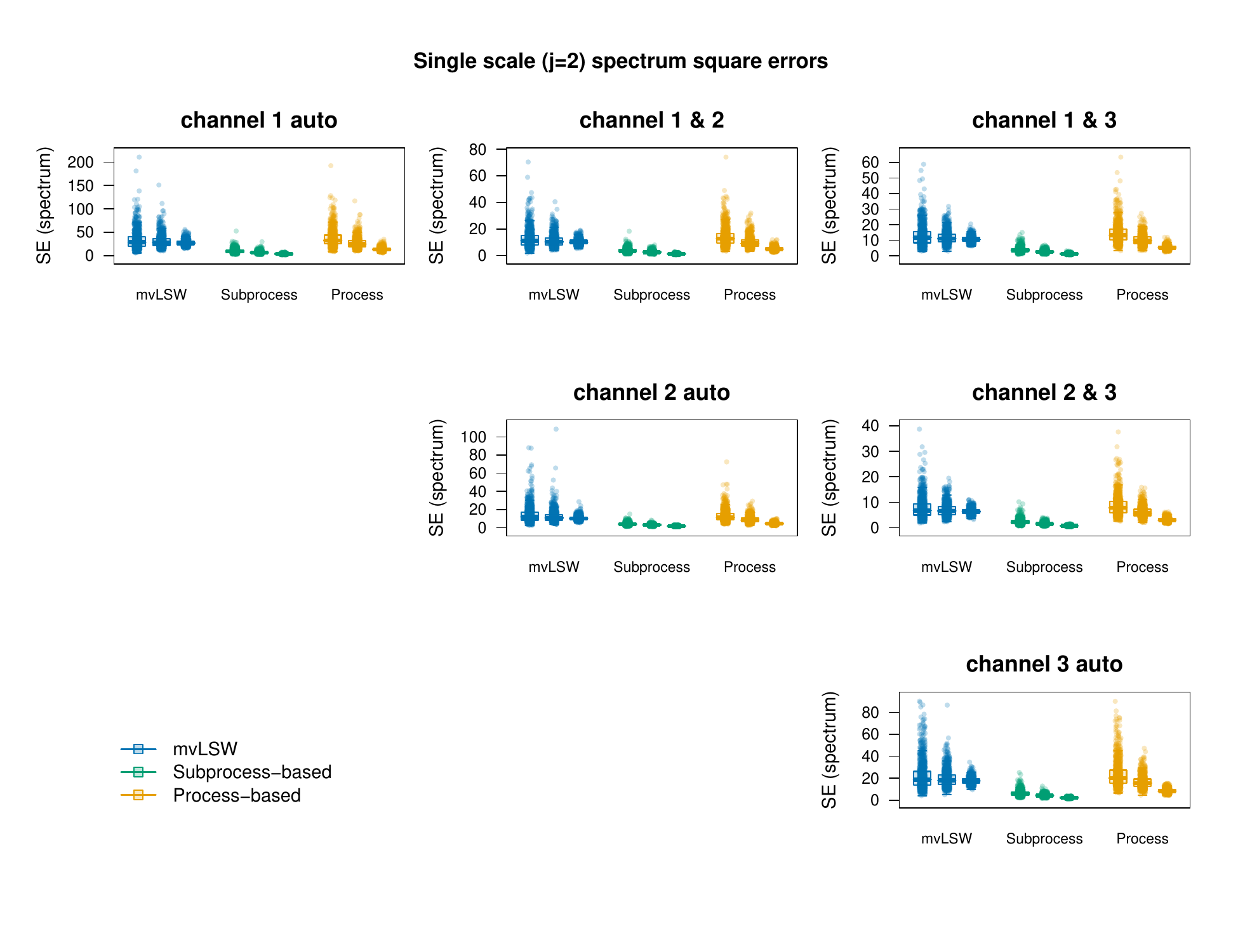}
    \hfill     \includegraphics[width=.9\linewidth,height=3.5in]{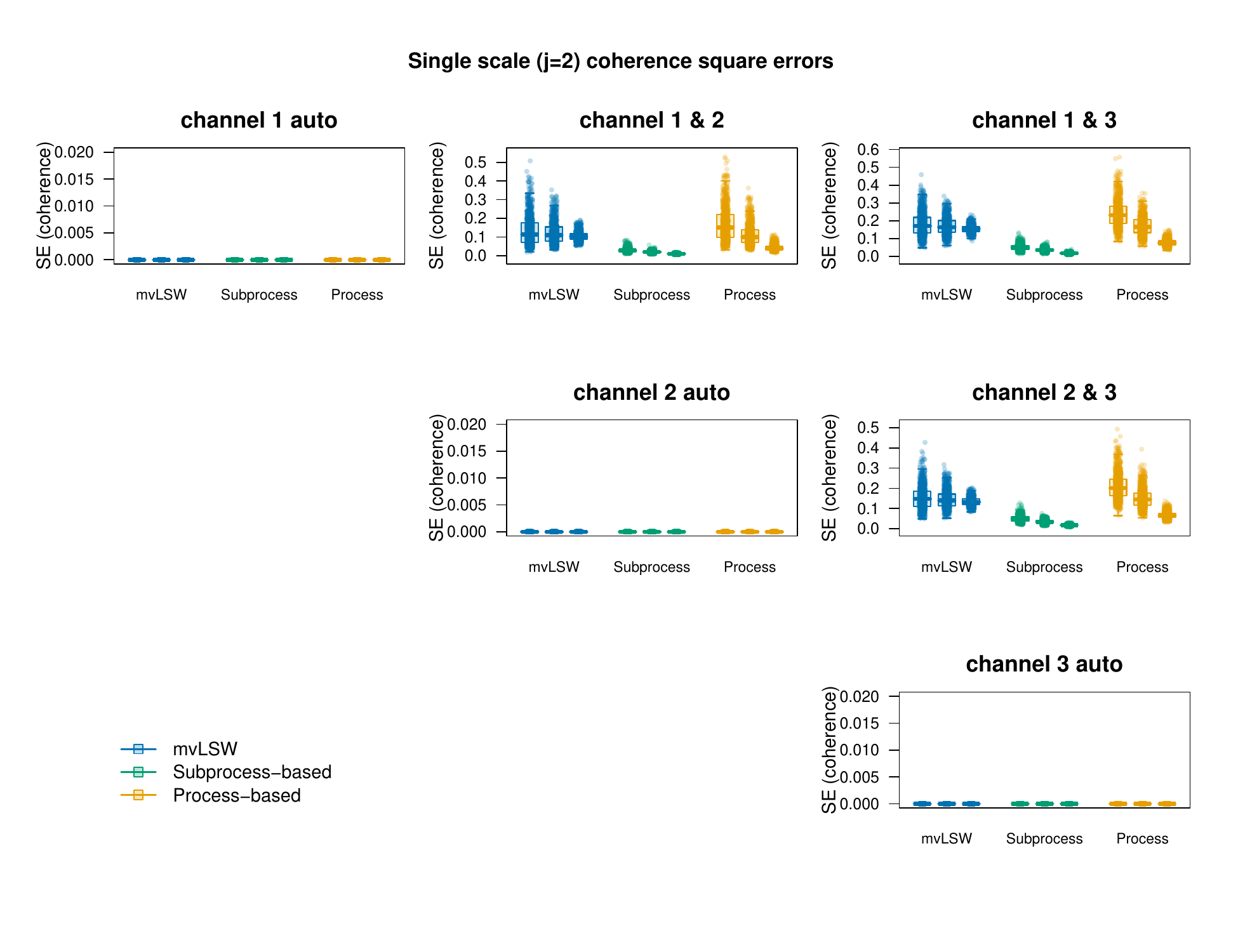}
  \caption{Time-averaged squared errors of the single-scale spectrum (top) and coherence (bottom) estimators at scale $j=2$ (Scenario 1). Each panel corresponds to an auto- or cross-channel pair. Boxplots summarize $R=1,000$ Monte Carlo replicates for mvLSW, subprocess-based (approximated), and process-based estimators. Within each estimator group, the three adjacent boxplots (left to right) correspond to $T=512$, $1024$, and $4096$.}
  \label{fig:sin_se_j2}
\end{figure}

When no cross-scale connections exist (Scenario 1), Figures \ref{fig:sin_se_j1} and \ref{fig:sin_se_j2} report time-averaged squared errors for spectrum and coherence at scales $j=1$ and $j=2$, respectively. 
Each panel corresponds to one entry of the spectral matrix (or the corresponding coherence), so that performance can be compared uniformly across auto- and cross-channel pairs. 
These summaries confirm the improvement of the proposed estimators over the classical MvLSW estimator, with the subprocess-based method providing the largest reduction in error.

\begin{figure}[htbp]
  \centering
    \includegraphics[width=.75\linewidth]{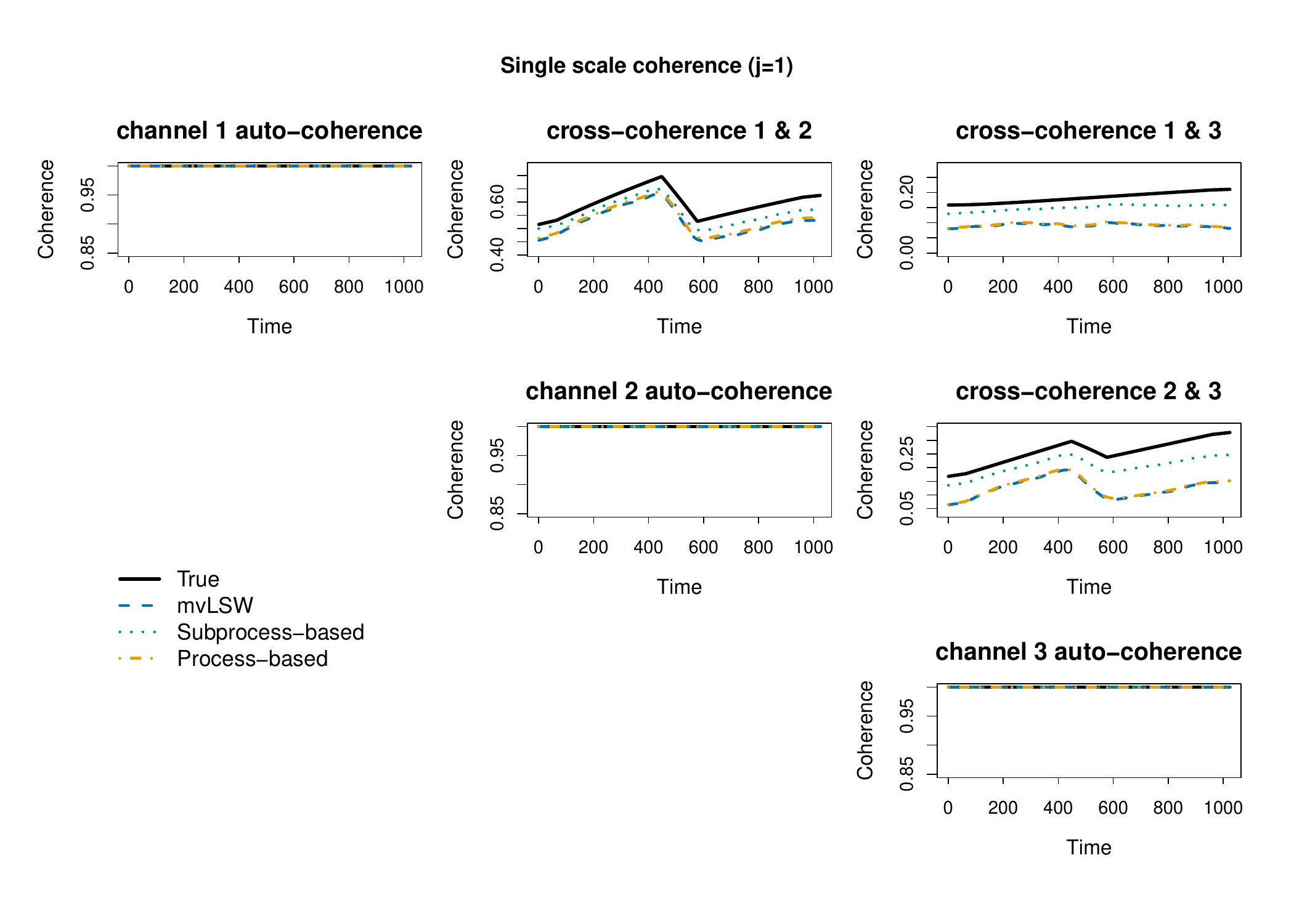}
    \hfill 
    \includegraphics[width=.75\linewidth]{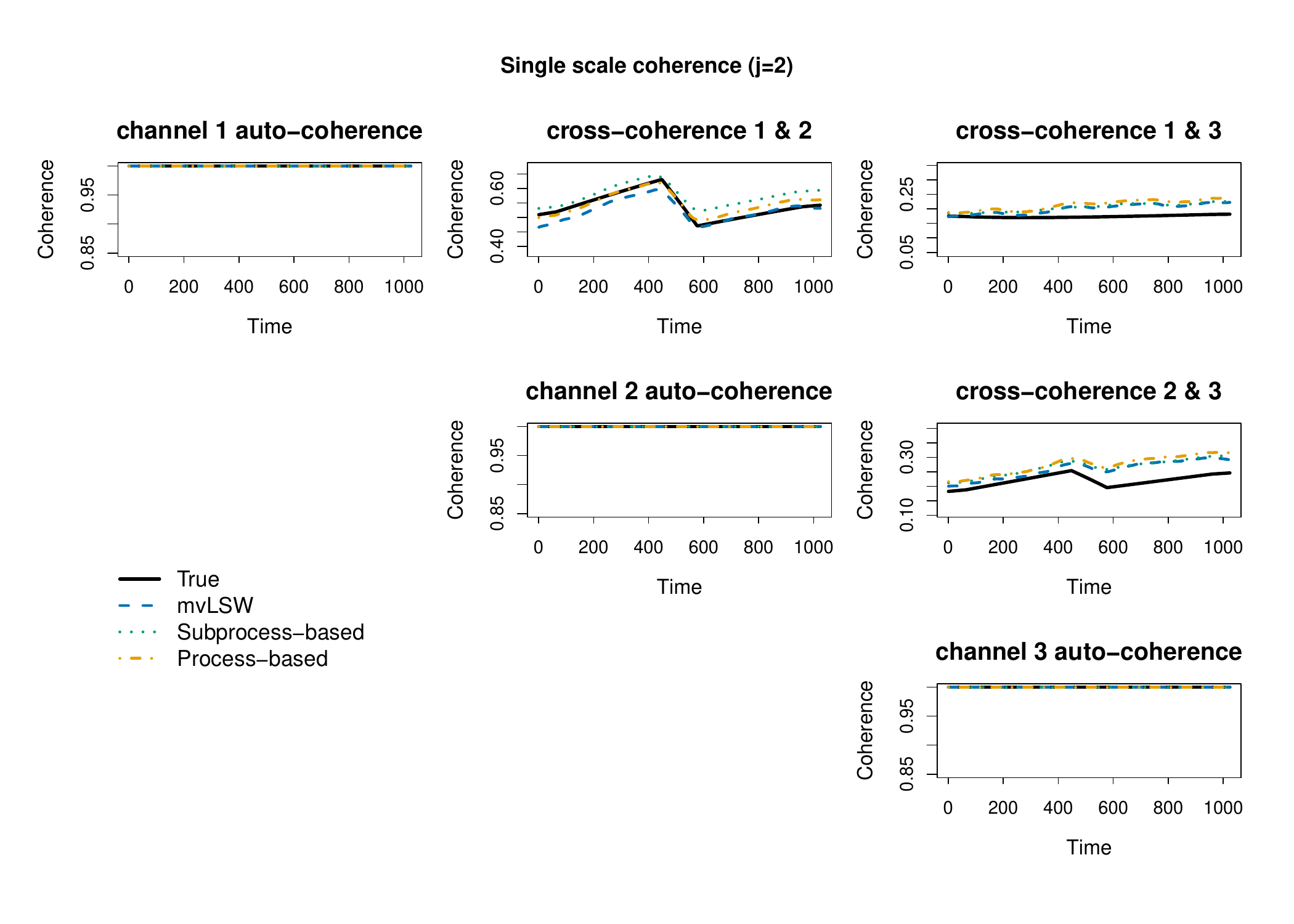}
  \caption{Single-scale coherence estimates at $j=1$ (top) and $j=2$ (bottom) for $T=1024$, with cross-scale structure present (Scenario 2, Case 1), averaged over 1{,}000 replicates. Black: truth; coloured: mvLSW, subprocess-based (approximated),  process-based estimates.}
  \label{fig:sin_coh_s2c1}
\end{figure}

\begin{figure}[htbp]
  \centering
    \includegraphics[width=.75\linewidth]{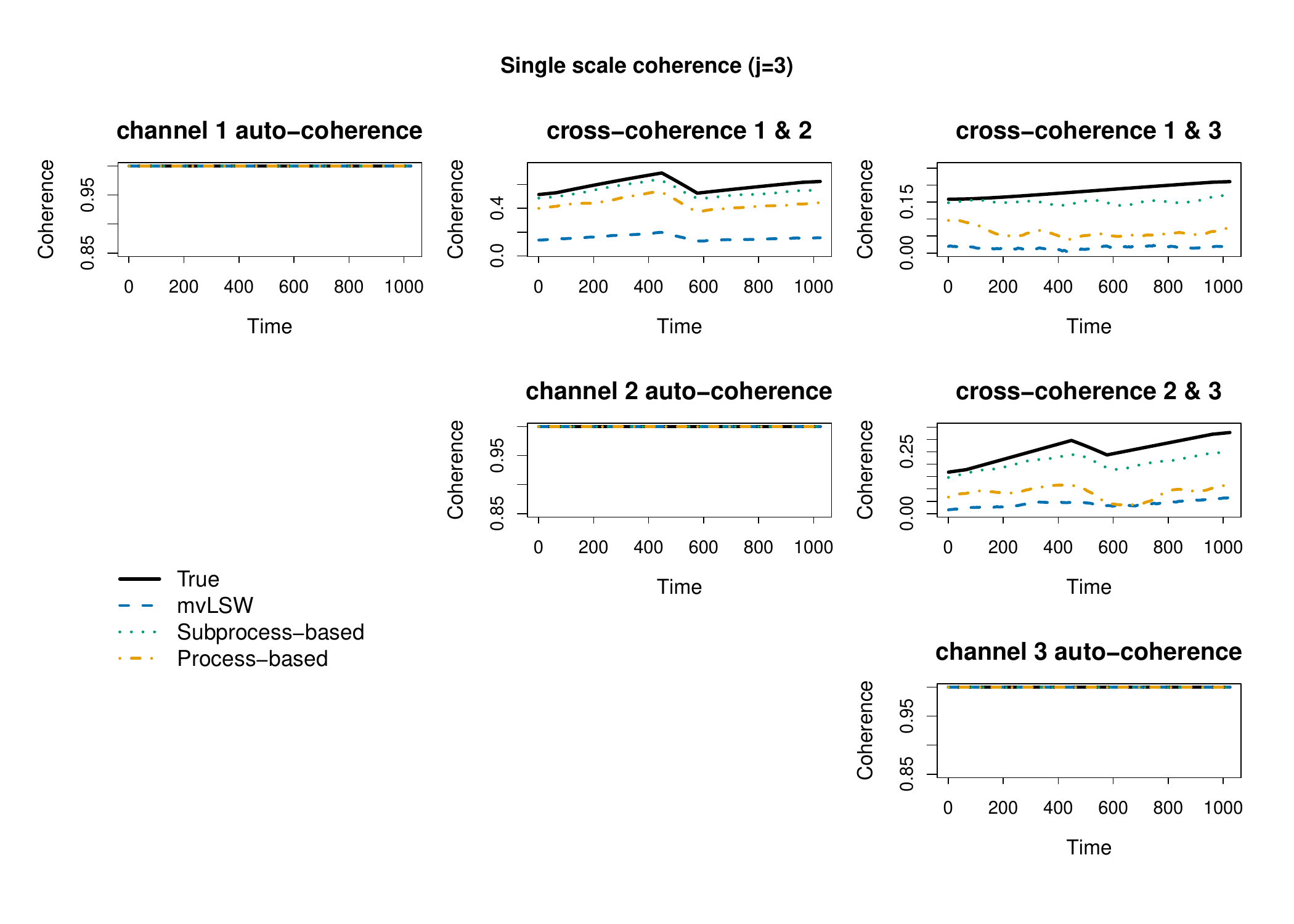}
    \hfill 
    \includegraphics[width=.75\linewidth]{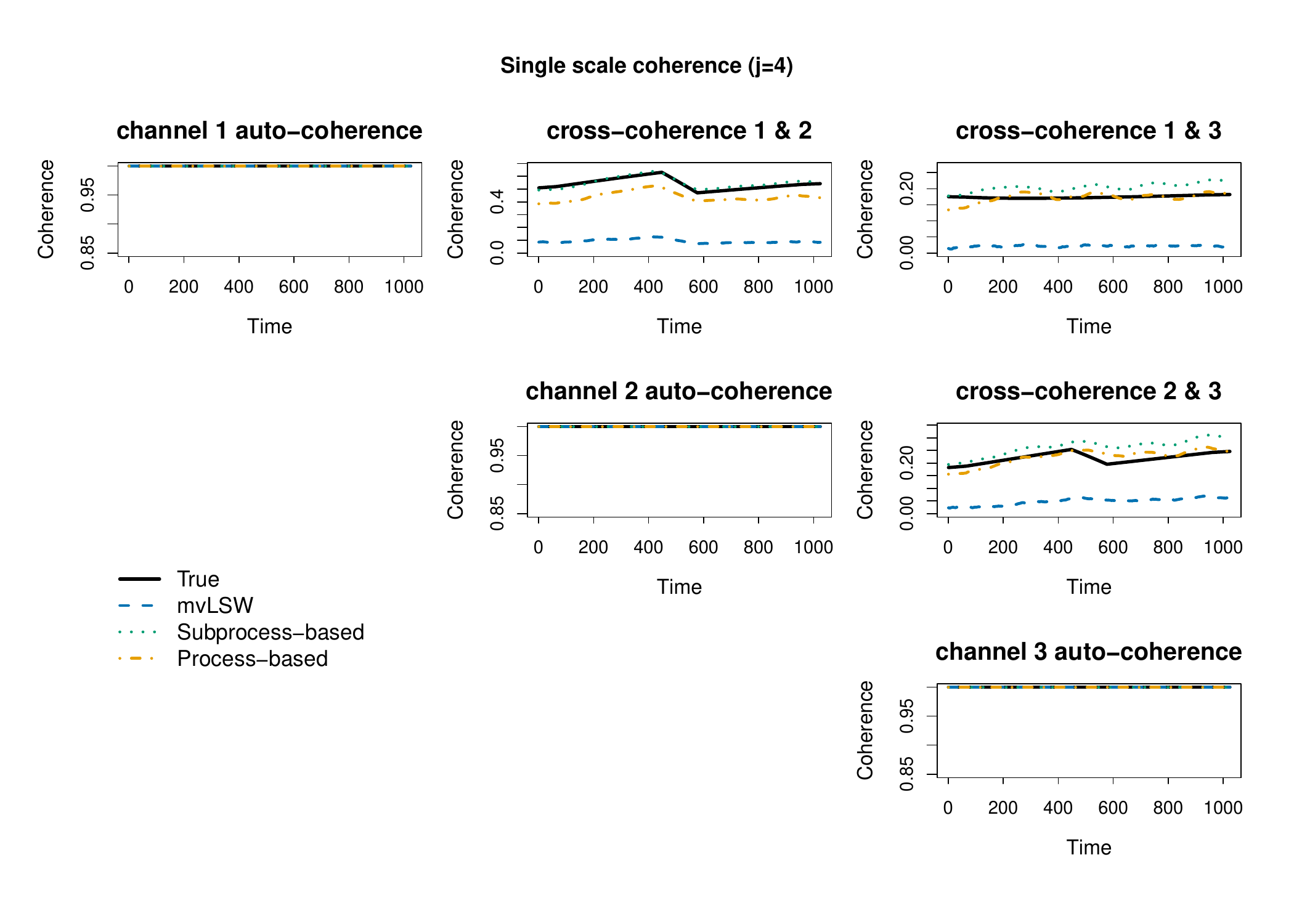}
  \caption{Single-scale coherence estimates at $j=3$ (top) and $j=4$ (bottom) for $T=1024$, with cross-scale structure present (Scenario 2, Case 2), averaged over 1{,}000 replicates. Black: truth; coloured: mvLSW, subprocess-based (approximated),  process-based estimates.}
  \label{fig:sin_coh_s2c2}
\end{figure}

\begin{figure}[htbp]
  \centering
    \includegraphics[width=.75\linewidth]{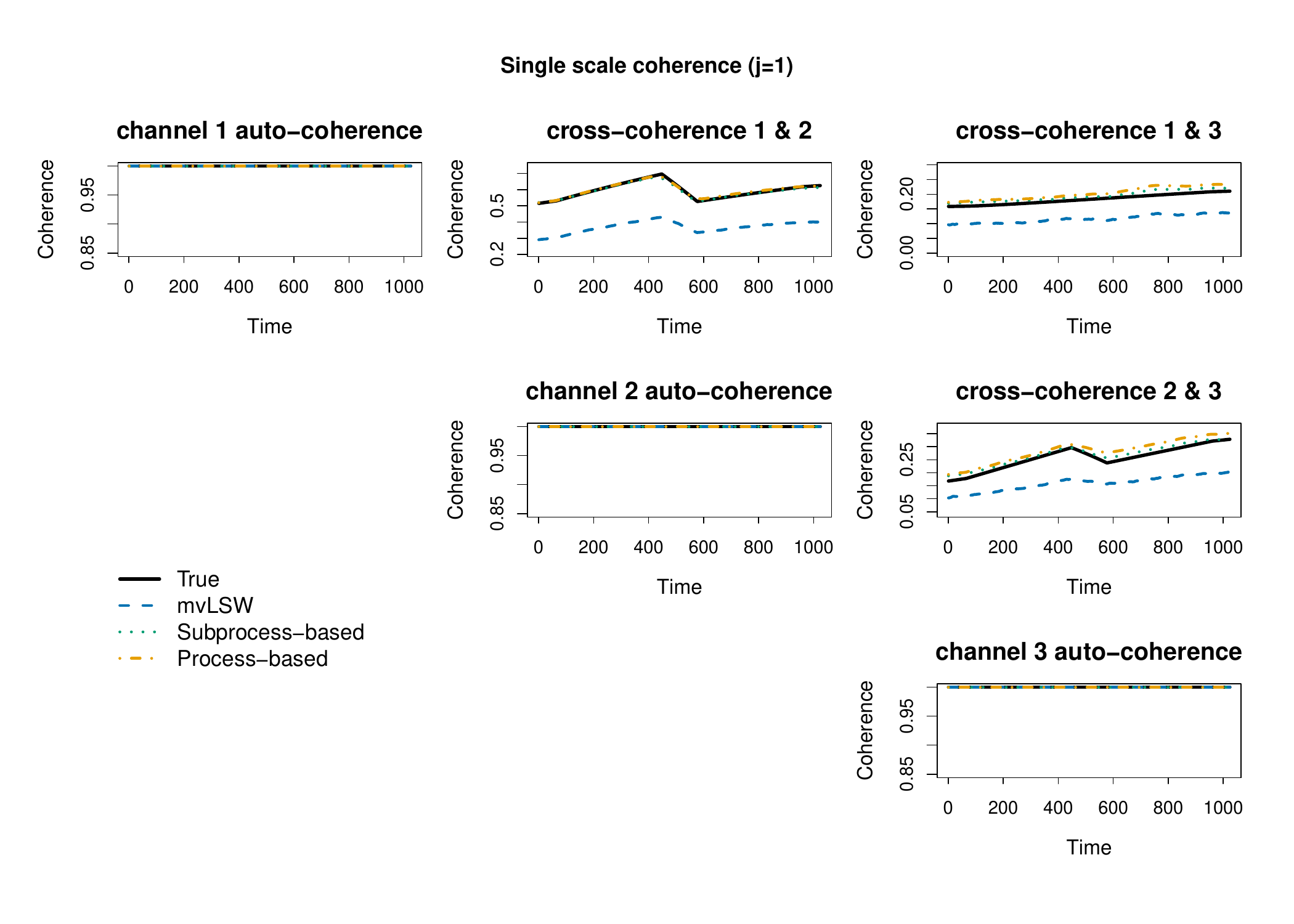}
    \hfill 
    \includegraphics[width=.75\linewidth]{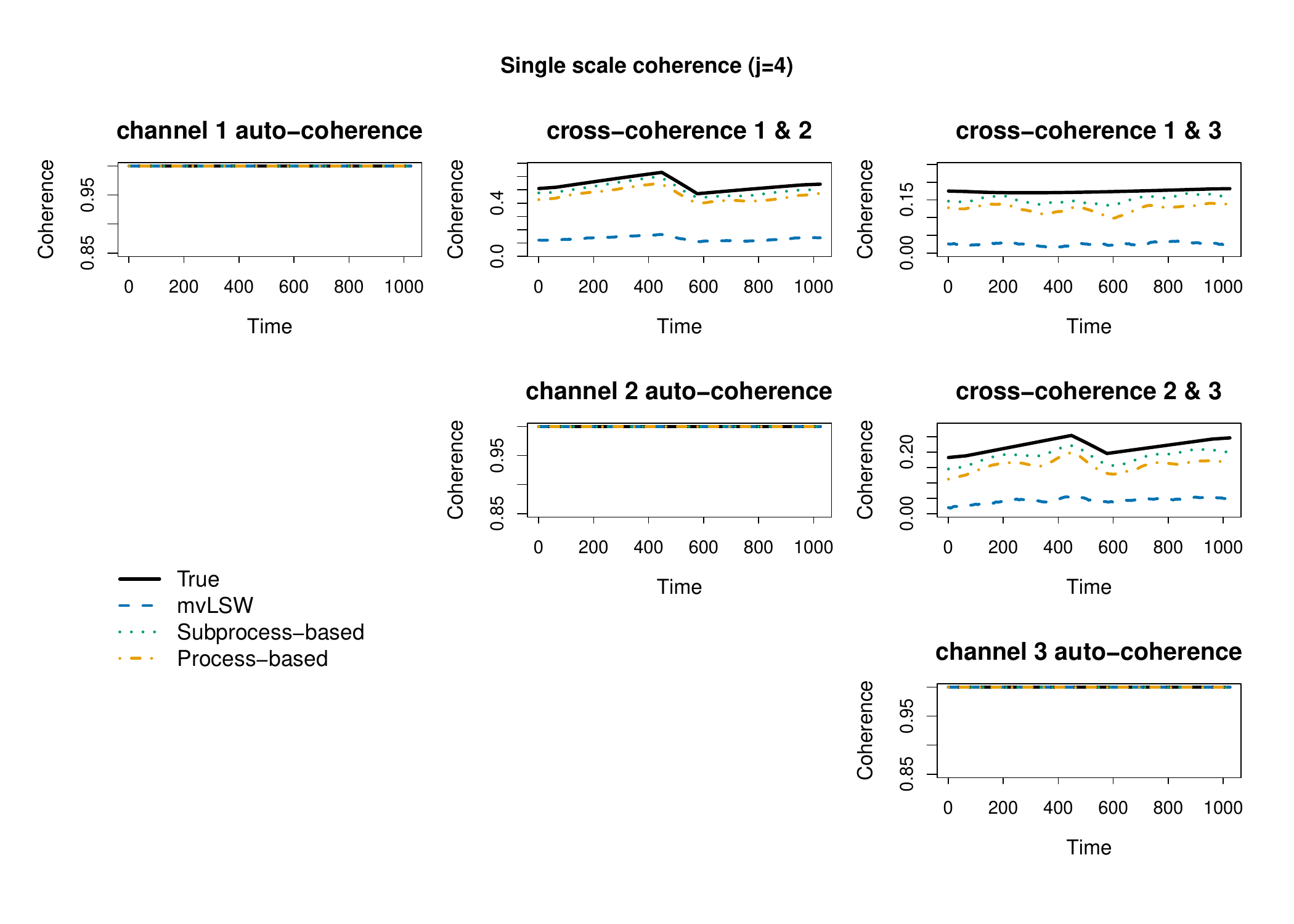}
  \caption{Single-scale coherence estimates at $j=1$ (top) and $j=4$ (bottom) for $T=1024$, with cross-scale structure present (Scenario 2, Case 3), averaged over 1{,}000 replicates. Black: truth; coloured: mvLSW, subprocess-based (approximated), process-based estimates.}
  \label{fig:sin_coh_s2c3}
\end{figure}

When cross-scale activity is present (Scenario 2), Figures~\ref{fig:sin_coh_s2c1} -- \ref{fig:sin_coh_s2c3} show the single-scale coherence estimates for the three cases considered under this setting. The results indicate that, in the presence of cross-scale structure, the classical MvLSW estimator is less effective at recovering the true coherence, particularly at coarser scales. In contrast, the proposed process-based estimator, and especially the subprocess-based estimator, more closely track the true coherence. Notably, the performance of the subprocess-based estimator remains stable across the three cases and appears largely insensitive to the increasing complexity of the cross-scale structure. This is consistent with the MSE comparisons in Tables~\ref{tab:coh_msb_mse_s2c1} -- \ref{tab:coh_msb_mse_s2c3}.

\newcolumntype{L}[1]{>{\raggedright\arraybackslash}p{#1}}

\newcommand{\DiagCell}[3]{
\begin{tikzpicture}[baseline=(c.base)]
  \node[inner sep=2pt, minimum width=#1, minimum height=1.6cm] (c) {};
  \draw (c.north west) -- (c.south east);
  \node[anchor=south west, xshift=1pt, yshift=1pt] at (c.south west) {\textbf{#2}};
  \node[anchor=north east, xshift=-1pt, yshift=-1pt] at (c.north east) {\textbf{#3}};
\end{tikzpicture}%
}

\begin{sidewaystable}[htbp]
\centering
\begin{threeparttable}
\caption{\textbf{Scenario 1:} Coherence accuracy under different scales and time lengths $T$ for channel $(p,q)=(1,2)$.}
\label{tab:coh_msb_mse_s1}

\small
\setlength{\tabcolsep}{2.5pt}
\renewcommand{\arraystretch}{1.15}

\begin{tabular}{L{2.1cm}!{\vrule width 1.1pt} *{12}{S[table-format=2.3]}}
\toprule
 \multicolumn{1}{c!{\vrule width 1.1pt}}{\DiagCell{2.1cm}{Scale}{Estimator}} &
\multicolumn{3}{c}{\textbf{MvLSW}} &
\multicolumn{3}{c}{\textbf{Process-based}} &
\multicolumn{3}{c}{\textbf{Subprocess (approx)}}&
\multicolumn{3}{c}{\textbf{Subprocess (true)}} \\
\cmidrule(lr){2-4}\cmidrule(lr){5-7}\cmidrule(lr){8-10}\cmidrule(lr){11-13}
& \textbf{$T{=}512$} & \textbf{$T{=}1024$} & \textbf{$T{=}4096$}
& \textbf{$T{=}512$} & \textbf{$T{=}1024$} & \textbf{$T{=}4096$}
& \textbf{$T{=}512$} & \textbf{$T{=}1024$} & \textbf{$T{=}4096$}
& \textbf{$T{=}512$} & \textbf{$T{=}1024$} & \textbf{$T{=}4096$} \\
\midrule
\multicolumn{13}{l}{\textbf{Panel A: MSB} (reported in units of $10^{-4}$)}\\
\addlinespace[2pt]
$j_1=1$ & 9.24 & 4.35 & 2.13 & 8.95 & 3.93 & 1.67 &  11.40  & 8.07 & 5.31 & 5.07 & 3.14 & 1.51 \\
$j_2=2$ & 57.85 & 48.89 & 28.80 & 44.85 & 23.52 & 3.86 & 3.87 & 2.91 & 2.17 & 4.52 & 3.38 & 1.34 \\
\addlinespace[4pt]

\multicolumn{13}{l}{\textbf{Panel B: MSE} (original scale)}\\
\addlinespace[2pt]
$j_1=1$ & 0.039 & 0.034 & 0.029 & 0.047 & 0.029 & 0.013 & 0.020 & 0.013 & 0.006 & 0.031 & 0.020 & 0.009 \\
$j_2=2$ & 0.132 & 0.122 & 0.104 & 0.166 & 0.111 & 0.042 & 0.029 & 0.020 &0.010 & 0.038 & 0.026 & 0.012 \\
\bottomrule
\end{tabular}

\begin{tablenotes}[flushleft]
\footnotesize
\item Notes: Panel A reports mean square bias (MSB) in units of $10^{-4}$ (multiply by $10^{-4}$ to recover MSB on the original scale). Panel B reports mean square error (MSE) on the original scale. The Subprocess (true) columns serve as an oracle baseline, corresponding to the idealized case in which the true subprocess construction is known, and are included as a reference rather than as a competing implementable estimator.
\end{tablenotes}
\end{threeparttable}
\end{sidewaystable}

\begin{sidewaystable}[p]
\centering
\begin{threeparttable}
\caption{\textbf{Scenario 2 (Case 1):} Coherence accuracy under different scales and the number of time points $T$ for channel $(p,q)=(1,2)$.}
\label{tab:coh_msb_mse_s2c1}

\small
\setlength{\tabcolsep}{2.5pt}
\renewcommand{\arraystretch}{1.15}

\begin{tabular}{L{2.1cm}!{\vrule width 1.1pt} *{12}{S[table-format=1.4]}}
\toprule
\multicolumn{1}{c!{\vrule width 1.1pt}}{\DiagCell{2.1cm}{Scale}{Estimator}} &
\multicolumn{3}{c}{\textbf{MvLSW}} &
\multicolumn{3}{c}{\textbf{Process-based}} &
\multicolumn{3}{c}{\textbf{Subprocess (approx)}} &
\multicolumn{3}{c}{\textbf{Subprocess (true)}} \\
\cmidrule(lr){2-4}\cmidrule(lr){5-7}\cmidrule(lr){8-10}\cmidrule(lr){11-13}
& \textbf{$T{=}512$} & \textbf{$T{=}1024$} & \textbf{$T{=}4096$}
& \textbf{$T{=}512$} & \textbf{$T{=}1024$} & \textbf{$T{=}4096$}
& \textbf{$T{=}512$} & \textbf{$T{=}1024$} & \textbf{$T{=}4096$}
& \textbf{$T{=}512$} & \textbf{$T{=}1024$} & \textbf{$T{=}4096$} \\
\midrule

\multicolumn{13}{l}{\textbf{Panel A: MSB} (reported in units of $10^{-4}$)}\\
\addlinespace[2pt]
$j_1=1$      & {55.09} & {53.29} & {40.27} & {53.05} & {43.16} & {27.59} & {18.50} & {17.57} & {14.82} & {4.76} & {3.55} & {1.54} \\
$j_2=2$      & {18.94} & {8.46} & {3.63} & {8.87} & {7.26} & {5.56} & {13.37} & {17.28} & {13.85} & {4.45} & {2.75} & {1.32} \\
\addlinespace
$(j_1,j_2)=(1,2)$
& \multicolumn{3}{c}{\textit{N/A}}
& {29.94} & {6.56} & {4.68}
& {1282.88} & {1192.38} & {1157.92}
& {24.72} & {15.75} & {12.56} \\
\addlinespace[4pt]

\multicolumn{13}{l}{\textbf{Panel B: MSE} (original scale)}\\
\addlinespace[2pt]
$j_1=1$      & {0.074} & {0.066} & {0.052}
          & {0.082} & {0.051} & {0.021}
          & {0.015} & {0.014} & {0.007}
          & {0.029} & {0.018} & {0.009} \\
$j_2=2$      & {0.108} & {0.094} & {0.078}
          & {0.126} & {0.077} & {0.030}
          & {0.022} & {0.015} & {0.011}
          & {0.043} & {0.028} & {0.014} \\
\addlinespace
$(j_1,j_2)=(1,2)$
& \multicolumn{3}{c}{\textit{N/A}}
& {0.853} & {0.477} & {0.200}
& {0.181} & {0.153} & {0.132}
& {0.106} & {0.061} & {0.030} \\
\bottomrule
\end{tabular}

\begin{tablenotes}[flushleft]
\footnotesize
\item Notes: Panel A reports MSB in units of $10^{-4}$ (multiply by $10^{-4}$ to recover MSB on the original scale). Panel B reports MSE on the original scale. The Subprocess (true) columns serve as an oracle baseline, corresponding to the idealized case in which the true subprocess construction is known, and are included as a reference rather than as a competing implementable estimator. 
\end{tablenotes}
\end{threeparttable}
\end{sidewaystable}

\begin{sidewaystable}[p]
\centering
\begin{threeparttable}
\caption{\textbf{Scenario 2 (Case 2):} Coherence accuracy under different scales and the number of time points $T$ for channel $(p,q)=(1,2)$.}
\label{tab:coh_msb_mse_s2c2}

\small
\setlength{\tabcolsep}{2.5pt}
\renewcommand{\arraystretch}{1.15}

\begin{tabular}{L{2.1cm}!{\vrule width 1.1pt} *{12}{S[table-format=1.4]}}
\toprule
\multicolumn{1}{c!{\vrule width 1.1pt}}{\DiagCell{2.1cm}{Scale}{Estimator}} &
\multicolumn{3}{c}{\textbf{MvLSW}} &
\multicolumn{3}{c}{\textbf{Process-based}} &
\multicolumn{3}{c}{\textbf{Subprocess (approx)}} &
\multicolumn{3}{c}{\textbf{Subprocess (true)}} \\
\cmidrule(lr){2-4}\cmidrule(lr){5-7}\cmidrule(lr){8-10}\cmidrule(lr){11-13}
& \textbf{$T{=}512$} & \textbf{$T{=}1024$} & \textbf{$T{=}4096$}
& \textbf{$T{=}512$} & \textbf{$T{=}1024$} & \textbf{$T{=}4096$}
& \textbf{$T{=}512$} & \textbf{$T{=}1024$} & \textbf{$T{=}4096$}
& \textbf{$T{=}512$} & \textbf{$T{=}1024$} & \textbf{$T{=}4096$} \\
\midrule

\multicolumn{13}{l}{\textbf{Panel A: MSB} (reported in units of $10^{-4}$)}\\
\addlinespace[2pt]
$j_1=3$      & {1905.39} & {1971.43} & {2186.95} & {379.33} & {245.56} & {118.10} & {28.89} & {30.71} & {18.81} & {7.62} & {3.05} & {1.89} \\
$j_2=4$      & {1922.06} & {2019.68} & {2064.86} & {154.62} & {107.27} & {9.65} & {10.22} & {6.52} & {3.52} & {7.76} & {7.26} & {2.66} \\
\addlinespace
$(j_1,j_2)=(3,4)$
& \multicolumn{3}{c}{\textit{N/A}}
& {2165.13} & {500.82} & {80.72}
& {933.24} & {912.30} & {763.79}
& {49.74} & {22.39} & {7.85} \\
\addlinespace[4pt]

\multicolumn{13}{l}{\textbf{Panel B: MSE} (original scale)}\\
\addlinespace[2pt]
$j_1=3$      & {0.612} & {0.628} & {0.681}
          & {0.337} & {0.257} & {0.144}
          & {0.071} & {0.047} & {0.022}
          & {0.113} & {0.070} & {0.030} \\
$j_2=4$      & {0.658} & {0.674} & {0.696}
          & {0.403} & {0.323} & {0.162}
          & {0.135} & {0.090} & {0.042}
          & {0.224} & {0.149} & {0.062} \\
\addlinespace
$(j_1,j_2)=(3,4)$
& \multicolumn{3}{c}{\textit{N/A}}
& {1.292} & {0.881} & {0.672}
& {0.271} & {0.202} & {0.125}
& {0.503} & {0.233} & {0.096} \\
\bottomrule
\end{tabular}

\begin{tablenotes}[flushleft]
\footnotesize
\item Notes: Panel A reports MSB in units of $10^{-4}$ (multiply by $10^{-4}$ to recover MSB on the original scale). Panel B reports MSE on the original scale. The Subprocess (true) columns serve as an oracle baseline, corresponding to the idealized case in which the true subprocess construction is known, and are included as a reference rather than as a competing implementable estimator.
\end{tablenotes}
\end{threeparttable}
\end{sidewaystable}

\begin{sidewaystable}[p]
\centering
\begin{threeparttable}
\caption{\textbf{Scenario 2 (Case 3):} Coherence accuracy under different scales and the number of time points $T$ for channel $(p,q)=(1,2)$.}
\label{tab:coh_msb_mse_s2c3}

\small
\setlength{\tabcolsep}{2.5pt}
\renewcommand{\arraystretch}{1.15}


\begin{tabular}{L{2.1cm}!{\vrule width 1.1pt} *{12}{S[table-format=1.4]}}
\toprule
\multicolumn{1}{c!{\vrule width 1.1pt}}{\DiagCell{2.1cm}{Scale}{Estimator}} &
\multicolumn{3}{c}{\textbf{MvLSW}} &
\multicolumn{3}{c}{\textbf{Process-based}} &
\multicolumn{3}{c}{\textbf{Subprocess (approx)}} &
\multicolumn{3}{c}{\textbf{Subprocess (true)}} \\
\cmidrule(lr){2-4}\cmidrule(lr){5-7}\cmidrule(lr){8-10}\cmidrule(lr){11-13}
& \textbf{$T{=}512$} & \textbf{$T{=}1024$} & \textbf{$T{=}4096$}
& \textbf{$T{=}512$} & \textbf{$T{=}1024$} & \textbf{$T{=}4096$}
& \textbf{$T{=}512$} & \textbf{$T{=}1024$} & \textbf{$T{=}4096$}
& \textbf{$T{=}512$} & \textbf{$T{=}1024$} & \textbf{$T{=}4096$} \\
\midrule

\multicolumn{13}{l}{\textbf{Panel A: MSB} (reported in units of $10^{-4}$)}\\
\addlinespace[2pt]
$j_1=1$      & {454.43} & {514.18} & {567.31} & {3.81} & {3.25} & {2.26} & {4.18} & {3.30} &{1.46} & {4.69} & {3.67} & {1.59}  \\
$j_2=4$      & {1614.51} & {1678.38} & {1727.72} & {128.68} & {68.20} & {21.44} & {22.88} & {16.31} & {8.08} & {10.44} & {6.88} & {3.00} \\
\addlinespace
$(j_1,j_2)=(1,4)$
& \multicolumn{3}{c}{\textit{N/A}}
& {587.39} & {197.81} & {38.19}
& {624.32} & {691.40} & {678.93}
& {78.44} & {93.10} & {51.61} \\
\addlinespace[4pt]

\multicolumn{13}{l}{\textbf{Panel B: MSE} (original scale)}\\
\addlinespace[2pt]
$j_1=1$      & {0.286} & {0.299} & {0.308}
          & {0.034} & {0.021} & {0.011}
          & {0.031} & {0.018} & {0.008}
          & {0.033} & {0.020} & {0.008} \\
$j_2=4$      & {0.573} & {0.589} & {0.599}
          & {0.284} & {0.200} & {0.088}
          & {0.154} & {0.099} & {0.048}
          & {0.104} & {0.070} & {0.035} \\
\addlinespace
$(j_1,j_2)=(1,4)$
& \multicolumn{3}{c}{\textit{N/A}}
& {1.415} & {0.648} & {0.532}
& {0.723} & {0.437} & {0.210}
& {0.781} & {0.450} & {0.175} \\
\bottomrule
\end{tabular}%
 
\begin{tablenotes}[flushleft]
\footnotesize
\item Notes: Panel A reports MSB in units of $10^{-4}$ (multiply by $10^{-4}$ to recover MSB on the original scale). Panel B reports MSE on the original scale. The Subprocess (true) columns serve as an oracle baseline, corresponding to the idealized case in which the true subprocess construction is known, and are included as a reference rather than as a competing implementable estimator.
\end{tablenotes}
\end{threeparttable}
\end{sidewaystable}

\begin{figure}[t]
\centering

\begin{minipage}{0.48\textwidth}
  \centering
  \includegraphics[width=\linewidth]{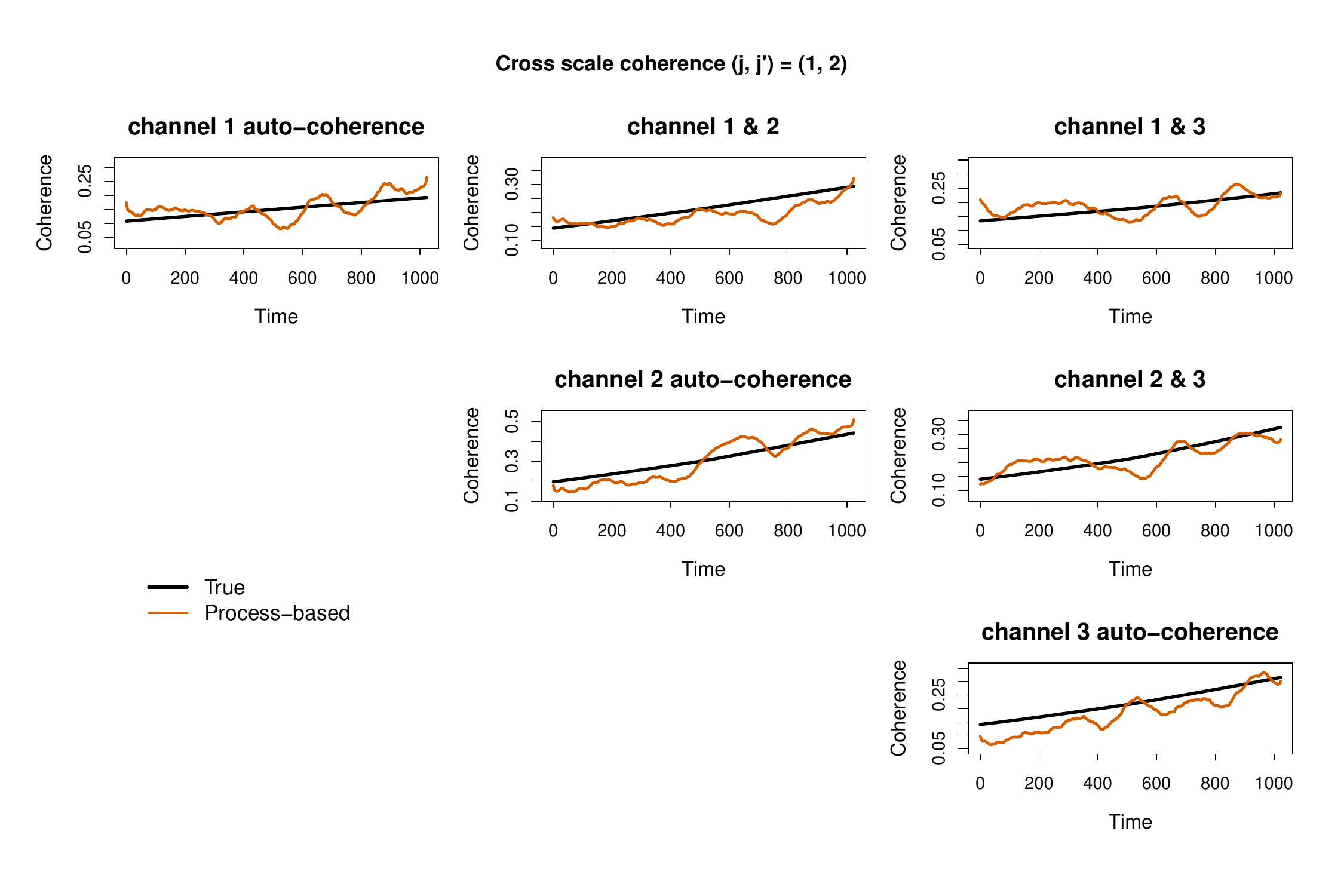}
  \par\smallskip
\end{minipage}\hfill
\begin{minipage}{0.48\textwidth}
  \centering
  \includegraphics[width=\linewidth]{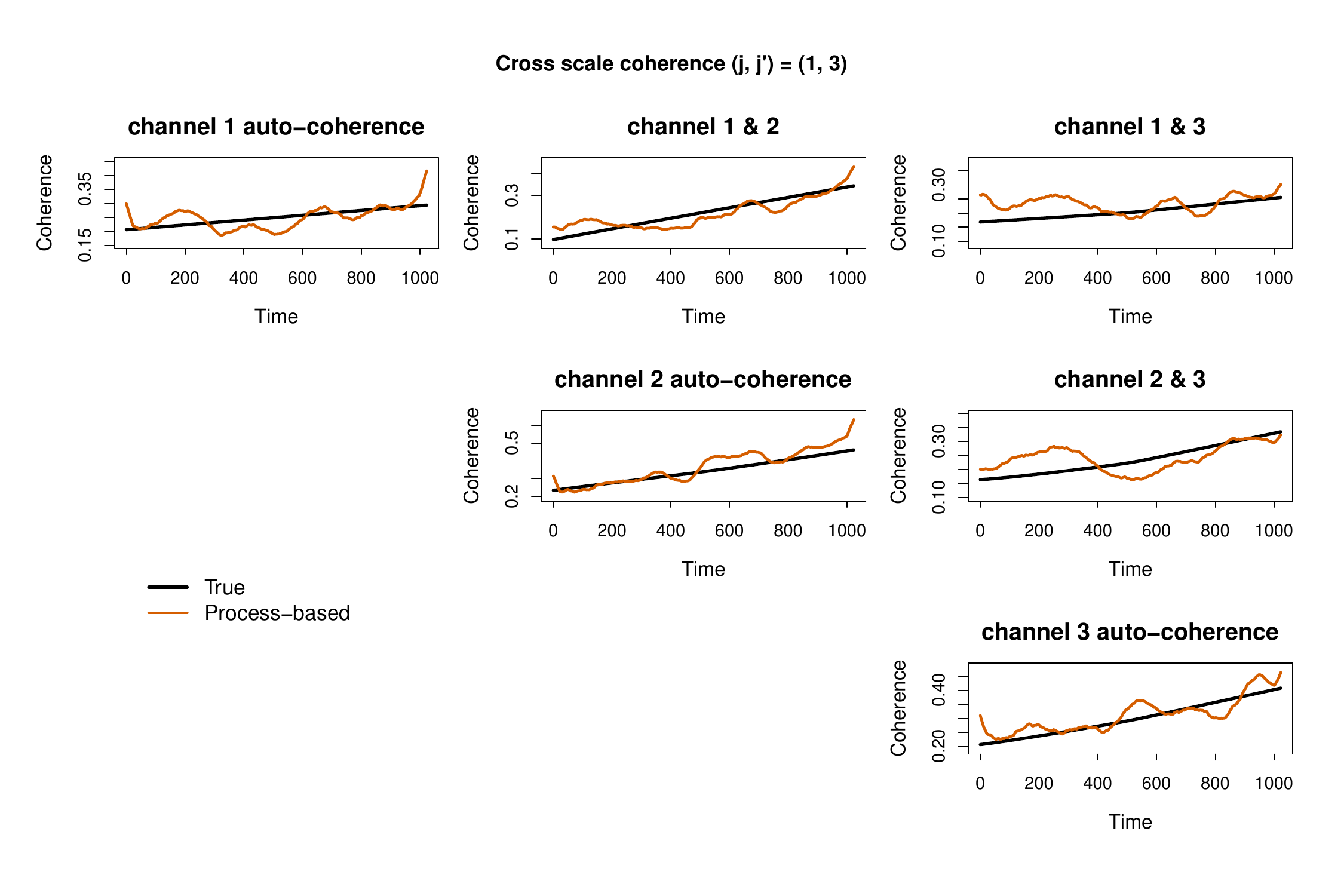}
  \par\smallskip
\end{minipage}

\vspace{0.6em}

\begin{minipage}{0.48\textwidth}
  \centering
  \includegraphics[width=\linewidth]{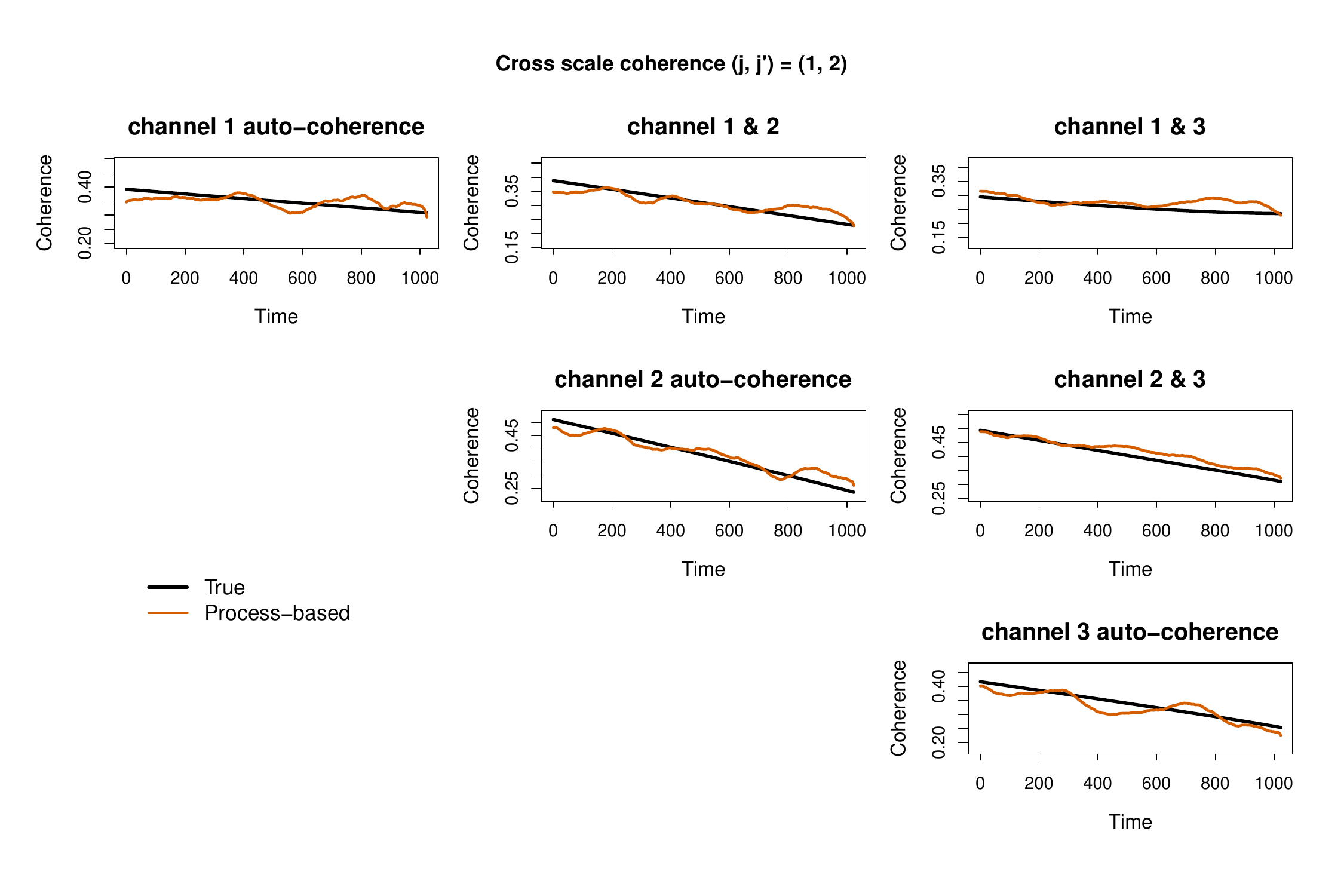}
  \par\smallskip
\end{minipage}\hfill
\begin{minipage}{0.48\textwidth}
  \centering
  \includegraphics[width=\linewidth]{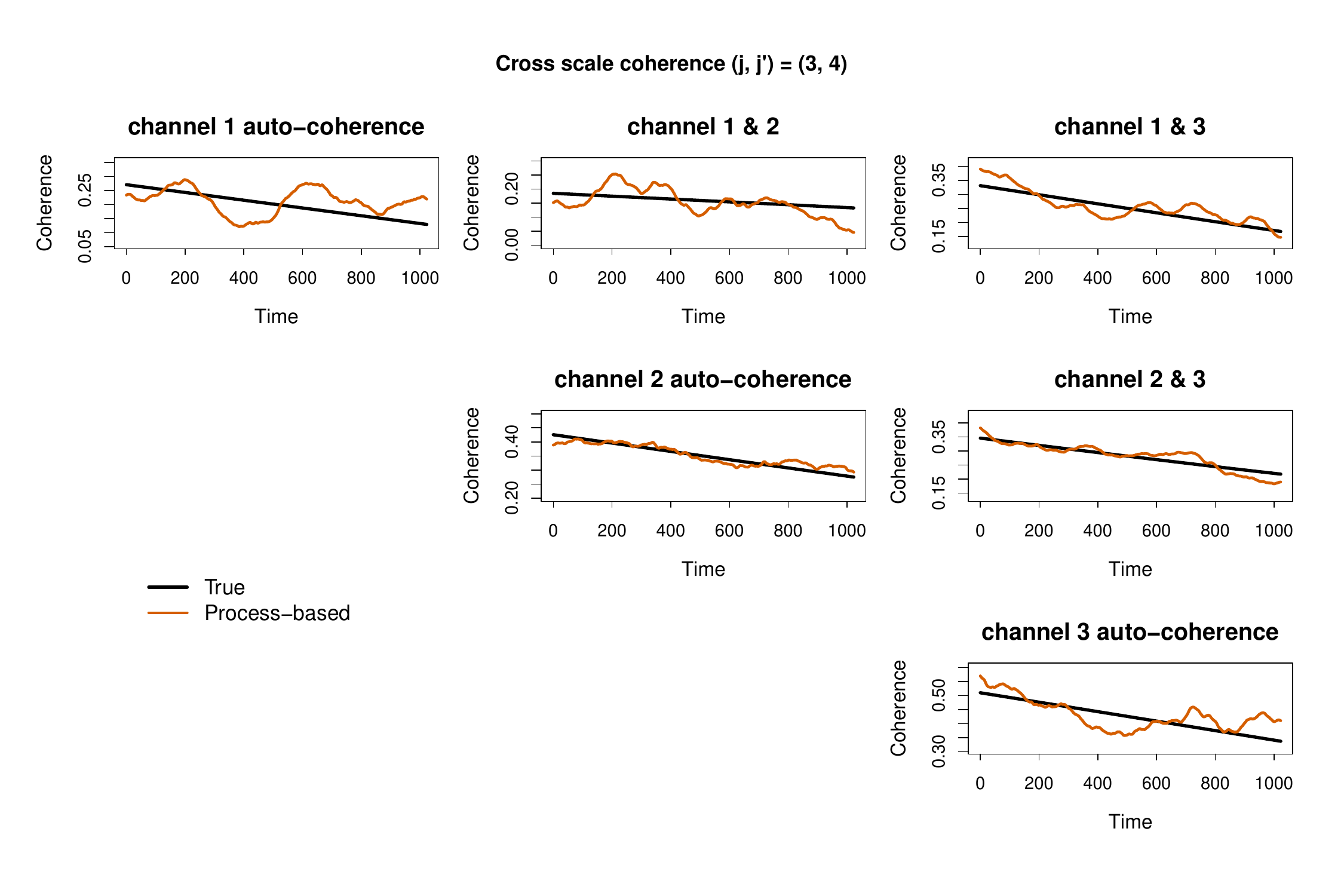}
  \par\smallskip
\end{minipage}

\caption{Cross-scale coherence estimates under the multiscale settings in detailed in Appendix~\ref{app:sim}. 
Top row: Scenario 3 (three-scale; $(j,j')=(1,2)$ and $(1,3)$). 
Bottom row: Scenario 4 (four-scale; $(j,j')=(1,2)$ and $(3,4)$). 
Black curves show the truth; coloured curves show the process-based estimates.}
\label{fig:multiple_scales}
\end{figure}

To further assess performance beyond the two-scale designs in the main text, we consider the multiscale settings described in Appendix \ref{app:sim}: Scenario 3 with non-zero spectra at scales $j=1,2,3$ and concurrent cross-scale links $(1,2)$ and $(1,3)$, and Scenario 4 with non-zero spectra at $j=1,2,3,4$ and separated cross-scale links $(1,2)$ and $(3,4)$. Figure \ref{fig:multiple_scales} shows the resulting estimated cross-scale coherence curves closely follow the truth across time and channel pairs, demonstrating that the estimator continues to perform well when multiple cross-scale links coexist.

\clearpage
\section{Appendix C (Supplementary EEG data analysis)} \label{app: add eeg results}
\begin{figure}[htbp]
  \centering
    \includegraphics[width=.82\linewidth]{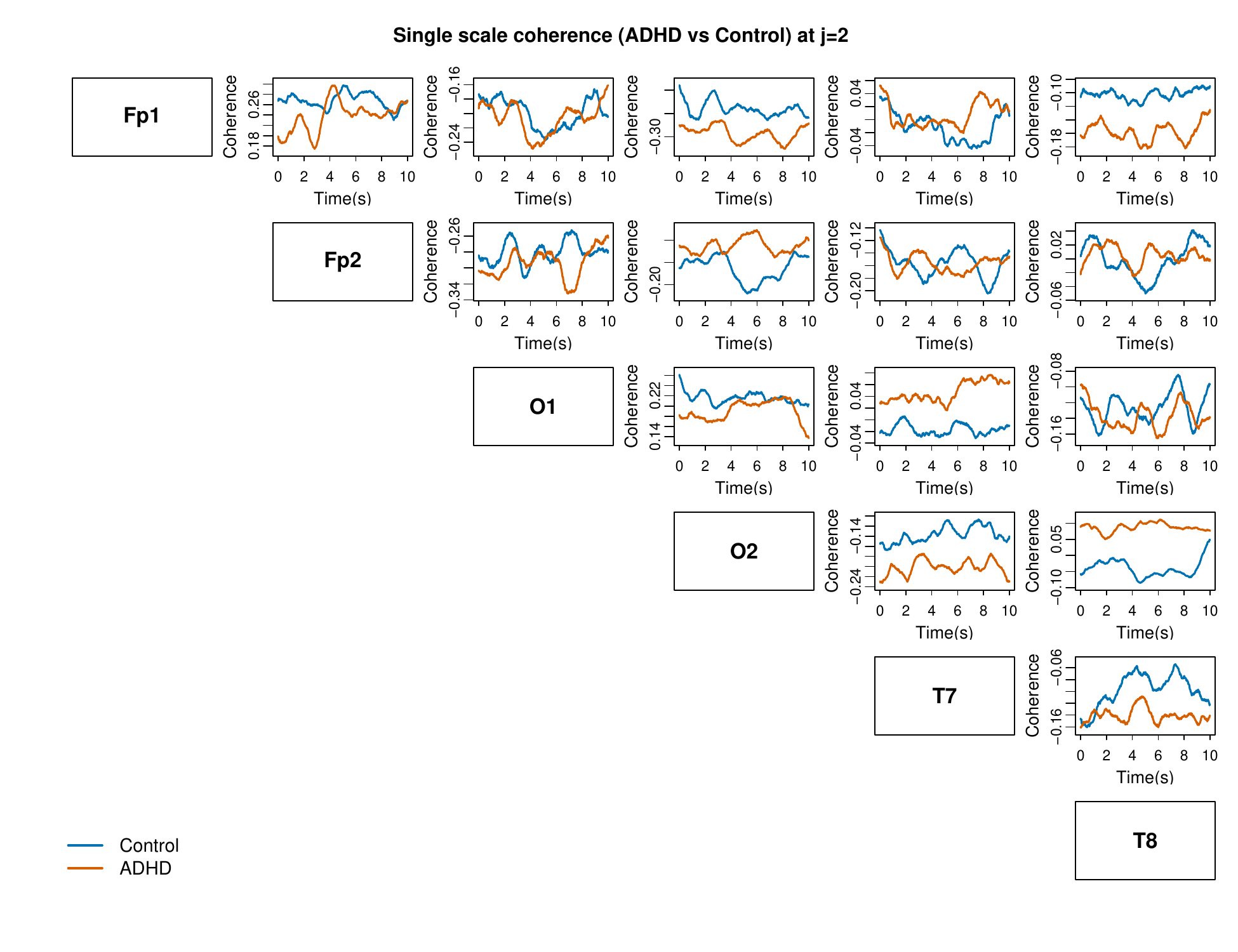}
    \hfill 
    \includegraphics[width=.82\linewidth]{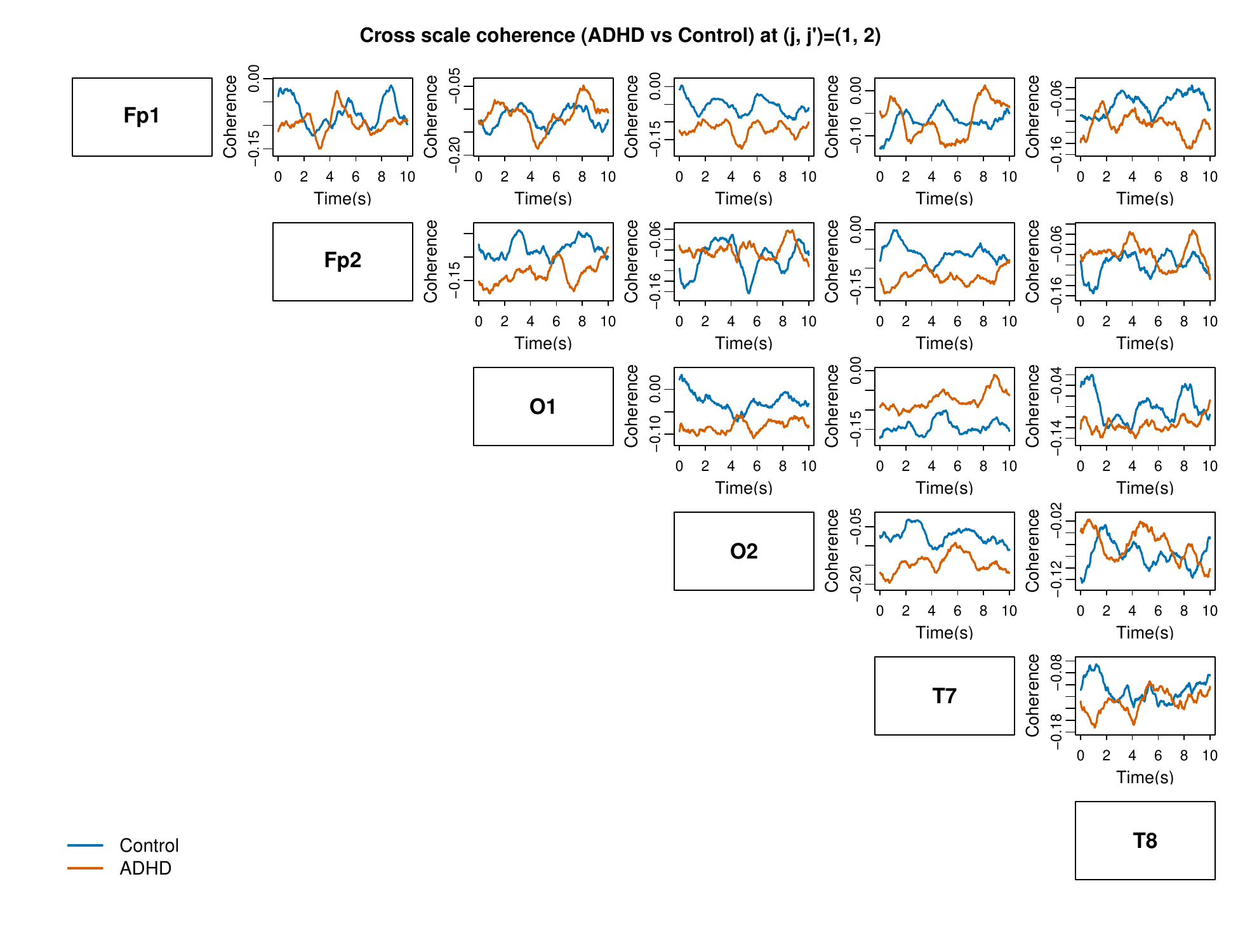}
  \caption{Averaged EEG time-varying single-scale (subprocess-based estimator) wavelet coherence at $j=2$ (approximately 16--32  Hz; top) and cross-scale (process-based estimator) wavelet coherence between $(j,j')=(1,2)$, corresponding to the 32--64  Hz and 16--32  Hz bands, respectively (bottom), comparing ADHD and control subjects.}
  \label{fig:eeg_coh}
\end{figure}

In this section, we provide visualisations of single- and cross-scale coherence across the six EEG channels. The results in both panels of Figure~\ref{fig:eeg_coh} are discussed in the main text. 

\begin{figure}[htbp]
\centering

\begin{minipage}{0.48\textwidth}
  \centering
  \includegraphics[width=\linewidth]{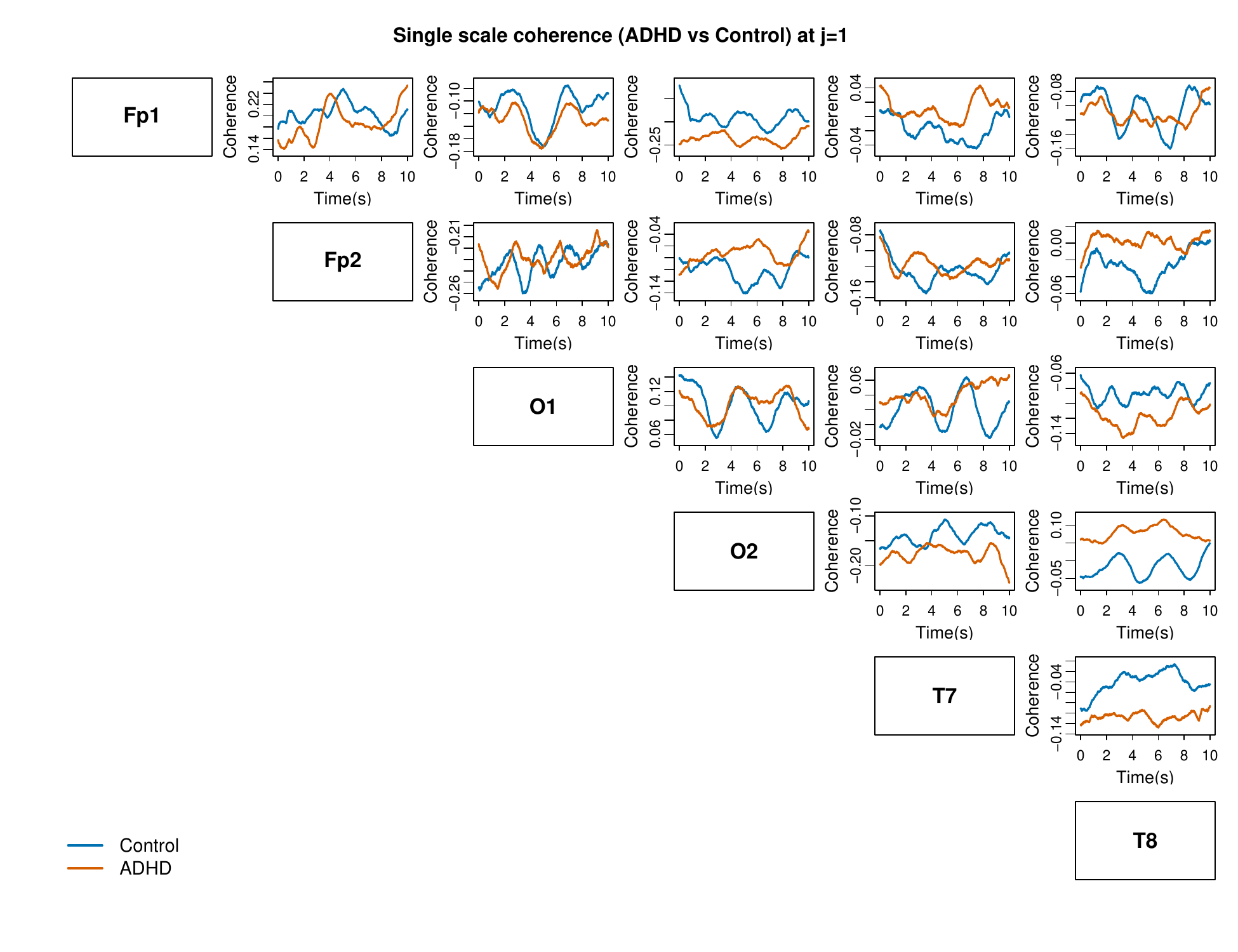}
  \par\smallskip
\end{minipage}\hfill
\begin{minipage}{0.48\textwidth}
  \centering
  \includegraphics[width=\linewidth]{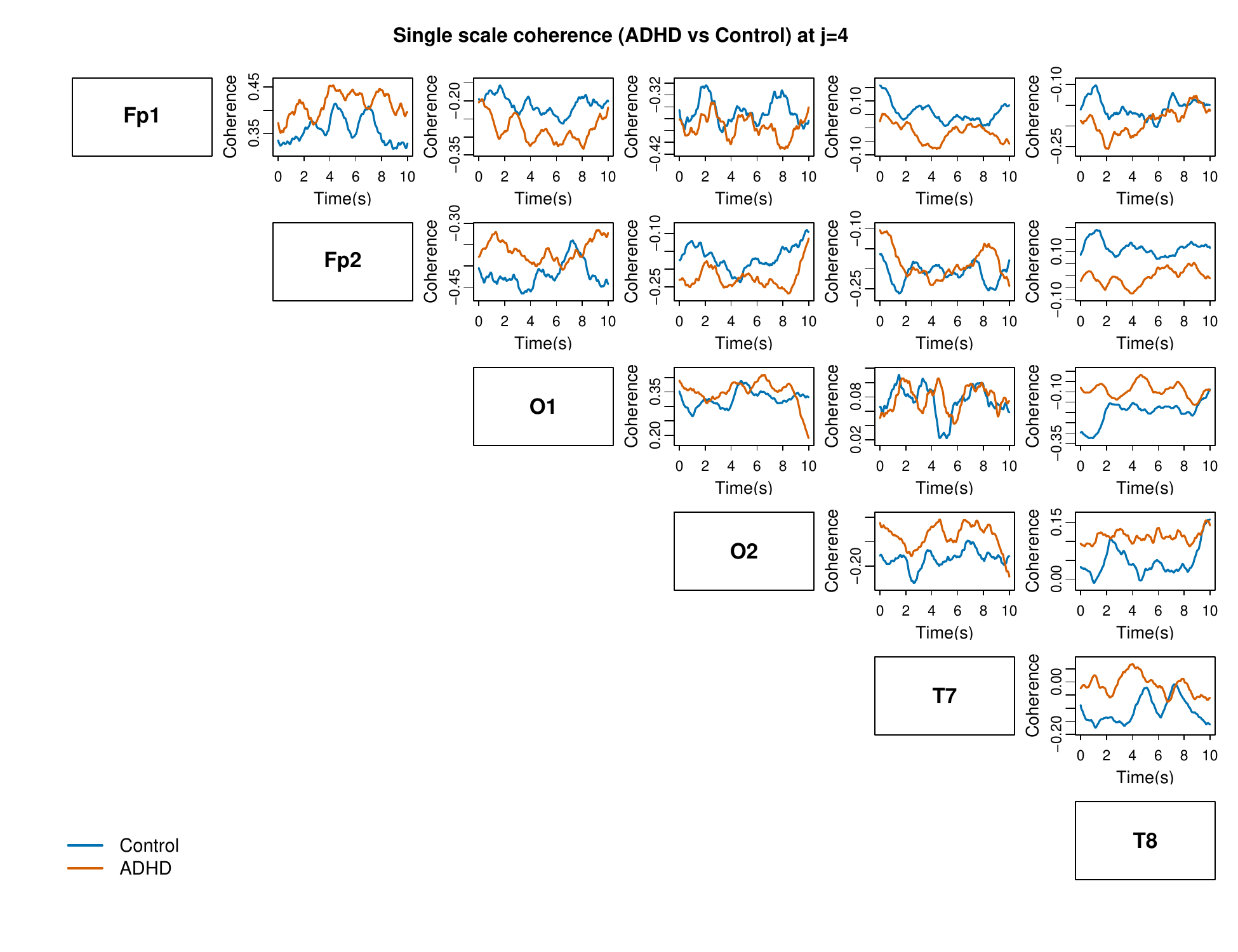}
  \par\smallskip
\end{minipage}

\vspace{0.6em}

\begin{minipage}{0.48\textwidth}
  \centering
  \includegraphics[width=\linewidth]{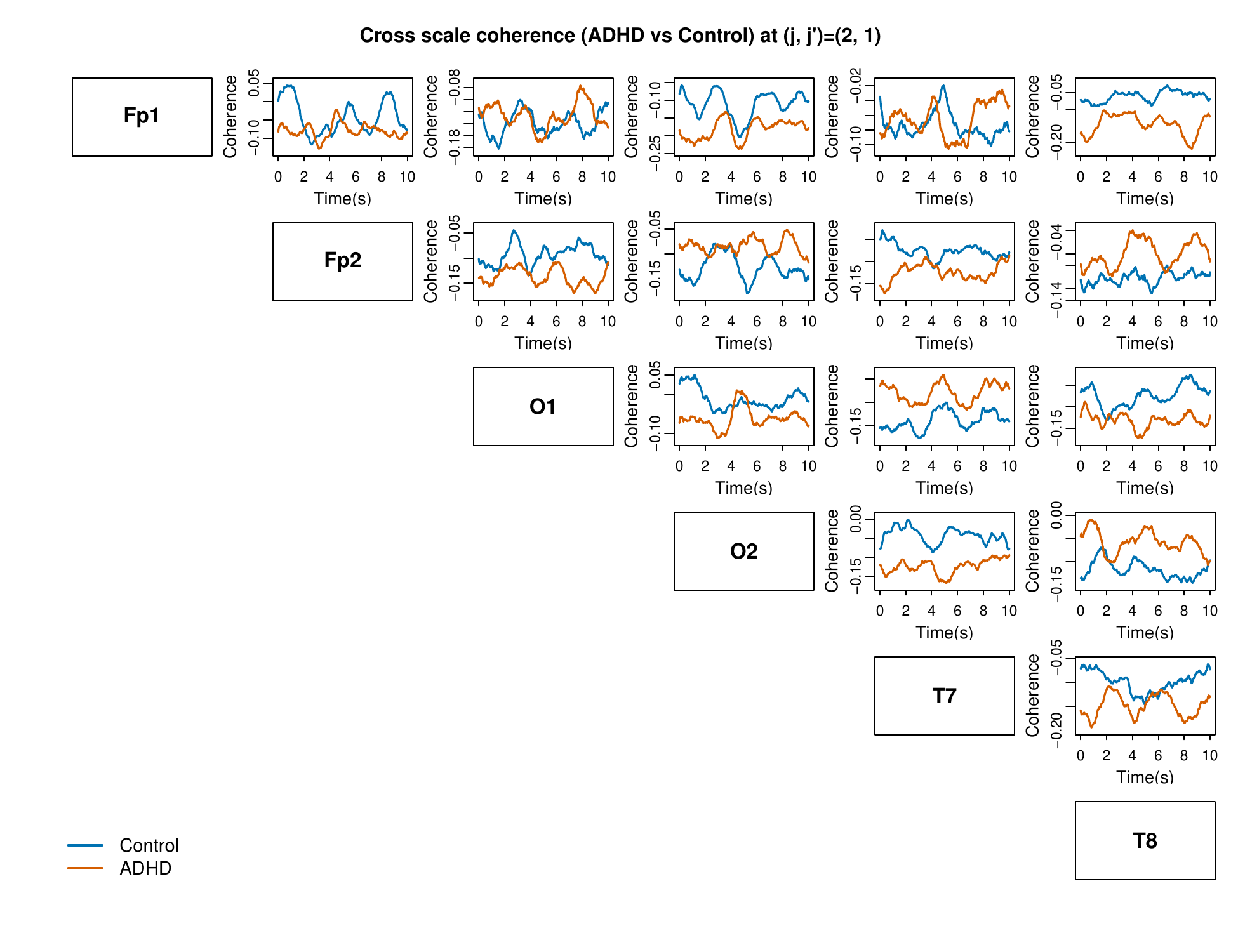}
  \par\smallskip
\end{minipage}\hfill
\begin{minipage}{0.48\textwidth}
  \centering
  \includegraphics[width=\linewidth]{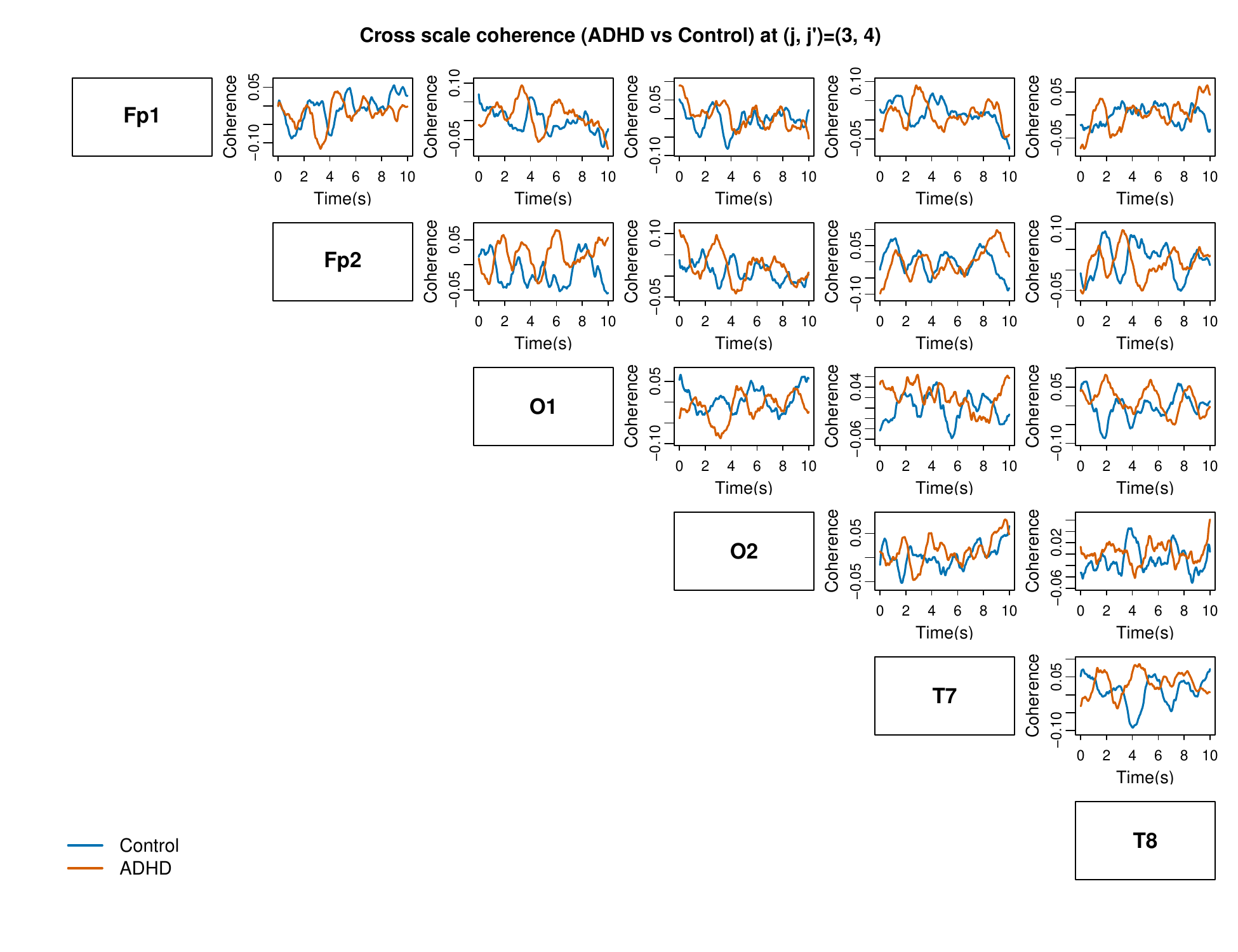}
  \par\smallskip
\end{minipage}

\caption{Estimated EEG coherence. Top row: averaged time-varying single-scale (subprocess-based estimator) wavelet coherence at $j=1$ (approximately 32--64 Hz; left) and $j=4$ (approximately 4--8 Hz; right); Bottom row: cross-scale (process-based estimator) wavelet coherence between $(j,j')=(2,1)$, corresponding to the 16--32 Hz and 32--64 Hz bands, respectively (left), $(j,j')=(3,4)$, corresponding to the 8--16 Hz and 4--8 Hz bands, respectively (right).}
\label{fig:eeg_addition}
\end{figure}

To complement Section \ref{EEG analysis}, we report additional group-averaged coherence curves for scale pairs not shown in Figure \ref{fig:eeg_coh}.
Figure \ref{fig:eeg_addition} (top row) displays single-scale coherence at the highest-frequency scale $j=1$ (approximately 32--64 Hz) and the lowest-frequency scale $j=4$ (approximately 4--8 Hz), providing a broader view of single-scale dependence across the selected channels. Compared with the beta band results emphasized in the main text, the single-scale $j=1$ coherence is generally more intermittent and rapidly varying, whereas the $j=4$ coherence exhibits smoother temporal evolution, consistent with slower low-frequency neural dynamics.
Figure \ref{fig:eeg_addition} (bottom row) presents cross-scale coherence for two additional scale pairs. The panel with $(j,j')=(2,1)$ corresponds to the same beta--gamma interaction as in Figure~\ref{fig:eeg_coh} but with reversed scale ordering, serving as a robustness check that the observed cross-scale group differences are not an artefact of presenting $(j,j')=(1,2)$. The panel with $(j,j')=(3,4)$ summarizes lower-frequency cross-scale coupling (approximately 8--16 Hz versus 4--8 Hz), showing that cross-frequency dependence is also present at slower rhythms, although the group separation is typically less pronounced than that observed for beta--gamma interactions in the main analysis.

We provide additional information about the permutation test conducted in Section \ref{EEG analysis}.
For each channel pair and each dual-scale pair $(j,j')$, we test whether the ADHD and control groups (labelled below as `(A)' and `(C)', respectively) have the same population mean cross-scale coherence curve. Let
$\hat{\rho}_{i,jj'}(t/T)$ denote the estimated cross-scale coherence curve for subject $i$ in group `G' at dual scales $(j,j')$, and $\hat{\overline{\rho}}^{(G)}_{jj'}(t/T)$ denote the group mean. 

With $n_A=n_C=50$, the null and alternative hypotheses are
\begin{align*}
H_0:\ \rho^{(A)}_{jj'}(u)=\rho^{(C)}_{jj'}(u)\ \text{for all } u\in(0,1),
\qquad
H_1:\ \rho^{(A)}_{jj'}(u)\neq \rho^{(C)}_{jj'}(u)\ \text{for some } u\in(0,1),  
\end{align*}
where $\rho^{(A)}_{jj'}(u)$ and $\rho^{(C)}_{jj'}(u)$ denote the population mean cross-scale coherence curves for the ADHD and control groups, respectively.

The observed test statistic is defined as the integrated squared difference between the two group-average coherence curves,
\begin{align*}
T_{\mathrm{obs}}
=
\sum_t
\left\{
\hat{\overline{\rho}}^{(A)}_{jj'}(t/T)
-
\hat{\overline{\rho}}^{(C)}_{jj'}(t/T)
\right\}^2 .
\end{align*}
Under $H_0$, the group labels are exchangeable across subjects. We therefore pool the 100 subject-level coherence curves and randomly reassign 50 curves to the ADHD group and 50 curves to the control group, while keeping each subject's entire coherence curve intact. For the $r$th permutation, we recompute the two group-average curves and calculate
\begin{align*}
T_{\mathrm{perm}}^{(r)}
=
\sum_t
\left\{
\hat{\overline{\rho}}^{(A,r)}_{jj'}(t/T)
-
\hat{\overline{\rho}}^{(C,r)}_{jj'}(t/T)
\right\}^2 ,
\qquad r=1,\ldots,R,
\end{align*}
with $R=10{,}000$. The permutation $p$-value is then computed as
\begin{align*}
p
=
\frac{
1+\sum_{r=1}^{R}
\mathbf{1}\{T_{\mathrm{perm}}^{(r)}\geq T_{\mathrm{obs}}\}
}{
R+1
}.   
\end{align*}
This procedure evaluates whether the observed between-group difference in the full time-varying cross-scale coherence curve is larger than expected under random group assignment. The resulting permutation-based $p$-values are adjusted across multiple channel-pair comparisons using the FDR procedure. The statistic $T_{\mathrm{obs}}$ is a two-sided global discrepancy measure and therefore does not by itself indicate which group has stronger coherence. For channel pairs and dual-scale pairs showing a significant group difference after FDR correction, we determine the direction of the effect by comparing the time-averaged coherence magnitude between the two groups. Specifically, we compute
\begin{align*}
D_{jj'}
=
\frac{1}{T}
\sum_t
\left\{
\left|
\hat{\overline{\rho}}^{(A)}_{jj'}(t/T)
\right|
-
\left|
\hat{\overline{\rho}}^{(C)}_{jj'}(t/T)
\right|
\right\}.
\end{align*}
A positive value of $D_{jj'}$ indicates that the ADHD group has larger average dual-scale coherence magnitude for that channel pair and dual-scale pair, whereas a negative value indicates stronger coherence in the control group. Thus, the permutation test is used to assess whether a statistically significant group difference exists, while $D_{jj'}$ is used to summarize the direction and relative strength of that difference, see findings  encapsulated in Figure~\ref{fig:dynamics}. The results reveal several significant differences between the ADHD and control groups, involving both single-scale and cross-scale coherency across different brain regions, suggesting that, when compared to healthy controls, the ADHD group is associated with altered within-scale and dual-scale connectivity patterns that vary across regional networks.
\end{document}